\documentclass{article}

\usepackage{arxiv}

\usepackage[utf8]{inputenc} 
\usepackage[T1]{fontenc}    
\usepackage{hyperref}       

\hypersetup{
    colorlinks=true,
    linkcolor=blue,
    urlcolor=blue,
    citecolor=blue,
    anchorcolor=blue
}

\usepackage{url}            
\usepackage{booktabs}       
\usepackage{amsfonts}       
\usepackage{nicefrac}       
\usepackage{microtype}      
\usepackage{lipsum}		
\usepackage{graphicx}
\usepackage{natbib}
\usepackage{doi}

\usepackage{graphicx}
\usepackage{amsmath}
\usepackage{float}
\usepackage{longtable}
\usepackage{lipsum}
\usepackage{algorithm}
\usepackage{tabularx}

\title{Annealed Stein Variational Gradient Descent for Improved Uncertainty Estimation in Full-Waveform Inversion}

\author{
    \begin{minipage}[t]{0.45\textwidth}
        \centering
        {\hypersetup{hidelinks}\href{https://orcid.org/0000-0002-2332-8311}{\includegraphics[scale=0.06]{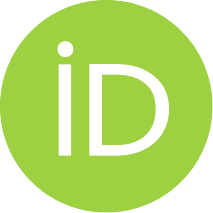}\hspace{1mm}\textbf{Miguel~Corrales}}}\thanks{Equal contribution} \\
        \normalfont
        Physical Science and Engineering Division\\
        KAUST\\
        Thuwal, Saudi Arabia \\
        \texttt{miguel.corrales@kaust.edu.sa} \\
        \vspace{1em} 
        {\includegraphics[scale=0.06]{orcid.pdf}\hspace{1mm}\textbf{Bertrand~Denel}} \\
        \normalfont
        OneTech R\&D\\
        TotalEnergies\\
        Pau, France \\
        \texttt{bertrand.denel@totalenergies.com} \\
        \vspace{1em}
        {\hypersetup{hidelinks}\href{https://orcid.org/0000-0003-1433-0281}{\includegraphics[scale=0.06]{orcid.pdf}\hspace{1mm}\textbf{Mattia~Aleardi}}} \\
        \normalfont
        Department of Earth Sciences\\
        University of Pisa\\
        Pisa, Italy \\
        \texttt{mattia.aleardi@unifi.it}
    \end{minipage}%
    \hfill 
    \begin{minipage}[t]{0.45\textwidth}
        \centering
        {\hypersetup{hidelinks}\href{https://orcid.org/0009-0003-7894-5450}{\includegraphics[scale=0.06]{orcid.pdf}\hspace{1mm}\textbf{Sean~Berti}}}\footnotemark[1] \\
        \normalfont
        Department of Earth Sciences\\
        University of Pisa\\
        Pisa, Italy \\
        \texttt{sean.berti@unifi.it} \\
        \vspace{1em} 
        {\hypersetup{hidelinks}\href{https://orcid.org/0000-0001-8109-4632}{\includegraphics[scale=0.06]{orcid.pdf}\hspace{1mm}\textbf{Paul~Williamson}}} \\
        \normalfont
        OneTech R\&D\\
        TotalEnergies\\
        Pau, France \\
        \texttt{paul.williamson@totalenergies.com} \\
        \vspace{1em}
        {\hypersetup{hidelinks}\href{https://orcid.org/0000-0003-1433-0281}{\includegraphics[scale=0.06]{orcid.pdf}\hspace{1mm}\textbf{Matteo~Ravasi}}} \\
        Physical Science and Engineering Division\\
        KAUST\\
        Thuwal, Saudi Arabia \\
        \texttt{matteo.ravasi@kaust.edu.sa}
    \end{minipage}
}

\renewcommand{\shorttitle}{Annealed SVGD for Improved Uncertainty Estimation in FWI}

\hypersetup{
pdftitle={Annealed SVGD for Improved Uncertainty Estimation in FWI},
pdfsubject={Geophysics},
pdfauthor={Miguel Corrales, Sean Berti, Bertrand Denel, Paul Williamson, Mattia Aleardi, Matteo Ravasi},
pdfkeywords={Inverse Theory, Waveform Inversion, Probability distributions},
}

\begin{document}
\maketitle

\begin{abstract}
In recent years, Full-Waveform Inversion (FWI) has been extensively used to derive high-resolution subsurface velocity models from seismic data. However, due to the nonlinearity and ill-posed nature of the problem, FWI requires a good starting model to avoid producing non-physical solutions (i.e., being trapped in local minima). Moreover, conventional optimization methods fail to quantify the uncertainty associated with the recovered solution, which is critical for decision-making processes.
Bayesian inference offers an alternative approach as it directly or indirectly evaluates the posterior probability density function using Bayes' theorem. For example, Markov Chain Monte Carlo (MCMC) methods generate multiple sample chains to characterize the solution's uncertainty. Despite their ability to theoretically handle any form of distribution, MCMC methods require many sampling steps; this limits their usage in high-dimensional problems with computationally intensive forward modeling, as is the FWI case. Variational Inference (VI), on the other hand, provides an approximate solution to the posterior distribution in the form of a parametric or non-parametric proposal distribution. Among the various algorithms used in VI, Stein Variational Gradient Descent (SVGD) is recognized for its ability to iteratively refine a set of samples (commonly defined as particles) to approximate the target distribution through an optimization process. However, mode and variance-collapse issues affect SVGD in high-dimensional inverse problems. This study aims to improve the performance of SVGD within the context of FWI by utilizing, for the first time, an annealed variant of the SVGD algorithm and combining it with a multi-scale strategy, a common practice in deterministic FWI settings. Additionally, we demonstrate that Principal Component Analysis (PCA) can be used to evaluate the performance of the optimization process and gain insights into the behavior of the produced particles and their overall distribution. Clustering techniques are also employed to provide more rigorous and meaningful statistical analysis of the particles in the presence of multi-modal distributions (as is usually the case in FWI). Numerical tests, performed on a portion of the acoustic Marmousi model using both single and multi-scale frequency ranges, reveal the benefits of annealed SVGD compared to vanilla SVGD to enhance uncertainty estimation using a limited number of particles and thus address the challenges of dimensionality and computational constraints. 

\end{abstract}


\keywords{Inverse theory \and Waveform Inversion \and Probability distributions}



\section{Introduction}
Full-Waveform Inversion (FWI) is a high-resolution imaging technique for estimating subsurface parameters from recorded seismic waveform data. Unlike methods that rely solely on the kinematic component of the recorded seismic waveforms (i.e., traveltimes), FWI exploits the entire wavefield information to invert for detailed subsurface models \citep{virieux_overview_2009}. However, the complex and nonlinear relationships between model parameters and seismic data --- coupled with the oscillatory nature of the seismic waveforms, incomplete data coverage, and noise in the data --- renders FWI an ill-posed inverse problem with a non-unique solution. In other words, many sets of model parameters can fit the data equally well within their inherent uncertainties; therefore, it is crucial to quantify the uncertainties to assess the confidence in the inverted models \citep{tarantola_inverse_2005}.

FWI is typically addressed through optimization by minimizing a misfit function (e.g. L2 norm) between the observed and predicted seismograms \citep{lailly_migration_1984, tarantola_inversion_1984}. Due to the highly nonlinear nature of the problem and the multimodal landscape of the objective function, local optimization algorithms often get trapped in local minima. This challenge can be mitigated by enforcing specific requirements on the observed data, such as the presence of low frequencies or long offset, and/or a good starting model. \cite{bozdag_misfit_2011} and \cite{guo_bayesian_2020} have shown that a poor starting model can easily lead the inversion into a local minimum of the objective function, compromising the inversion outcome. Various misfit functions have been proposed to mitigate this dependency \citep{luo_waveequation_1991, brossier_which_2010, warner_adaptive_2014, metivier_measuring_2016, sambridge_monte_2002}. Yet, when deterministic optimization algorithms are used alongside these misfit functions, no information about the uncertainty of the solution is provided.

Bayesian inference represents an alternative to deterministic optimization methods as it provides a probabilistic approach to quantify the uncertainty of the inverted models \citep{mosegaard_16_2002, sambridge_monte_2002}. Bayesian methods embody Bayes' theorem to update our prior knowledge about the model parameters with new information obtained from the observed data. In this case, the posterior probability distribution (PPD), or the number of samples taken from such a distribution, represents the solution of the inversion process, thus Bayesian methods offer a more comprehensive solution describing all parameter values consistent with the observed data and quantifying their relative probabilities. Markov Chain Monte Carlo (MCMC) methods are commonly employed to characterize the PPD by constructing multiple chains of successive samples from the target posterior distribution through structured random walks in the parameter space. These samples form the basis for inferring valuable statistics of the PPD, and thereby enable the estimation of uncertainties that affect the recovered solution. One such method, the random walk Metropolis algorithm, has been applied across various geophysical problems, including electrical resistivity inversion \citep{malinverno_parsimonious_2002}, traveltime tomography \citep{bodin_seismic_2009}, and gravity inversion \citep{mosegaard_16_2002}. However, this algorithm faces significant computational challenges, as the curse of dimensionality \citep{curtis_prior_2001} restricts its applicability in high-dimensional problems and computationally expensive forward modeling operators, such as those encountered in FWI.

Over the past decade, by leveraging rapid advances in computing capabilities, researchers have revisited sampling-based methods to solve Bayesian FWI problems, developing sophisticated algorithms that aim to improve the efficiency for large-scale inversions. However, while more computationally feasible than standard MCMC approaches, these methods may also risk losing information due to implicit undersampling in high-dimensional spaces. These include Hamiltonian Monte Carlo (HMC) \citep{fichtner_hamiltonian_2019, gebraad_bayesian_2020}, stochastic Newton MCMC \citep{martin_stochastic_2012}, parallel tempering \citep{sambridge_parallel_2014}, and gradient-based MCMC \citep{aleardi_gradientbased_2021, zhao_gradient-based_2021, berti_computationally_2024, berti_elastic_2024}. Finally, trans-dimensional MCMC represent another class of MCMC techniques in which the number of model parameters is treated as an additional unknown \citep{bodin_seismic_2009, ray_frequency_2016, sen_transdimensional_2017, guo_bayesian_2020}. Despite their robustness and efficiency, when applied to problems such as FWI, MCMC methods typically require a large number of sampling steps ($>10^6$) and a long burn-in period to achieve precise uncertainty estimations. 

Variational Inference (VI) has emerged as an appealing alternative as it offers greater adaptability for approximating the posterior distributions with significantly lower computational overhead than MCMC \citep{jordan_introduction_1998, blei_variational_2017, zhang_advances_2019}. In VI, a set of simple probability distributions is defined (often called the variational family), and an optimal member of such a family is sought to approximate the true PPD. The Kullback-Leibler (KL) divergence measures the disparity between the two distributions, which enables potentially efficient and parallelizable optimization processes with well-understood convergence criteria. This can be achieved by either directly estimating the free parameters of the chosen distribution that best approximate the true PPD \citep{kucukelbir_automatic_2017, kingma_improved_2016} or by deterministically modifying a set of samples from the proposal distribution to match the PPD \citep{gallego_stochastic_2020, liu_stein_2016}. Variational approaches have been applied to various problems in geophysics, including traveltime tomography \citep{zhang_seismic_2020, zhao_bayesian_2022}, seismic denoising \citep{siahkoohi_preconditioned_2021}, seismic interpolation \citep{ravasi_2023}, earthquake hypocenter inversion \citep{smith_hyposvi_2022}, 2D FWI \citep{zhang_variational_2020, urozayev_reduced-order_2022}, and 3D FWI \citep{lomas_3d_2023, zhang_3-d_2023}.

Recently, particle-based VI methods have emerged to bridge the gap between parametric VI and MCMC techniques. These methods utilize a specific number of samples, or particles, to represent the approximate distribution, akin to MCMC, whilst updating these particles through an optimization process similar to VI. This hybrid approach offers greater flexibility than parametric VI and is more particle-efficient than MCMC, as it fully leverages particle interactions. A notable example of this category is the Stein Variational Gradient Descent (SVGD) method \citep{liu_stein_2016}, which has already been applied to post-stack seismic inversion \citep{izzatullah_posterior_2024}, petrophysical inversion \citep{corrales_bayesian_2022}, and FWI \citep{zhang_3-d_2023, izzatullah_physics-reliable_2024}. This deterministic sampling algorithm iteratively minimizes the Kullback-Leibler (KL) divergence between the chosen approximate distribution and the target density to ensure that the final set of particles is distributed according to the desired posterior distribution. Despite the empirical successes of SVGD, its application to high-dimensional problems remains challenging, as it becomes computationally demanding to sample more particles than there are unknowns. SVGD can suffer from mode- and variance-collapse issues as the dimensionality of the problem increases. More specifically, variance collapse refers to the scenario in which the variance estimated by SVGD is significantly smaller than the true variance of the target distribution \citep{zhuo_message_2018}. This is undesirable because 
underestimation of the variance leads to a failure in explaining the uncertainty of the model predictions, which is a key benefit of Bayesian inference.

This study aims to enhance the performance of SVGD within the FWI framework by replacing the standard SVGD algorithm with an annealed variant. We conduct numerical experiments on a portion of the Marmousi model using both single- and multi-scale frequency approaches to evaluate the effectiveness of these methods to improve uncertainty estimation when working with a limited number of particles. Furthermore, we propose a number of additional strategies to augment our analysis of the particles after the SVGD process. First, Principal Component Analysis (PCA) is used to evaluate the performance of SVGD and gain deeper insights into the behavior and distribution of the particles. In addition to reducing the dimensionality of the solution space, PCA highlights the least confident model combinations, enabling more precise analysis of model uncertainties compared to simply examining the parameter variances (the diagonal of the covariance matrix). However, interpreting these results can be more challenging. Additionally, we employ clustering techniques to identify whether particles converge to distinct modes, allowing for more rigorous and meaningful statistical insights by grouping particles into geological and non-geological clusters. Overall, this research addresses the challenges of high dimensionality and computational constraints by bridging the gap between theoretical potential and practical applications and, thereby,  faster and more informed decision-making.

\section{Theoretical Framework}

FWI aims to estimate subsurface model parameters, such as P-wave velocity, represented by \(\mathbf{m} \in \mathbb{R}^m\), from observed seismic data \(\mathbf{d} \in \mathbb{R}^d\), where \(m\) and \(d\) denote the dimensions of model and data spaces, respectively. In order to capture the uncertainties inherent in this estimation process, any Bayesian estimation algorithms formulates this inverse problem using Bayes' rule:
\begin{equation}
    p(\mathbf{m}|\mathbf{d}) = \frac{p(\mathbf{d}|\mathbf{m}) p(\mathbf{m})}{p(\mathbf{d})}
    \label{posterior_pdf}
\end{equation}
where the probability density function (PDF) of the posterior, \(p(\mathbf{m}|\mathbf{d})\), is determined by the likelihood \(p(\mathbf{d}|\mathbf{m})\), describing the conditional probability of successfully modeling the seismic data given a seismic velocity model, and by our prior knowledge \(p(\mathbf{m})\) of the model parameters, which reflects our initial confidence in the unknown model based on any available prior information. Lastly, a normalization constant, also known as the evidence, \(p(\mathbf{d})\), ensures that the posterior distribution properly integrates to one over the entire parameter space.

\subsection{Variational Inference}
At the core of Variational Inference lies the idea of approximating this posterior distribution with a simpler, surrogate distribution, denoted as \(q(\mathbf{m})\). This distribution should be selected from a family (called variational family) that is easy to sample and evaluate; a common choice is therefore the Gaussian distribution. The essence of this optimization process lies in minimizing the Kullback-Leibler (KL) divergence \citep{kullback_information_1951, blei_variational_2017}, which quantifies the discrepancy between the surrogate (approximate) distribution and the target (posterior) distribution. The KL divergence is expressed as:

\begin{equation}
    \mathrm{KL} (q(\mathbf{m})|p(\mathbf{m}|\mathbf{d})) = \mathrm{E}_{\mathbf{m} \sim q} [-\log p(\mathbf{m}|\mathbf{d}) + \log q(\mathbf{m})]
    \label{KL_div}
\end{equation}

By minimizing Equation \(\ref{KL_div}\), we are able to compute the expectation of the model parameters sampled from the surrogate distribution. Consequently, VI achieves an approximation of the posterior distribution through the minimization outlined below:

\begin{equation}
    q^* =  \underset{q}{\mathrm{argmin}} \mathrm{KL} (q(\mathbf{m})|p(\mathbf{m}|\mathbf{d}))
    \label{q_surrogate}
\end{equation}

\subsection{Stein Variational Gradient Descent}

Stein Variational Gradient Descent represents a distinctive approach in VI. Unlike traditional VI techniques, SVGD is a deterministic, particle-based inference algorithm that iteratively minimizes the KL divergence between the chosen approximate distribution and the target density. This innovative method leverages the concept of functional gradients to effectively transport a predefined set of particles toward the target distribution. The transport occurs within the Reproducing Kernel Hilbert Space (RKHS), guided by the gradient of the KL divergence (for a detailed derivation of the SVGD formulation, we refer the reader to \cite{liu_stein_2016}). 

Given a collection of particles, the optimal update direction $\phi^{*}$ of Equation \(\ref{q_surrogate}\), for each particle, is given by:

\begin{equation}
    \phi^*(\cdot) = \mathrm{E}_{\mathbf{m} \sim q} \underbrace{[k(\mathbf{m}, \cdot) \nabla_{\mathbf{m}} \log p(\mathbf{m}|\mathbf{d})}_{{\mathrm{driving~force}}} ~+ \underbrace{\nabla_{\mathbf{m}} k(\mathbf{m}, \cdot)]}_{\mathrm{{repulsive~force}}}
    \label{svgd_eq}
\end{equation}

If we denote the particles that we are using to represent \(\textit{q}\) as \(\{\textbf{m}_{i}\}_{n=1}^{N}\), the expectation in Equation (\(\ref{svgd_eq}\)) can be approximated using the sample mean over the particles. Thus, the KL divergence can be iteratively minimized as follows:

\begin{equation}
    \phi_{q_l,p}^*(\mathbf{m}) = \frac{1}{N} \sum_{j=1}^N [k(\mathbf{m}_j^l,\mathbf{m})\nabla_{\mathbf{m}_j^l}\log p(\mathbf{m}_j^l | \mathbf{d}) + \nabla_{\mathbf{m}_j^l}k(\mathbf{m}_j^l,\mathbf{m})] \label{svgd_update}
\end{equation}

\begin{equation*}
    \mathbf{m}_i^{l+1} = \mathbf{m}_i^l + \epsilon_l \phi_{q_l,p}^*(\mathbf{m}_i^l)    
\end{equation*}

where \(\textit{l}\) denotes the current iteration, \(\textit{N}\) is the number of particles, and \(\epsilon_{l}\) is the step size. Assuming the step size to be sufficiently small, the process asymptotically converges to the target posterior as the number of particles tends to infinity.

Equation $\ref{svgd_eq}$ comprises two distinct terms: the $\emph{driving force}$ and the $\emph{repulsive force}$. The driving force aims to direct the particles toward higher probability regions. Conversely, the repulsive force has the crucial role of maintaining particle diversity, and actively prevents particle collapse by dispersing particles across the parameter space. This balance enables a comprehensive exploration and characterization of the target distribution.

Various types of kernel functions have been proposed over the last few years (\citep{liu_stein_2016, gorham_measuring_2017, zhang_variational_2020}.
In our study, we adopt the two most commonly employed kernels, namely the radial basis function (RBF) and the inverse multi-quadratic (IMQ).
The RBF kernel is defined as:

\begin{equation}
    k(\mathbf{m},\mathbf{m}') = \exp\left(-\frac{||\mathbf{m}-\mathbf{m}'||^{2}}{2h^2}\right) \label{rbf}
\end{equation}

where $\textit{h}$ is the bandwidth, a scaling factor that controls the strength of the interaction between different particles based on their distances. As suggested by \cite{liu_stein_2016}, we set $\textit{h}=\tilde{d}^{2}/log N$, where $\tilde{d}$ is the median of pairwise distances between all particles. It is worth noting that this parameter is recalculated at each iteration with limited heuristic justification, and there is some evidence that this non-linearity can generate instability.
The IMQ kernel, instead, is defined as:

\begin{equation}
    k(\mathbf{m},\mathbf{m}') = \left( \textit{c}^2 + \frac{||\mathbf{m}-\mathbf{m}'||^{2}}{2h^2} \right)^{\beta} \label{imq}
\end{equation}

where $\textit{c}$ and $\beta$ are two user-defined parameters. We set $\textit{c}$ and $\beta$  to 1 and $-\frac{1}{2}$, respectively, as suggested by \cite{gorham_measuring_2017}.

\subsection{Annealed Stein Variational Gradient Descent}

Despite the empirical success of SVGD, convergence guarantees are absent --- except in the mean-field limit (where the number of particles \(n \to \infty\), while the dimensionality \(d\) is kept fixed).
\cite{zhuo_message_2018} showed that SVGD encounters degeneracy issues under finite particle conditions, which cause the particles to collapse into a small number of modes --- the so-called mode collapse issue. On the other hand, as dimensionality increases (such that \(d>n\) ), the variance estimated by SVGD may significantly underestimates the variance of the target distribution — a phenomenon known as variance collapse; this is an important concern in high-dimensional problems like FWI.

\cite{ba_understanding_2021} compared the SVGD update to the application of gradient descent to a Maximum Mean Discrepancy (MMD) objective function. In a high-dimensional example, they found that SVGD and MMD descent differ primarily in the driving force term that becomes increasingly problematic in higher dimensions. They have empirically demonstrated that removing the bias introduced by the deterministic update present in the driving force leads to more accurate estimation of the variance. Consequently, \cite{ba_understanding_2021} proposed modifying the driving force term of SVGD with a damped version, resulting in the damped SVGD approach. At a high level, this modification mirrors the approach taken by \cite{dangelo_annealed_2021} under the name of Annealed SVGD (A-SVGD), where a heuristic temperature parameter \(\alpha(l) \in [0,1]\) is introduced to adjust the intensity of the driving force. 
The updated rule is then simply expressed as follows:

\begin{equation}
    \phi_{q_l,p}^*(\mathbf{m}) = \frac{1}{N} \sum_{j=1}^N [\alpha(l) k(\mathbf{m}_j^l,\mathbf{m})\nabla_{\mathbf{m}_j^l}\log p(\mathbf{m}_j^l) + \nabla_{\mathbf{m}_j^l}k(\mathbf{m}_j^l,\mathbf{m})]. \label{asvgd_update}
\end{equation}

Varying the $\alpha$ parameter within the $[0,1]$ range induces two distinct phases. A first exploratory phase, dominated by a strong repulsive force ($\alpha$ close to 0) that disperses the particles from their initial positions, facilitating broad coverage of the target distribution. This is followed by a second exploitative phase, where the driving force dominates ($\alpha$ close to 1) and concentrates the particle distribution around different modes.
The selection of the temperature parameter $\alpha(l)$ is crucial to maintain the convergence properties of SVGD, in order to ensure that the final iterations operate effectively on the target density, i.e. $\lim_{l \to \infty}\alpha(l)=1$. 
In this work we will employ and compare the performances of two different annealing schedules proposed by \cite{dangelo_annealed_2021}.
The first schedule is the hyperbolic tangent, defined as:

\begin{equation}
    \alpha(l) =  \tanh [(1.3\frac{l}{N})^{p}] \label{tanh} 
\end{equation}

where $l$ is the current iteration, $N$ is the total number of iterations, and $p$ is a user-defined parameter that controls the rate of transition between the two phases.
The second is the cyclic schedule, which allows a sequence of exploratory and converging phases and can be defined as:

\begin{equation}
    \alpha(l) =  (\frac{mod(l,N/C)}{N/C})^{p} \label{cyclic} 
\end{equation}

where $C$ is the number of cycles, where the $\alpha$ value ranges from 0 to 1.

\subsection{Principal Component Analysis}
Principal Component Analysis is a dimensionality reduction technique, particularly effective when dealing with high-dimensional and highly correlated data. The main objective of PCA is to find a smaller set of features that can accurately represent the original data in a lower-dimensional space, while preserving as much information as possible \citep{hotelling_analysis_1933}. PCA operates under the assumptions of linearity (principal components are orthogonal to each other), and can be summarized as follows. 

First, let us consider a set of velocity particles \(\mathbf{X}\) of size \(n \times m \), where \(n\) is the number of particles (observations) and \(m\) is the number of dimensions (variables). PCA begins by computing the mean vector \(\bar{\mathbf{X}} = \frac{1}{n} \sum_{i=1}^{n} \mathbf{X}_i\), which contains the mean of each column and allows us to standardize the data to have a zero mean:
\begin{equation}
    \mathbf{Z} = \mathbf{X} - \mathbf{I}_{n \times 1} \bar{\mathbf{X}} \label{data_standarized}
\end{equation}
Next, we compute the covariance matrix \(\mathbf{C}\) of the mean-centered data \(\mathbf{Z}\):
\begin{equation}
    \mathbf{C} = \frac{1}{n-1} \mathbf{Z}^T \mathbf{Z} 
    \label{cov_matrix}
\end{equation}
We then perform eigenvalue decomposition on the covariance matrix \(\mathbf{C}\):
\begin{equation}
    \mathbf{C} \mathbf{V} = \mathbf{V} \mathbf{\Lambda}
    \label{eigen_decom}
\end{equation}
where \(\mathbf{V}\) is the \(m \times m\) matrix of eigenvectors (principal components) and \(\mathbf{\Lambda}\) is the \(m \times m\) diagonal matrix of eigenvalues. Finally, we sort the eigenvalues in descending order and reorder the eigenvectors accordingly to project the mean-centered data onto the new principal components:
\begin{equation}
    \mathbf{Y} = \mathbf{Z} \mathbf{V}
    \label{pca_proj}
\end{equation}
where \(\mathbf{Y}\) is the \(n \times m\) matrix of the transformed data.

In the present study, we apply PCA to the entire set of particles collected per iteration (after SGVD is performed) for a number of components equal to $n-1$, where $n$ is the number of particles. We aim to obtain the explained variance per component, which quantifies the proportion of the total variability in the data that each principal component captures. Given a user-defined number of components, the explained variance for each component is determined by its respective eigenvalue:

\begin{equation}
\text{\textit{Explained Variance of the i-th component}} = \frac{\lambda_i}{\sum_{j=1}^{n} \lambda_j}
\end{equation}

where \(\lambda_i\) represents the eigenvalue of the \(i\)-th component, and \(\sum_{j=1}^{n} \lambda_j\) denotes the sum of all eigenvalues, which accounts for the total variance. Analyzing the explained variance per component helps us understanding the behavior and distribution of the particles.

\subsection{Clustering}

Given the highly non-linear nature of FWI, the particles obtained at the end of the SVGD iterations may have converged to different modes (i.e., become trapped in various local minima). Whilst the particles that have reached the global minimum are likely to be geologically meaningful, this may not be the case for those that reached other basins of attraction. Consequently, it is important to identify clusters of particles within our high-dimensional model space and distinguish those that are geologically meaningful from the other ones. 

In order to do so, we have employed the Hierarchical Density-Based Spatial Clustering of Applications with Noise (HDBSCAN) to perform such a clustering operation \citep{campello_density-based_2013}. HDBSCAN is an advanced clustering algorithm that extends Density-Based Spatial Clustering of Applications with Noise (DBSCAN) \citep{ester1996density}. In short, HDBSCAN uses the Mutual Reachability Distance, a metric that combines a density-based measure with pairwise distances to facilitate meaningful clustering. The algorithm creates a Minimum Spanning Tree (MST) from the mutual reachability distances, assembling the basis of the hierarchical cluster tree. Through Condensed Clustering, HDBSCAN extracts significant clusters by systematically removing edges in the MST. Finally, Stability-based Clustering selects the most stable clusters from the hierarchical tree. Unlike other clustering methods, HDBSCAN is well-suited for high-dimensional data and does not require the user to specify the number of clusters in advance. This flexibility makes HDBSCAN an ideal candidate for our analysis, as it allows for robust identification of clusters and noise.

\section{Numerical examples}
\subsection{Synthetic data}
\begin{figure}
    \centering
    \includegraphics[width=\textwidth]{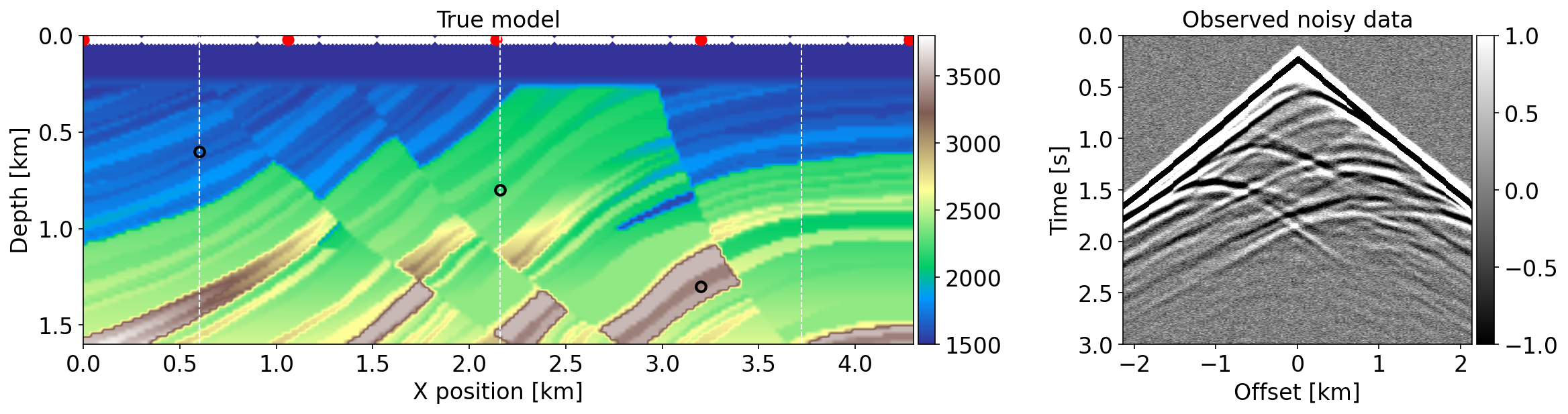}
    \vspace{-0.5cm}
    \caption{Portion of the Marmousi model used in our numerical experiments. The left panel illustrates the velocity model in meters per second (m/s), with red dots indicating shot locations and white dots representing receivers. Vertical white lines represent pseudo-well logs, and black circles indicate pixel locations for marginal plots discussed in the results section and appendix. The right panel shows the shot gather for the source in the middle of the model. }
    \label{fig:true_model_data}
\end{figure}

In this section, we present a series of numerical experiments using a portion of the Marmousi model \citep{brougois_marmousi_1990} to evaluate the capability of the previously described SVGD approaches to estimate the uncertainty associated with FWI when working with a limited number of particles. Synthetic data are modeled using Deepwave \citep{richardson_alan_2023} in a rectangular uniform grid with dimensions of $n_z=81$ and $n_x=216$ and spacing of 20m in both directions, using five shots evenly distributed along the horizontal axis and recorded by 201 receivers placed at 1-meter intervals. The seismic source is modeled using a Ricker wavelet with a peak frequency of 7 Hz. The registration time is set to 3s and the sample interval used for the forward modelling is 1ms. To simulate more realistic conditions, colored noise matching the frequency spectrum of the signal is added to the synthetic data, resulting in noisy observed data with a Signal-to-Noise Ratio (SNR) of approximately 17 dB (see Figure \ref{fig:true_model_data}).

\subsection{Selection of hyperparameters }
The primary objective of this work is to investigate the uncertainty associated with the modeling operator (the likelihood is assumed to be Gaussian) by deliberately excluding any influence of our prior knowledge, rather than imposing a uniform bound limits on velocities.
Since the uncertainty associated with the FWI problem is expected to encapsulate both the scattering (high wavenumber) and transmission (low wavenumber) components of the model, the choice of perturbations of the initial particles is crucial. Drawing inspiration from \cite{izzatullah_physics-reliable_2024}, we aim to explore the uncertainties of these components by generating Gaussian Random Field (GRF) perturbations that introduce variability in both amplitude and scale. 

In terms of the hyperparameters employed in this study, we opt for the Gaussian RBF and IMQ kernels. Bandwidth selection is performed using both the median trick and a fixed constant value. The constant value of the bandwidth is determined after a meticulous analysis of the bandwidth evolution in the scenario where the median trick is used. Finally, we use the Adam optimizer to update the particles at each iteration, given their gradient in Equation \ref{svgd_update}, with a constant learning rate of 100 and a fixed number of iterations equal to 600. 

The experiments are conducted in two particle sets: one with a small number of particles (50) and the other with a larger number (200).
To mitigate the effects of mode and variance collapse, we evaluate both the vanilla and annealed SVGD approaches. For the annealed SVGD, we consider two formulations for the temperature parameter, namely hyperbolic and cyclic. For the hyperbolic formulation (Eq. \ref{tanh}), we set \( p=3 \) and maintained \( \alpha = 1 \) for the last 20\% of iterations. For the cyclic formulation (Eq. \ref{cyclic}), we set \( p=2 \) and \( C=8 \), ensuring that the temperature remains at \( \alpha = 1 \) during the last two cycles.

The overall performance in terms of data misfit (L2 norm) and SNR for the various experiments with different hyperparameters is presented in Figures \ref{fig:loss_comparison} and \ref{fig:snr_comparison}. The subsequent results and discussion focus on the vanilla and annealed SVGD (tanh) formulations, utilizing the RBF kernel with the median trick and a set of 200 particles, as these configurations showed the best performance.

\subsection{Single-scale experiments}\label{single_freq_exp}

In the first set of experiments, we perform FWI using a single frequency band with a peak frequency of 7 Hz. Initial particles are generated by applying GRF perturbations with different variances to a highly smoothed version of the true Marmousi model. During the inversion process, we impose lower and upper-velocity bounds, allowing velocity values between 1500 m/s and 4370 m/s. 

\begin{figure}
    \centering
    \includegraphics[width=\textwidth]{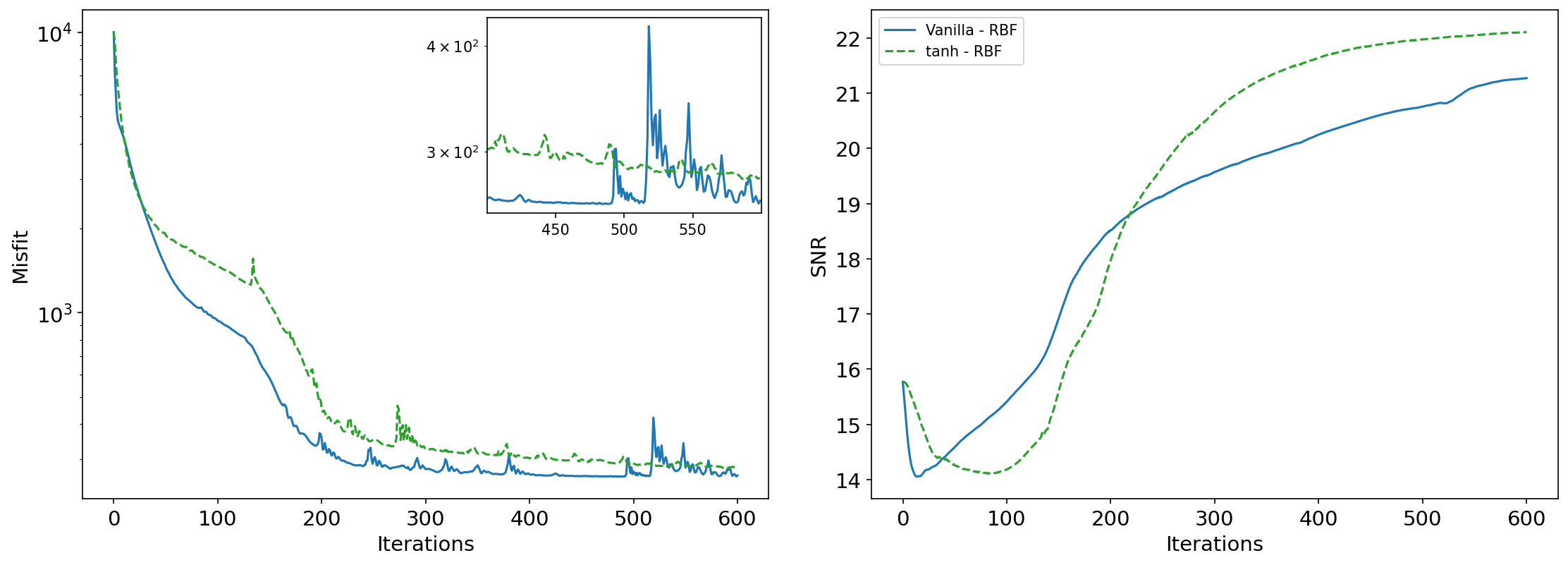}
    \vspace{-0.5cm}
    \caption{Data misfit (left) and SNR with respect to the true model (right) for vanilla and annealed (tanh) SVGD in the single-frequency scenario using 200 particles. The zoom window provides a clearer misfit comparison for the final 150 iterations.}
    \label{fig:misfit_snr_comparison}
\end{figure}

Figure \ref{fig:misfit_snr_comparison} illustrates the performances of the different methods in terms of data misfit and SNR with respect to the true model, computed from the mean over iterations for the various experiments. The vanilla and annealed SVGD with hyperbolic tangent demonstrate superior performance compared to the other methods (see Appendix \ref{appendix_A}). Notably, the annealed SVGD using the cyclic formulation results in poorer data misfit and SNR; this may be due to its design, which aims to explore the parameter space better and identify widely separated high-probability regions. As a result, the mean model may be perturbed, fitting the data less precisely compared to a more compact, but less comprehensive, posterior distribution.

\begin{figure}
    \centering
    \includegraphics[width=\textwidth]{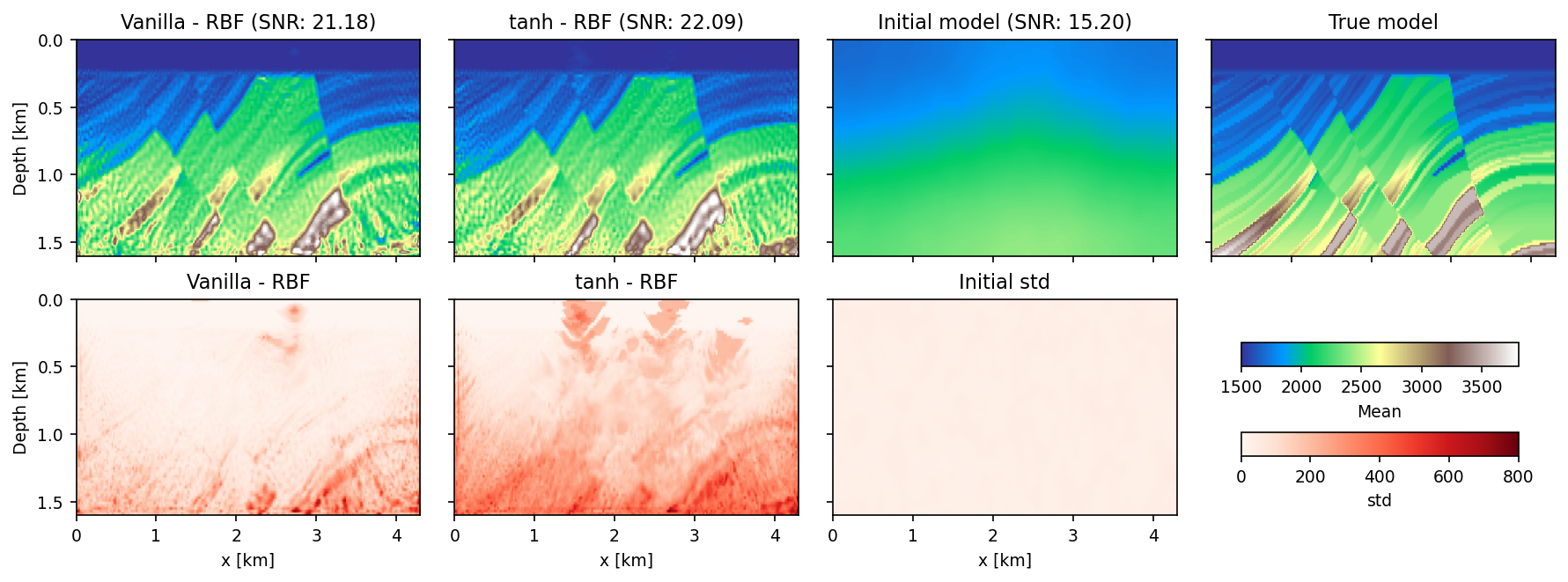}
    \vspace{-0.5cm}
    \caption{Mean and standard deviation comparison of the experiments using 200 particles for vanilla SVGD and annealed SVGD using RBF kernel and median trick after 600 iterations in the single-scale scenario. The velocity values are expressed in m/s.}
    \label{fig:mean_comparison_200p_vanilla_annealed}
\end{figure}

A more detailed comparison of the mean and standard deviation for the 200-particle experiments is presented in Figures \ref{fig:mean_comparison_200p} and \ref{fig:std_comparison_200p}, respectively. Figure \ref{fig:mean_comparison_200p_vanilla_annealed} shows the mean obtained after 600 iterations for the vanilla and annealed versions for the 200-particle experiment, along with the mean of the initial distribution. All predicted models are relatively similar in the shallower part ($<$1 km) and reproduce most of the main features of the true model. However, the model reconstruction is poorer in the deeper part, likely due to a lack of illumination (especially near the edges of the model). As stated before, the results obtained using the vanilla SVGD and the annealed SVGD with the hyperbolic tangent formulation (both with the RBF kernel) exhibit higher SNR and are less affected by artifacts. In contrast, the models predicted using other approaches show errors in terms of velocity magnitudes, particularly in the last 500 meters of depth.

In addition, Figure \ref{fig:mean_comparison_200p_vanilla_annealed} presents the standard deviation maps (expressed in m/s) obtained after 600 iterations using 200 particles. Both maps exhibit a similar expected pattern, with very low values in the shallower parts where illumination is greater and the values increasing towards the deeper portions of the model. These deeper regions, along with the lateral edges of the model, are expected to have higher uncertainties due to limitations in acquisition geometry and the physics of wave propagation within the subsurface. Higher uncertainties are also observed in areas of high velocity and near the main velocity contrasts.

The standard deviation associated with the initial distribution is relatively low ($<$ 150 m/s) across the entire model. We select a narrow proposal for the initial samples to prevent SVGD from rapidly repelling particles into undesirable modes at the early stages (particles converging to suboptimal local minima). By maintaining an initial low standard deviation, we aim to direct the convergence toward fewer and more geologically consistent local minima. In contrast, the values associated with the predicted models are substantially higher, reaching over 800 m/s. Notably, the standard deviation maps associated with the annealed version of SVGD show significantly higher values throughout the model compared to those associated with the vanilla SVGD. This indicates that the annealed approach allows for better exploration of the model space and reduces the variance collapse phenomenon affecting the vanilla SVGD.

\begin{figure}
    \centering
    \includegraphics[width=\textwidth]{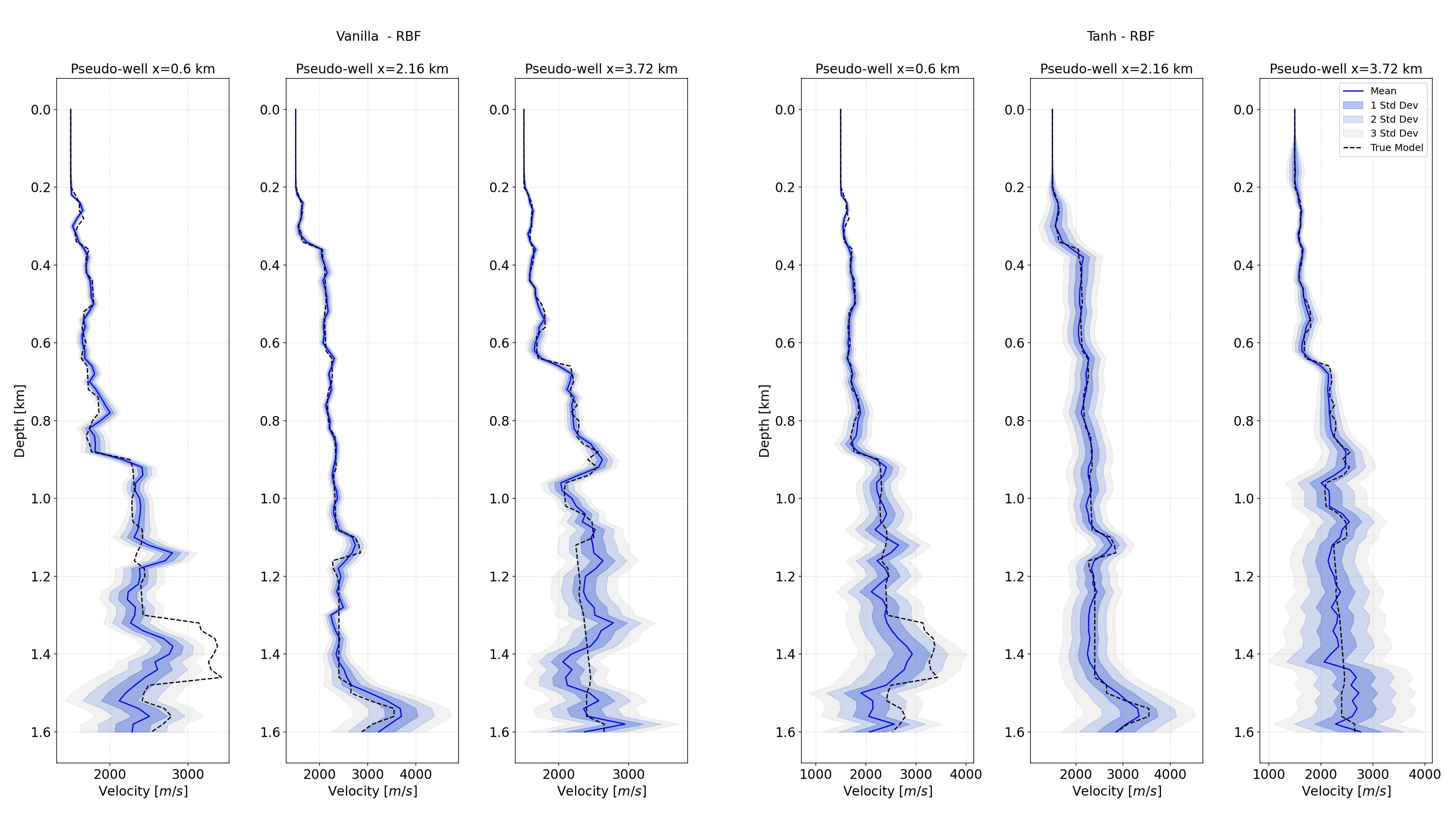}
    \vspace{-0.5cm}
    \caption{Pseudo-well marginal distributions showing mean and three confidence intervals for the experiment using vanilla SVGD (left) and annealed SVGD (right), using the RBF kernel with median trick and 200 particles in the single-frequency scenario.}
    \label{fig:appx_marginal_wells_vanilla_rbf}
\end{figure}

\begin{figure}
    \centering
    \includegraphics[width=0.9\textwidth]{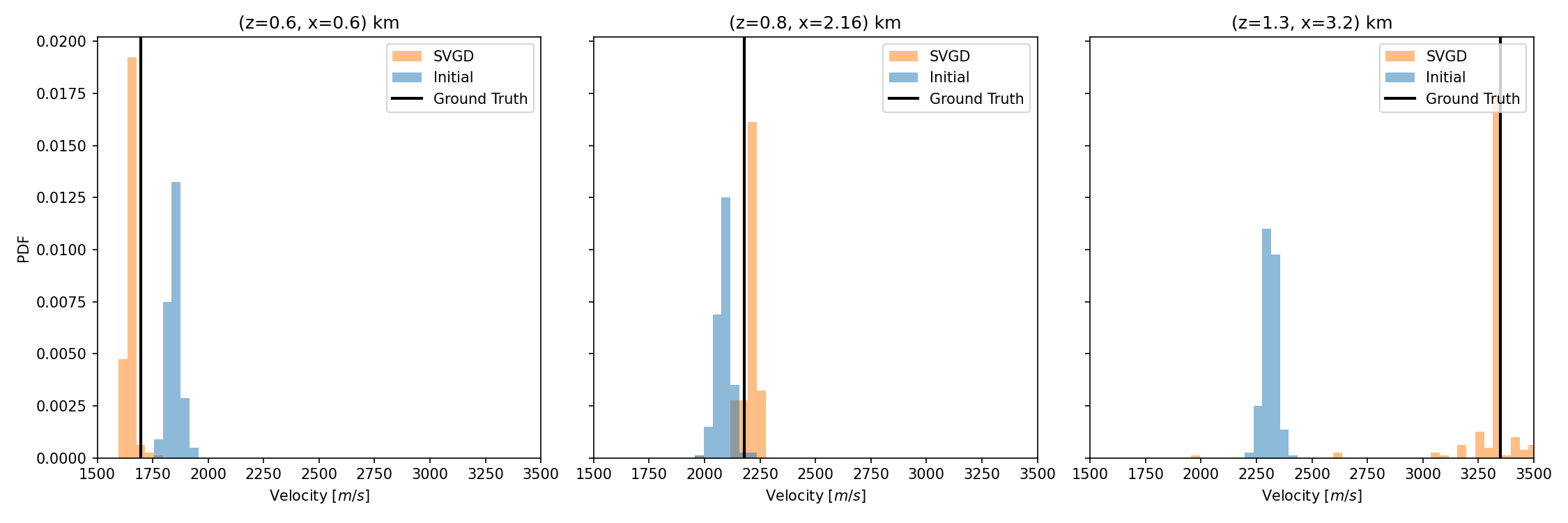}
    \vspace{-0.5cm}
    \caption{Single-scale scenario:, pixels marginals for vanilla SVGD with RBF kernel and 200 particles.}
    \label{appx_marginal_pixels_vanilla_rbf}
\end{figure}

\begin{figure}
    \centering
    \includegraphics[width=0.9\textwidth]{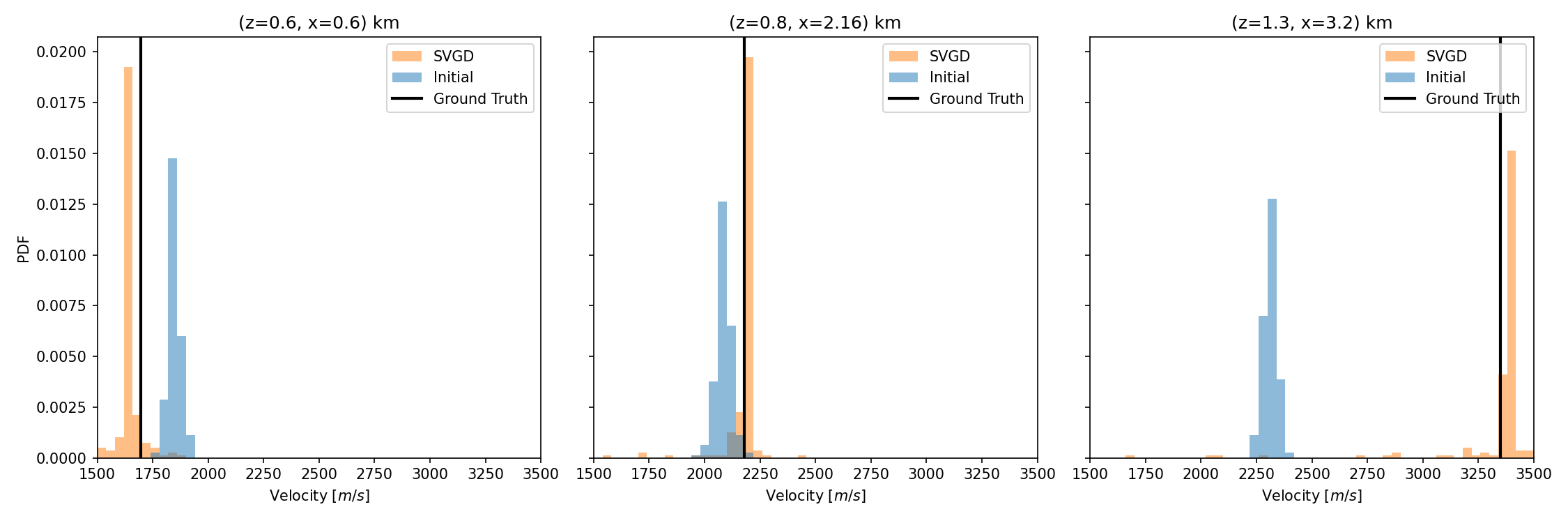}
    \vspace{-0.5cm}
    \caption{Single-scale scenario: pixels marginals for annealed SVGD (tanh) with RBF kernel, and 200 particles.  }
    \label{appx_marginal_pixels_tanh_rbf}
\end{figure}

Moreover, to better highlight the differences between the vanilla and annealed approaches, in Figures \ref{fig:appx_marginal_wells_vanilla_rbf} (left) and \ref{fig:appx_marginal_wells_vanilla_rbf} (right) we present three pseudo well logs corresponding to three distinct spatial locations, as indicated by the white vertical lines in Figure \ref{fig:true_model_data}. For each position, we illustrate the true velocity varying with depth, the velocities extracted from the mean model obtained using both approaches, and three confidence intervals (corresponding to one, two, and three standard deviations) based on the standard deviation maps shown in Figure \ref{fig:std_comparison_200p}. We observe that in both cases, the width of the confidence intervals increases with depth, indicating larger uncertainty at greater depths, as expected. Within the first 800 meters of depth, the mean model closely resembles the true model, with standard deviation values close to zero. The main differences between the two approaches become more evident at greater depths, where discrepancies between the predicted and true models are more significant. Specifically, the logs associated with the annealed SVGD show larger confidence intervals, with the values extracted from the true model falling almost entirely within these bounds. In contrast, the vanilla SVGD approach yields smaller confidence intervals, with the true values occasionally falling outside these bounds (e.g., the pseudo well at the spatial position of 0.6 km and around 1.4 km of depth). This comparison indicates that, while the annealed approach captures a broader range of uncertainty, the vanilla approach may sometimes underestimate the true variability at greater depths. Similar results are also obtained at three different pixel locations for vanilla (see figure \ref{appx_marginal_pixels_vanilla_rbf}) and annealed formulations (see figure \ref{appx_marginal_pixels_tanh_rbf}).

To further investigate the behavior and distribution of particles during the SVGD optimization process, we performed PCA. This analysis helps us understand the explained variance of the components, providing insights into how the particles evolve throughout the optimization. By examining the explained variance for \(n-1\) components (where \(n\) is the number of particles), it is possible to identify the dominant directions in which the particle positions vary the most. This allows us to understand the main variances captured by components and the structural dynamics of the particle distribution.

\begin{figure}
    \centering
    \includegraphics[width=\textwidth]{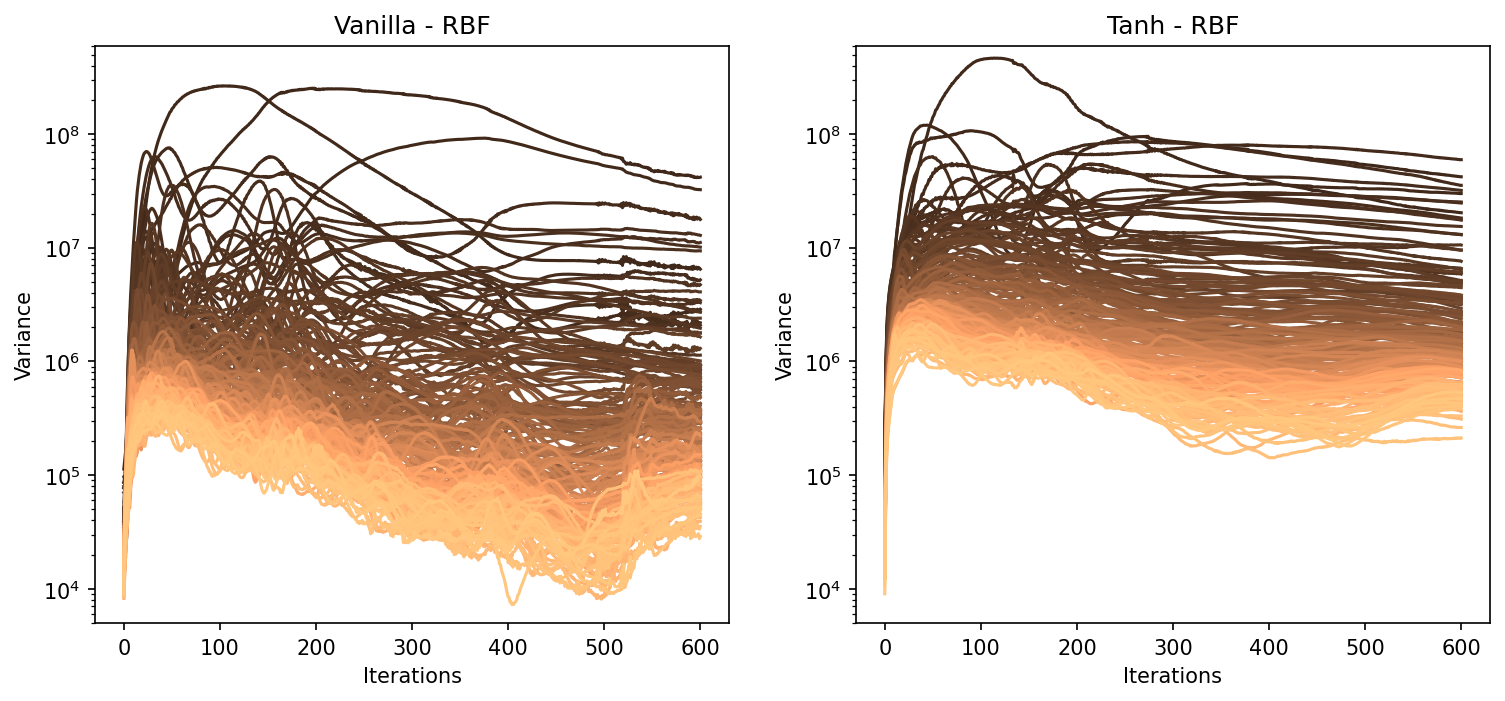}
    \vspace{-0.5cm}
    \caption{Explained variance per component (using $n-1$ components, where $n$ is the particle number) for the experiments using (left) vanilla SVGD with the RBF kernel and (right) annealed SVGD with the tanh formulation and RBF kernel with 200 particles.}
    \label{fig:explained_var_vanilla_tanh_200p}
\end{figure}

Figure \ref{fig:explained_var_vanilla_tanh_200p} (left) illustrates the PCA variances for the vanilla SVGD case with 200 particles using the RBF kernel. We observe a relatively uniform spread of the components over the variance but erratic convergence, with a significant portion of the components converging towards mid- and low-variance values. In contrast, Figure \ref{fig:explained_var_vanilla_tanh_200p} (right) shows the PCA for the annealed SVGD with the tanh formulation, which demonstrates a more stable convergence pattern and a more balanced distribution of variance between high and low variance directions. 
These PCA results highlight the effectiveness of the annealed SVGD to provide a more stable and informative representation of the particle dynamics compared to the vanilla SVGD approach. 

\subsection{Multi-scale experiments}
\begin{figure}
    \centering
    \includegraphics[width=\textwidth]{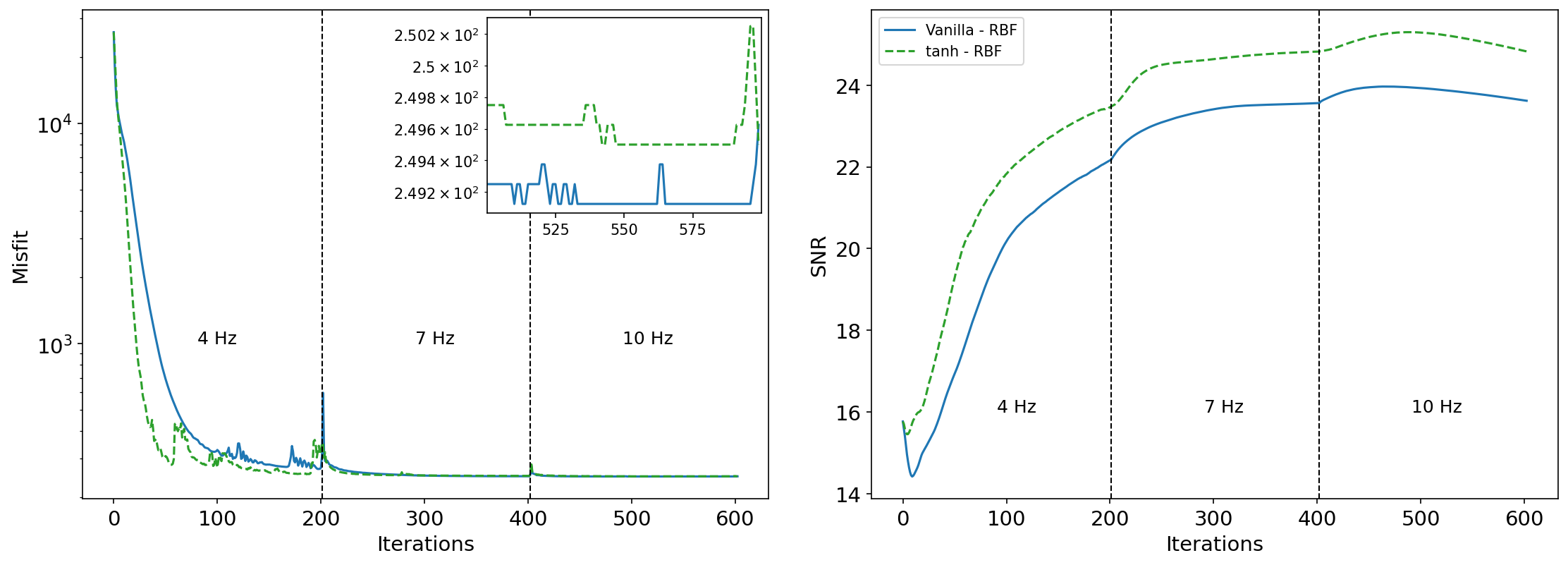}
    \vspace{-0.5cm}
    \caption{Data misfit (left) and SNR (right) across different experiments for 200 particles in the multi-scale scenario. The close-up window provide a clearer comparison of the final 150 iterations.}
    \label{fig:multi_scale_loss_comparison}
\end{figure}

In this section, we present results for the multi-scale FWI experiments, focusing exclusively on the Vanilla and Annealed variants of SVGD (see Figure \ref{fig:multi_scale_loss_comparison}), as these have demonstrated superior performance in our single-frequency experiments. The multi-scale approach (introduced by \cite{bunks_multiscale_1995}) has become a standard practice in FWI due to its ability to improve the inversion's convergence and accuracy, mitigating the cycle-skipping issue. It begins with the lowest frequencies available in the observed data and progressively incorporates higher frequencies, thereby mitigating the non-linearity of the inversion process by initially targeting large-scale features, which are more sensitive to low frequencies, and then refining the model with higher frequencies to capture higher-resolution details. Specifically, the inversion process is conducted in three stages: 200 iterations at a peak frequency of 4 Hz, 200 iterations at 7 Hz, and 200 iterations at 10 Hz, with learning rates of 100, 10, and 10, respectively. This step-wise frequency escalation ensures that each scale of the model is accurately resolved before moving on to the next, providing a robust framework for the inversion process.

As in the single-frequency experiment, the multi-scale approach using annealed SVGD demonstrates superior performance compared to the vanilla formulation in terms of SNR with respect to the true model (see Figure \ref{fig:multi_scale_loss_comparison}). It is essential to highlight that multi-scale SVGD not only achieves higher SNR values but also yields more meaningful and representative statistics compared to the single-frequency approach. This improvement is primarily due to the enhanced control over the particle refinement process, which prevents some particles from diverging and causing high standard deviation values. Consequently, the optimization process is more likely to converge toward geologically representative models, leading to more accurate and reliable results.

\begin{figure}
    \centering
    \includegraphics[width=\textwidth]{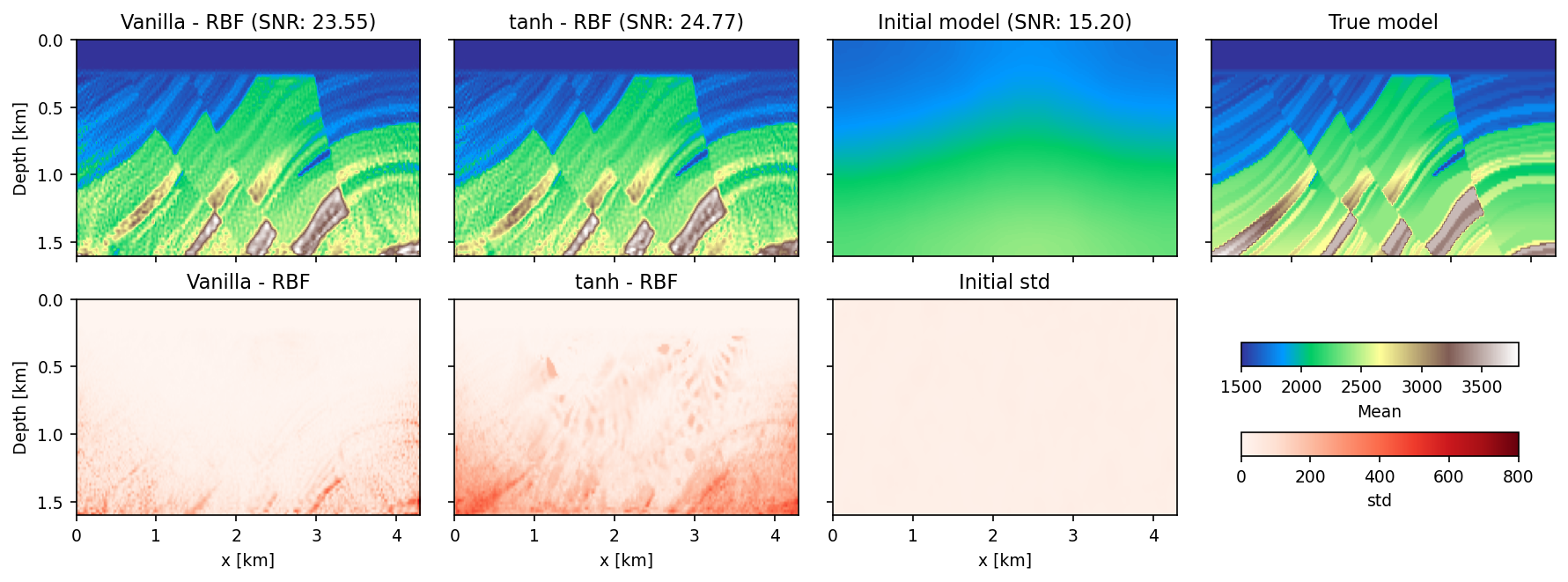}
    \vspace{-0.5cm}
    \caption{Mean and standard deviation comparison of the experiments using 200 particles for vanilla SVGD and annealed SVGD using RBF kernel and median trick after 600 iterations in the multi-scale scenario. The velocity values are expressed in m/s.}
    \label{fig:multi_scale_mean_comparison_200p}
\end{figure}

Appendices \ref{appendix_A} and \ref{appendix_B} provide supplementary results for both the single-scale and multi-scale approaches, respectively, which further strengthen the evidence supporting our findings.

\subsection{Cluster Analysis and Statistical Evaluation}
Our single-frequency experiments yield several key observations. A small number of principal components explain the majority of the variance (Figure \ref{fig:explained_var_vanilla_tanh_200p}, left). Consequently, many components capture small variances rather than meaningful patterns in the data. This is further supported by the presence of abnormal individual particles (see Appendix \ref{appendix_A}) and artifacts in the shallower parts of the particles, which are evident in the respective mean and standard deviation maps (Figures \ref{fig:mean_comparison_200p} and \ref{fig:std_comparison_200p}). These observations suggest that the particles tend to converge to different modes, some of which may not be geologically meaningful. Although these particles fit the data term, they do not accurately represent the subsurface structure. Therefore, it is crucial to perform clustering to identify the presence of different modes and conduct statistical analysis within each cluster, rather than assuming all samples have converged to the same global minima.

\begin{figure}
    \centering
    \includegraphics[width=\textwidth]{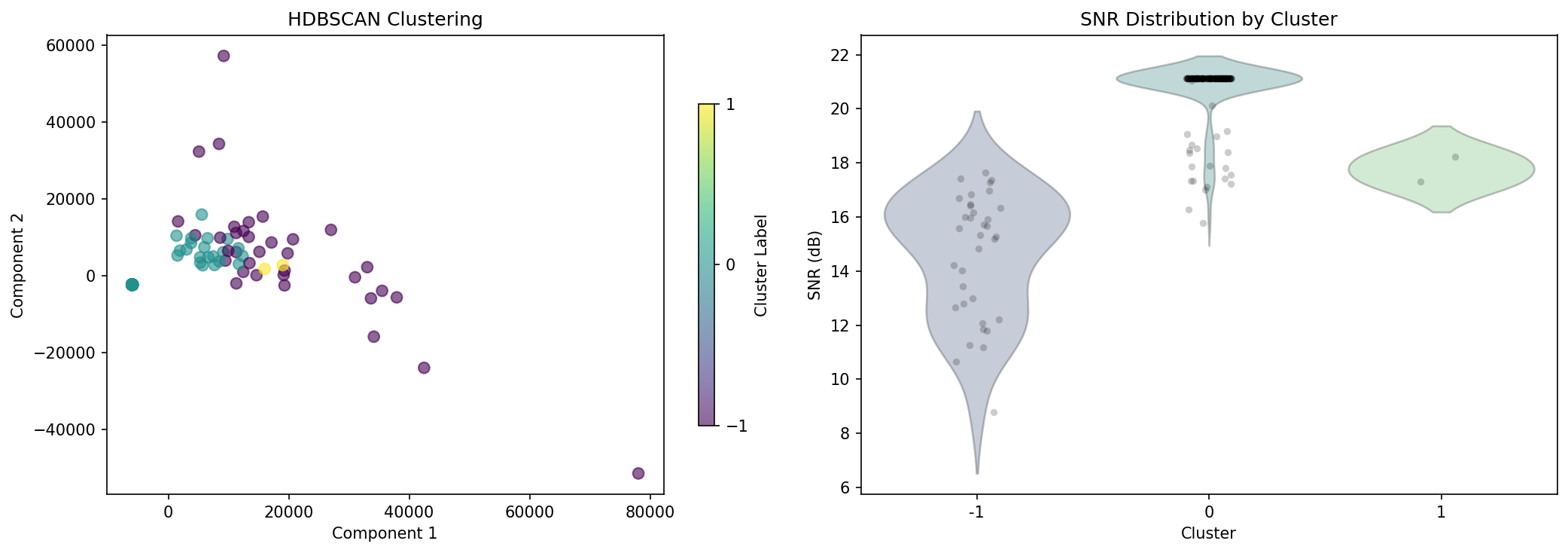}
    \vspace{-0.5cm}
    \caption{HDBSCAN clustering for the experiment using annealed SVGD with tanh formulation and RBF kernel with 200 particles. The left panel displays clusters obtained in high-dimensional data, plotted after dimensionality reduction to two components. The right panel shows the distribution of each cluster after computing the SNR.}
    \label{fig:hdbscan_snr_clusters_tanh_rbf_200p}
\end{figure}

\begin{figure}
    \centering
    \includegraphics[width=\textwidth]{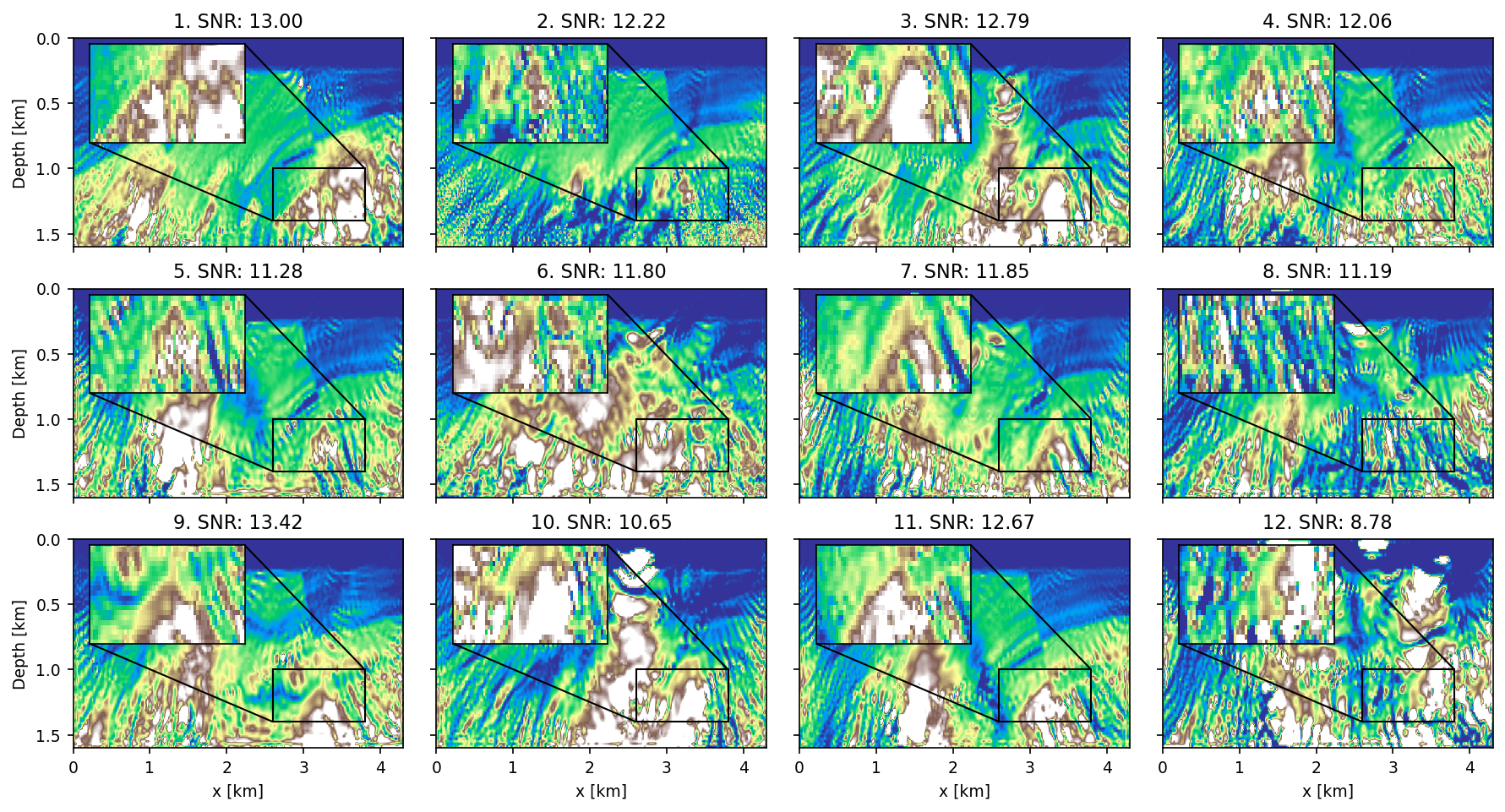}
    \vspace{-0.5cm}
    \caption{Noisy particles corresponding to cluster -1 for the experiment with annealed SVGD using the tanh formulation and RBF kernel with 200 particles.}
    \label{fig:noisy_particles_cluster_tanh_rbf_200p}
\end{figure}

To illustrate the importance of identifying clusters in the final particles and use such information for any subsequent statistical analysis, we consider the experiment with annealed SVGD employing a tanh temperature parameter and RBF kernel for a set of 200 particles. After the optimization process (600 iterations), we apply HDBSCAN to the final particles. In this case, the clustering algorithm produces three distinct groups, labeled as -1, 0, and 1, of sizes 35, 163, and 2, respectively. For visualization purposes, we plot the particles in the 2D space defined by the first two components of the PCA and color-code the particles to indicate which cluster they belong to. (Figure \ref{fig:hdbscan_snr_clusters_tanh_rbf_200p} left). We then compute the SNR for each individual particle in the different clusters and display their distribution in Figure \ref{fig:hdbscan_snr_clusters_tanh_rbf_200p} (right). Overall, cluster 0 and cluster 1 contain particles with higher SNR values, although some of the particles in cluster 0 converge to similar SNR values. More importantly, cluster -1 is identified as the noisy cluster, representing particles not assigned to any other cluster, to be considered as outliers and geologically implausible. (Figure \ref{fig:noisy_particles_cluster_tanh_rbf_200p}).

\begin{figure}
    \centering
    \includegraphics[width=0.9\textwidth]{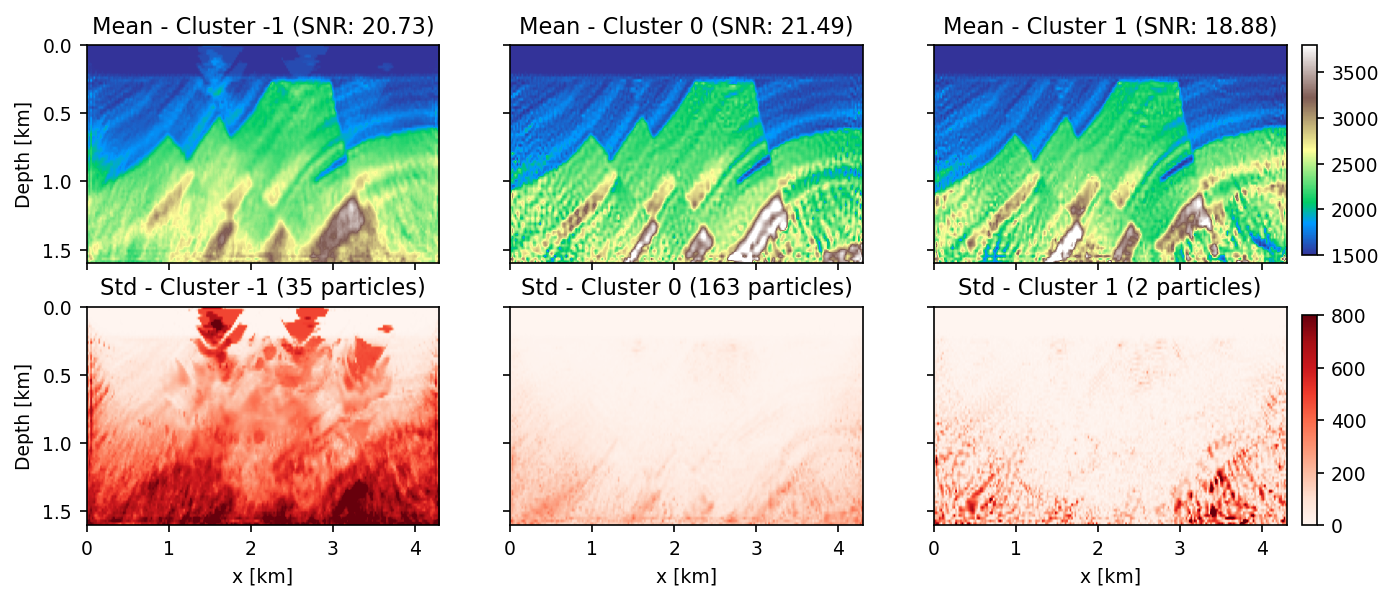}
    \vspace{-0.5cm}
    \caption{Cluster statistics for the experiment with annealed SVGD using the tanh formulation and RBF kernel with 200 particles.}
    \label{fig:statistics_clusters_tanh_rbf_200p}
\end{figure}

Further analysis of the mean and standard deviation for each cluster (Figure \ref{fig:statistics_clusters_tanh_rbf_200p}) reveals that while the mean of cluster -1 appears reasonable, the individual particles are not representative, which leads to significant variations in the shallow parts of the model and, therefore, high uncertainty. This observation correlates with the artifacts detected in the shallow regions of the results in subsection \ref{single_freq_exp}. In contrast, clusters 0 and 1 display more consistent and plausible mean and standard deviation patterns, similar to those obtained with the multi-scale approach. This underscores the importance of performing clustering analysis of the particles produced by SVGD in order to discard non-representative particles and thereby obtain more accurate and geologically plausible outcomes.

\section{Discussion}

In the context of FWI, addressing uncertainty quantification presents significant computational challenges. Given the high-dimensionality of our model space, we operate in a regime where the number of particles is much smaller than the number of unknown parameters, and therefore, it is only possible to provide a low-rank approximation of the posterior covariance, with the rank limited to at most \(n\) (particles) – 1. A key aspect of variance collapse occurs when the actual rank falls below this theoretical limit. This limitation prevents us from fully capturing uncertainty across all directions, thus our analyses produce only relative --- but still meaningful --- uncertainty estimates.

The SVGD algorithm stands out due to its flexibility, enabling optimization problems to be solved through standard gradient descent algorithms for a certain number of particles, whilst introducing inter-particle communication. However, it is crucial --- yet laborious --- to address the complexities of hyperparameter tuning. In our study, we experimented with two variants of SVGD (vanilla and annealed), Gaussian Random Field (GRF) perturbations to build initial particles, Gaussian kernels (RBF and IMQ), different bandwidth selection strategies (median trick and constant value), a constant learning rate, and varying particle counts. Moreover, our experimental evaluations were confined to single-scale and multi-scale scenarios and only assessed the uncertainties associated with the data misfit term (modeling operator). This methodology has the potential to produce nuanced standard deviation maps of velocities, though the integration of more informative prior information remains a subject for future exploration.

Our primary motivation was to apply annealed SVGD to mitigate mode- and variance-collapse issues that affect the vanilla SVGD approach. In the single-scale scenario, our findings reveal that annealed SVGD with the tanh formulation provide better model estimates (i.e., higher SNR, lower data misfit, and overall more meaningful uncertainty estimates) than the vanilla formulation. Higher standard deviation values are associated with high-velocity layers and areas of poor coverage due to the limited acquisition geometry.  The vanilla approach yields smaller confidence intervals, sometimes following a different trend than the true model. Conversely, the annealed approach with the tanh formulation captures a broader range of uncertainty and increases the standard deviation values throughout the model. The single-scale outcomes present undesirable artifacts in the shallower parts, which should theoretically be well-illuminated areas. Under such conditions, there is no guarantee that all particles will converge to a unique local minimum. When projecting the particles onto a subspace spanned by $n-1$ components, we observe that the annealed version recovers higher variances per component (reducing the variance collapse issue). Visualization of individual particles shows some particles fitting the data term but not representing the subsurface, which confirms the hypothesis that particles converge to different modes.

In the multi-scale scenario, both convergence and exploration are significantly enhanced. The mean and standard deviation estimates are improved and the shallower artifacts observed in the single-scale scenario are absent. This suggests that a sequential approach from low to high frequencies may mitigate, though not entirely eliminate, the mode-collapse issue. For what concerns the variance-collapse issue, the annealed version with the tanh formulation yields more reasonable standard deviation maps, as indicated by a more significant number of components explaining the majority of the data variance.

Given the non-linearity and high-dimensionality of the problem, it is challenging to ensure that all particles belong to a single mode. Therefore, it is essential to perform clustering analysis, regardless of the scenario. We opted for HDBSCAN due to its applicability in high dimensions. HDBSCAN enables the easy discovery of a noise cluster composed of non-geological particles. Independent statistical analysis per cluster in the single-scale scenario produced mean and standard deviation maps that are more realistic and do not suffer from shallower artifacts. The primary goal of using clustering analysis on the final set of particles as a post hoc technique is to quickly identify different modes and geologically meaningful particles. We prioritize this approach over incorporating a prior term, which can be mathematically challenging to formulate for filtering out geologically implausible features. A promising future direction could involve training a generative adversarial network (GAN) or variational autoencoder (VAE) to incorporate such a prior into the optimization process \citep{corrales_bayesian_2022}.

While annealed SVGD does not entirely solve the problem of variance collapse, it seems a promising method for FWI, where gradient computations are usually computationally expensive. Furthermore, when annealed SVGD is combined with a multi-scale FWI scenario, reasonable estimates could be obtained with fewer gradient evaluations (iterations). This study highlights the potential of advanced SVGD methods to improve the reliability of FWI.

One promising direction for future work involves applying SVGD in a reduced or projected space \citep{chen_projected_2020, liu_grassmann_2022} to decrease the number of unknowns in our inverse problem and assess the impact on computational efficiency, convergence rates and quality of the inversion results. This approach would align with the ideal conditions for SVGD, i.e., when the number of particles equals or exceeds the number of unknowns. For example, a potential solution could be to run the SVGD FWI algorithms in a DCT compressed domain, following the works of \cite{aleardi_gradientbased_2021}, \cite{berti_computationally_2024} and \cite{berti_probabilistic_2024}, who found that model compression through DCT effectively reduces the dimensionality of the problem. Also, we could use variational auto-encoders to compress both the model and data and apply SVGD in the compressed (latent) domain \citep{sun2024invertible}.
This exploration holds the potential to significantly advance the application of SVGD in FWI, which could pave the way for more accurate and computationally feasible seismic imaging techniques.

\section{Conclusions}
This study demonstrates that annealed SVGD can significantly improve convergence and performance compared to vanilla SVGD in FWI applications, in scenarios where the number of particles is much smaller than the number of unknown parameters. Specifically, the annealed SVGD with the tanh formulation enhances the accuracy of mean estimates, leading to higher SNR, lower data misfit, and more reasonable standard deviation maps and thereby mitigating --- but not eliminating --- the variance-collapse issue. Additionally, applying multi-scale FWI with SVGD yields better mean and standard deviation estimates compared to the single-frequency scenario. Also, combining multi-scale FWI with annealed SVGD yields superior performance. The use of Principal Component Analysis to explain variance by component provides valuable insights into the behavior of the samples during the optimization process. This method provides additional understanding of how different components contribute to the overall variance to give a clearer picture of the sample distribution and convergence patterns.
Finally, given the inherent complexity and high dimensionality of the problem, it is crucial to account for the possibility of multiple modes in the solution space. Consequently, statistical analysis should be conducted to identify clusters of particles; HDBSCAN is recommended for its ability to handle high-dimensional data and identify noise clusters composed of non-geological particles. These strategies collectively offer a more robust and insightful approach to uncertainty analysis in FWI as they enhance the reliability of the results, provide a deeper understanding of the subsurface, and ultimately aid more informed decision-making in industrial applications.

\section{Acknowledgments}
This research was supported by King Abdullah University of Science and Technology (KAUST) and the DeepWave consortium. MC gratefully acknowledges TotalEnergies for the opportunity to conduct an internship in Pau, France, and SB extends thanks to KAUST for hosting an internship in Saudi Arabia. MC and SB equally contributed for this work. 
\section*{DATA AVAILABILITY}
The Marmousi dataset utilized in this study is publicly accessible through the SEG Wiki page, available at \url{https://wiki.seg.org/wiki/Open_data}. 

\bibliographystyle{unsrtnat}
\bibliography{references}  

\begin{thebibliography}{59}
\providecommand{\natexlab}[1]{#1}
\providecommand{\url}[1]{\texttt{#1}}
\expandafter\ifx\csname urlstyle\endcsname\relax
  \providecommand{\doi}[1]{doi: #1}\else
  \providecommand{\doi}{doi: \begingroup \urlstyle{rm}\Url}\fi

\bibitem[Virieux and Operto(2009)]{virieux_overview_2009}
J.~Virieux and S.~Operto.
\newblock An overview of full-waveform inversion in exploration geophysics.
\newblock \emph{GEOPHYSICS}, 74\penalty0 (6):\penalty0 WCC1--WCC26, November 2009.
\newblock ISSN 0016-8033.
\newblock \doi{10.1190/1.3238367}.
\newblock URL \url{https://library.seg.org/doi/abs/10.1190/1.3238367}.

\bibitem[Tarantola(2005)]{tarantola_inverse_2005}
Albert Tarantola.
\newblock \emph{Inverse {Problem} {Theory} and {Methods} for {Model} {Parameter} {Estimation}}.
\newblock Society for Industrial and Applied Mathematics, January 2005.
\newblock ISBN 978-0-89871-572-9 978-0-89871-792-1.
\newblock \doi{10.1137/1.9780898717921}.
\newblock URL \url{http://epubs.siam.org/doi/book/10.1137/1.9780898717921}.

\bibitem[Lailly and Santosa(1984)]{lailly_migration_1984}
Patrick Lailly and F.~Santosa.
\newblock Migration methods: partial but efficient solutions to the seismic inverse problem.
\newblock In \emph{Inverse problems of acoustic and elastic waves}, volume~51, pages 1387--1403. 1984.
\newblock Publisher: SIAM Philadelphia.

\bibitem[Tarantola(1984)]{tarantola_inversion_1984}
Albert Tarantola.
\newblock Inversion of seismic reflection data in the acoustic approximation.
\newblock \emph{GEOPHYSICS}, 49\penalty0 (8):\penalty0 1259--1266, August 1984.
\newblock ISSN 0016-8033, 1942-2156.
\newblock \doi{10.1190/1.1441754}.
\newblock URL \url{https://library.seg.org/doi/10.1190/1.1441754}.

\bibitem[Bozdağ et~al.(2011)Bozdağ, Trampert, and Tromp]{bozdag_misfit_2011}
Ebru Bozdağ, Jeannot Trampert, and Jeroen Tromp.
\newblock Misfit functions for full waveform inversion based on instantaneous phase and envelope measurements.
\newblock \emph{Geophysical Journal International}, 185\penalty0 (2):\penalty0 845--870, May 2011.
\newblock ISSN 0956-540X.
\newblock URL \url{https://doi.org/10.1111/j.1365-246X.2011.04970.x}.

\bibitem[Guo et~al.(2020)Guo, Visser, and Saygin]{guo_bayesian_2020}
Peng Guo, Gerhard Visser, and Erdinc Saygin.
\newblock Bayesian trans-dimensional full waveform inversion: synthetic and field data application.
\newblock \emph{Geophysical Journal International}, 222\penalty0 (1):\penalty0 610--627, July 2020.
\newblock ISSN 0956-540X.
\newblock \doi{10.1093/gji/ggaa201}.
\newblock URL \url{https://doi.org/10.1093/gji/ggaa201}.

\bibitem[Luo and Schuster(1991)]{luo_waveequation_1991}
Y.~Luo and G.~T. Schuster.
\newblock Wave‐equation traveltime inversion.
\newblock \emph{GEOPHYSICS}, 56\penalty0 (5):\penalty0 645--653, May 1991.
\newblock ISSN 0016-8033.
\newblock \doi{10.1190/1.1443081}.
\newblock URL \url{https://library.seg.org/doi/abs/10.1190/1.1443081}.

\bibitem[Brossier et~al.(2010)Brossier, Operto, and Virieux]{brossier_which_2010}
Romain Brossier, Stéphane Operto, and Jean Virieux.
\newblock Which data residual norm for robust elastic frequency-domain full waveform inversion?
\newblock \emph{GEOPHYSICS}, 75\penalty0 (3):\penalty0 R37--R46, May 2010.
\newblock ISSN 0016-8033.
\newblock \doi{10.1190/1.3379323}.
\newblock URL \url{https://library.seg.org/doi/abs/10.1190/1.3379323}.

\bibitem[Warner and Guasch(2014)]{warner_adaptive_2014}
M.~Warner and L.~Guasch.
\newblock Adaptive {Waveform} {Inversion} - {FWI} {Without} {Cycle} {Skipping} - {Theory}.
\newblock volume 2014, pages 1--5. European Association of Geoscientists \& Engineers, June 2014.
\newblock \doi{10.3997/2214-4609.20141092}.
\newblock URL \url{https://www.earthdoc.org/content/papers/10.3997/2214-4609.20141092}.

\bibitem[Métivier et~al.(2016)Métivier, Brossier, Mérigot, Oudet, and Virieux]{metivier_measuring_2016}
L.~Métivier, R.~Brossier, Q.~Mérigot, E.~Oudet, and J.~Virieux.
\newblock Measuring the misfit between seismograms using an optimal transport distance: application to full waveform inversion.
\newblock \emph{Geophysical Journal International}, 205\penalty0 (1):\penalty0 345--377, April 2016.
\newblock ISSN 0956-540X.
\newblock \doi{10.1093/gji/ggw014}.
\newblock URL \url{https://doi.org/10.1093/gji/ggw014}.

\bibitem[Sambridge and Mosegaard(2002)]{sambridge_monte_2002}
Malcolm Sambridge and Klaus Mosegaard.
\newblock Monte {Carlo} {Methods} in {Geophysical} {Inverse} {Problems}.
\newblock \emph{Reviews of Geophysics}, 40\penalty0 (3):\penalty0 3--1--3--29, 2002.
\newblock ISSN 1944-9208.
\newblock \doi{10.1029/2000RG000089}.
\newblock URL \url{https://onlinelibrary.wiley.com/doi/abs/10.1029/2000RG000089}.

\bibitem[Mosegaard and Tarantola(2002)]{mosegaard_16_2002}
Klaus Mosegaard and Albert Tarantola.
\newblock 16 - {Probabilistic} {Approach} to {Inverse} {Problems}.
\newblock In William H.~K. Lee, Hiroo Kanamori, Paul~C. Jennings, and Carl Kisslinger, editors, \emph{International {Geophysics}}, volume~81 of \emph{International {Handbook} of {Earthquake} and {Engineering} {Seismology}, {Part} {A}}, pages 237--265. Academic Press, January 2002.
\newblock \doi{10.1016/S0074-6142(02)80219-4}.
\newblock URL \url{https://www.sciencedirect.com/science/article/pii/S0074614202802194}.

\bibitem[Malinverno(2002)]{malinverno_parsimonious_2002}
Alberto Malinverno.
\newblock Parsimonious {Bayesian} {Markov} chain {Monte} {Carlo} inversion in a nonlinear geophysical problem.
\newblock \emph{Geophysical Journal International}, 151\penalty0 (3):\penalty0 675--688, December 2002.
\newblock ISSN 0956-540X.
\newblock \doi{10.1046/j.1365-246X.2002.01847.x}.
\newblock URL \url{https://doi.org/10.1046/j.1365-246X.2002.01847.x}.

\bibitem[Bodin and Sambridge(2009)]{bodin_seismic_2009}
Thomas Bodin and Malcolm Sambridge.
\newblock Seismic tomography with the reversible jump algorithm.
\newblock \emph{Geophysical Journal International}, 178\penalty0 (3):\penalty0 1411--1436, September 2009.
\newblock ISSN 0956-540X.
\newblock \doi{10.1111/j.1365-246X.2009.04226.x}.
\newblock URL \url{https://doi.org/10.1111/j.1365-246X.2009.04226.x}.

\bibitem[Curtis and Lomax(2001)]{curtis_prior_2001}
Andrew Curtis and Anthony Lomax.
\newblock Prior information, sampling distributions, and the curse of dimensionality.
\newblock \emph{GEOPHYSICS}, 66\penalty0 (2):\penalty0 372--378, March 2001.
\newblock ISSN 0016-8033.
\newblock \doi{10.1190/1.1444928}.
\newblock URL \url{https://library.seg.org/doi/abs/10.1190/1.1444928}.

\bibitem[Fichtner et~al.(2019)Fichtner, Zunino, and Gebraad]{fichtner_hamiltonian_2019}
Andreas Fichtner, Andrea Zunino, and Lars Gebraad.
\newblock Hamiltonian {Monte} {Carlo} solution of tomographic inverse problems.
\newblock \emph{Geophysical Journal International}, 216\penalty0 (2):\penalty0 1344--1363, February 2019.
\newblock ISSN 0956-540X.
\newblock \doi{10.1093/gji/ggy496}.
\newblock URL \url{https://doi.org/10.1093/gji/ggy496}.

\bibitem[Gebraad et~al.(2020)Gebraad, Boehm, and Fichtner]{gebraad_bayesian_2020}
Lars Gebraad, Christian Boehm, and Andreas Fichtner.
\newblock Bayesian {Elastic} {Full}-{Waveform} {Inversion} {Using} {Hamiltonian} {Monte} {Carlo}.
\newblock \emph{Journal of Geophysical Research: Solid Earth}, 125\penalty0 (3):\penalty0 e2019JB018428, 2020.
\newblock ISSN 2169-9356.
\newblock \doi{10.1029/2019JB018428}.
\newblock URL \url{https://onlinelibrary.wiley.com/doi/abs/10.1029/2019JB018428}.

\bibitem[Martin et~al.(2012)Martin, Wilcox, Burstedde, and Ghattas]{martin_stochastic_2012}
James Martin, Lucas~C. Wilcox, Carsten Burstedde, and Omar Ghattas.
\newblock A {Stochastic} {Newton} {MCMC} {Method} for {Large}-{Scale} {Statistical} {Inverse} {Problems} with {Application} to {Seismic} {Inversion}.
\newblock \emph{SIAM Journal on Scientific Computing}, 34\penalty0 (3):\penalty0 A1460--A1487, January 2012.
\newblock ISSN 1064-8275.
\newblock \doi{10.1137/110845598}.
\newblock URL \url{https://epubs.siam.org/doi/abs/10.1137/110845598}.

\bibitem[Sambridge(2014)]{sambridge_parallel_2014}
Malcolm Sambridge.
\newblock A {Parallel} {Tempering} algorithm for probabilistic sampling and multimodal optimization.
\newblock \emph{Geophysical Journal International}, 196\penalty0 (1):\penalty0 357--374, January 2014.
\newblock ISSN 0956-540X.
\newblock \doi{10.1093/gji/ggt342}.
\newblock URL \url{https://doi.org/10.1093/gji/ggt342}.

\bibitem[Aleardi(2021)]{aleardi_gradientbased_2021}
Mattia Aleardi.
\newblock A gradient‐based {Markov} chain {Monte} {Carlo} algorithm for elastic pre‐stack inversion with data and model space reduction.
\newblock \emph{Geophysical Prospecting}, 69\penalty0 (5):\penalty0 926--948, May 2021.
\newblock ISSN 1365-2478.
\newblock \doi{10.1111/1365-2478.13081}.
\newblock URL \url{https://www.earthdoc.org/content/journals/10.1111/1365-2478.13081}.

\bibitem[Zhao and Sen(2021)]{zhao_gradient-based_2021}
Zeyu Zhao and Mrinal~K. Sen.
\newblock A gradient-based {Markov} chain {Monte} {Carlo} method for full-waveform inversion and uncertainty analysis.
\newblock \emph{GEOPHYSICS}, 86\penalty0 (1):\penalty0 R15--R30, January 2021.
\newblock ISSN 0016-8033.
\newblock \doi{10.1190/geo2019-0585.1}.
\newblock URL \url{https://library.seg.org/doi/abs/10.1190/geo2019-0585.1}.

\bibitem[Berti et~al.(2024{\natexlab{a}})Berti, Aleardi, and Stucchi]{berti_computationally_2024}
Sean Berti, Mattia Aleardi, and Eusebio Stucchi.
\newblock A computationally efficient {Bayesian} approach to full‐waveform inversion.
\newblock \emph{Geophysical Prospecting}, 72\penalty0 (2):\penalty0 580--603, January 2024{\natexlab{a}}.
\newblock ISSN 1365-2478.
\newblock \doi{10.1111/1365-2478.13437}.
\newblock URL \url{https://www.earthdoc.org/content/journals/10.1111/1365-2478.13437}.

\bibitem[Berti et~al.(2024{\natexlab{b}})Berti, Aleardi, and Stucchi]{berti_elastic_2024}
Sean Berti, Mattia Aleardi, and Eusebio Stucchi.
\newblock A bayesian approach to elastic full-waveform inversion: application to two synthetic near surface models.
\newblock \emph{Bulletin of Geophysics and Oceanography}, 65\penalty0 (2):\penalty0 291--308, 2024{\natexlab{b}}.
\newblock \doi{10.4430/bgo00442}.

\bibitem[Ray et~al.(2016)Ray, Sekar, Hoversten, and Albertin]{ray_frequency_2016}
Anandaroop Ray, Anusha Sekar, G.~Michael Hoversten, and Uwe Albertin.
\newblock Frequency domain full waveform elastic inversion of marine seismic data from the {Alba} field using a {Bayesian} trans-dimensional algorithm.
\newblock \emph{Geophysical Journal International}, 205\penalty0 (2):\penalty0 915--937, May 2016.
\newblock ISSN 0956-540X.
\newblock \doi{10.1093/gji/ggw061}.
\newblock URL \url{https://doi.org/10.1093/gji/ggw061}.

\bibitem[Sen and Biswas(2017)]{sen_transdimensional_2017}
Mrinal~K. Sen and Reetam Biswas.
\newblock Transdimensional seismic inversion using the reversible jump {Hamiltonian} {Monte} {Carlo} algorithm.
\newblock \emph{GEOPHYSICS}, 82\penalty0 (3):\penalty0 R119--R134, May 2017.
\newblock ISSN 0016-8033.
\newblock \doi{10.1190/geo2016-0010.1}.
\newblock URL \url{https://library.seg.org/doi/abs/10.1190/geo2016-0010.1}.

\bibitem[Jordan et~al.(1998)Jordan, Ghahramani, Jaakkola, and Saul]{jordan_introduction_1998}
Michael~I. Jordan, Zoubin Ghahramani, Tommi~S. Jaakkola, and Lawrence~K. Saul.
\newblock An {Introduction} to {Variational} {Methods} for {Graphical} {Models}.
\newblock In Michael~I. Jordan, editor, \emph{Learning in {Graphical} {Models}}, pages 105--161. Springer Netherlands, Dordrecht, 1998.
\newblock ISBN 978-94-010-6104-9 978-94-011-5014-9.
\newblock \doi{10.1007/978-94-011-5014-9_5}.
\newblock URL \url{http://link.springer.com/10.1007/978-94-011-5014-9_5}.

\bibitem[Blei et~al.(2017)Blei, Kucukelbir, and McAuliffe]{blei_variational_2017}
David~M. Blei, Alp Kucukelbir, and Jon~D. McAuliffe.
\newblock Variational {Inference}: {A} {Review} for {Statisticians}.
\newblock \emph{Journal of the American Statistical Association}, 112\penalty0 (518):\penalty0 859--877, April 2017.
\newblock ISSN 0162-1459.
\newblock \doi{10.1080/01621459.2017.1285773}.
\newblock URL \url{https://doi.org/10.1080/01621459.2017.1285773}.

\bibitem[Zhang et~al.(2019)Zhang, Bütepage, Kjellström, and Mandt]{zhang_advances_2019}
Cheng Zhang, Judith Bütepage, Hedvig Kjellström, and Stephan Mandt.
\newblock Advances in {Variational} {Inference}.
\newblock \emph{IEEE Transactions on Pattern Analysis and Machine Intelligence}, 41\penalty0 (8):\penalty0 2008--2026, August 2019.
\newblock ISSN 1939-3539.
\newblock \doi{10.1109/TPAMI.2018.2889774}.
\newblock URL \url{https://ieeexplore.ieee.org/abstract/document/8588399}.

\bibitem[Kucukelbir et~al.(2017)Kucukelbir, Tran, Ranganath, Gelman, and Blei]{kucukelbir_automatic_2017}
Alp Kucukelbir, Dustin Tran, Rajesh Ranganath, Andrew Gelman, and David~M. Blei.
\newblock Automatic {Differentiation} {Variational} {Inference}.
\newblock \emph{Journal of Machine Learning Research}, 18\penalty0 (14):\penalty0 1--45, 2017.
\newblock ISSN 1533-7928.
\newblock URL \url{http://jmlr.org/papers/v18/16-107.html}.

\bibitem[Kingma et~al.(2016)Kingma, Salimans, Jozefowicz, Chen, Sutskever, and Welling]{kingma_improved_2016}
Durk~P Kingma, Tim Salimans, Rafal Jozefowicz, Xi~Chen, Ilya Sutskever, and Max Welling.
\newblock Improved {Variational} {Inference} with {Inverse} {Autoregressive} {Flow}.
\newblock In \emph{Advances in {Neural} {Information} {Processing} {Systems}}, volume~29. Curran Associates, Inc., 2016.
\newblock URL \url{https://proceedings.neurips.cc/paper_files/paper/2016/hash/ddeebdeefdb7e7e7a697e1c3e3d8ef54-Abstract.html}.

\bibitem[Gallego and Insua(2020)]{gallego_stochastic_2020}
Victor Gallego and David~Rios Insua.
\newblock Stochastic {Gradient} {MCMC} with {Repulsive} {Forces}, February 2020.
\newblock URL \url{http://arxiv.org/abs/1812.00071}.

\bibitem[Liu and Wang(2016)]{liu_stein_2016}
Qiang Liu and Dilin Wang.
\newblock Stein {Variational} {Gradient} {Descent}: {A} {General} {Purpose} {Bayesian} {Inference} {Algorithm}.
\newblock In \emph{Advances in {Neural} {Information} {Processing} {Systems}}, volume~29. Curran Associates, Inc., 2016.
\newblock URL \url{https://proceedings.neurips.cc/paper_files/paper/2016/hash/b3ba8f1bee1238a2f37603d90b58898d-Abstract.html}.

\bibitem[Zhang and Curtis(2020{\natexlab{a}})]{zhang_seismic_2020}
Xin Zhang and Andrew Curtis.
\newblock Seismic {Tomography} {Using} {Variational} {Inference} {Methods}.
\newblock \emph{Journal of Geophysical Research: Solid Earth}, 125\penalty0 (4):\penalty0 e2019JB018589, 2020{\natexlab{a}}.
\newblock ISSN 2169-9356.
\newblock \doi{10.1029/2019JB018589}.
\newblock URL \url{https://onlinelibrary.wiley.com/doi/abs/10.1029/2019JB018589}.

\bibitem[Zhao et~al.(2022)Zhao, Curtis, and Zhang]{zhao_bayesian_2022}
Xuebin Zhao, Andrew Curtis, and Xin Zhang.
\newblock Bayesian seismic tomography using normalizing flows.
\newblock \emph{Geophysical Journal International}, 228\penalty0 (1):\penalty0 213--239, January 2022.
\newblock ISSN 0956-540X.
\newblock \doi{10.1093/gji/ggab298}.
\newblock URL \url{https://doi.org/10.1093/gji/ggab298}.

\bibitem[Siahkoohi et~al.(2021)Siahkoohi, Rizzuti, Louboutin, Witte, and Herrmann]{siahkoohi_preconditioned_2021}
Ali Siahkoohi, Gabrio Rizzuti, Mathias Louboutin, Philipp~A. Witte, and Felix~J. Herrmann.
\newblock Preconditioned training of normalizing flows for variational inference in inverse problems, January 2021.
\newblock URL \url{http://arxiv.org/abs/2101.03709}.

\bibitem[Ravasi(2023)]{ravasi_2023}
Matteo Ravasi.
\newblock Multi-realization seismic data processing with deep variational preconditioners.
\newblock 2023.
\newblock URL \url{https://dx.doi.org/10.1190/image2022-3745255.1}.

\bibitem[Smith et~al.(2022)Smith, Ross, Azizzadenesheli, and Muir]{smith_hyposvi_2022}
Jonthan~D Smith, Zachary~E Ross, Kamyar Azizzadenesheli, and Jack~B Muir.
\newblock {HypoSVI}: {Hypocentre} inversion with {Stein} variational inference and physics informed neural networks.
\newblock \emph{Geophysical Journal International}, 228\penalty0 (1):\penalty0 698--710, January 2022.
\newblock ISSN 0956-540X.
\newblock \doi{10.1093/gji/ggab309}.
\newblock URL \url{https://doi.org/10.1093/gji/ggab309}.

\bibitem[Zhang and Curtis(2020{\natexlab{b}})]{zhang_variational_2020}
Xin Zhang and Andrew Curtis.
\newblock Variational full-waveform inversion.
\newblock \emph{Geophysical Journal International}, 222\penalty0 (1):\penalty0 406--411, July 2020{\natexlab{b}}.
\newblock ISSN 0956-540X.
\newblock \doi{10.1093/gji/ggaa170}.
\newblock URL \url{https://doi.org/10.1093/gji/ggaa170}.

\bibitem[Urozayev et~al.(2022)Urozayev, Ait-El-Fquih, Hoteit, and Peter]{urozayev_reduced-order_2022}
Dias Urozayev, Boujemaa Ait-El-Fquih, Ibrahim Hoteit, and Daniel Peter.
\newblock A reduced-order variational {Bayesian} approach for efficient subsurface imaging.
\newblock \emph{Geophysical Journal International}, 229\penalty0 (2):\penalty0 838--852, May 2022.
\newblock ISSN 0956-540X.
\newblock \doi{10.1093/gji/ggab507}.
\newblock URL \url{https://doi.org/10.1093/gji/ggab507}.

\bibitem[Lomas et~al.(2023)Lomas, Luo, Irakarama, Johnston, Vyas, and Shen]{lomas_3d_2023}
A.~Lomas, S.~Luo, M.~Irakarama, R.~Johnston, M.~Vyas, and X.~Shen.
\newblock {3D} {Probabilistic} {Full} {Waveform} {Inversion}: {Application} to {Gulf} of {Mexico} {Field} {Data}.
\newblock In \emph{84th {EAGE} {Annual} {Conference} \& {Exhibition}}, pages 1--5, Vienna, Austria,, 2023. European Association of Geoscientists \& Engineers.
\newblock \doi{10.3997/2214-4609.202310720}.
\newblock URL \url{https://www.earthdoc.org/content/papers/10.3997/2214-4609.202310720}.

\bibitem[Zhang et~al.(2023)Zhang, Lomas, Zhou, Zheng, and Curtis]{zhang_3-d_2023}
Xin Zhang, Angus Lomas, Muhong Zhou, York Zheng, and Andrew Curtis.
\newblock 3-{D} {Bayesian} variational full waveform inversion.
\newblock \emph{Geophysical Journal International}, 234\penalty0 (1):\penalty0 546--561, July 2023.
\newblock ISSN 0956-540X.
\newblock \doi{10.1093/gji/ggad057}.
\newblock URL \url{https://doi.org/10.1093/gji/ggad057}.

\bibitem[Izzatullah et~al.(2024{\natexlab{a}})Izzatullah, Alkhalifah, Romero, Corrales, Luiken, and Ravasi]{izzatullah_posterior_2024}
Muhammad Izzatullah, Tariq Alkhalifah, Juan Romero, Miguel Corrales, Nick Luiken, and Matteo Ravasi.
\newblock Posterior sampling with convolutional neural network-based plug-and-play regularization with applications to poststack seismic inversion.
\newblock \emph{GEOPHYSICS}, 89\penalty0 (2):\penalty0 R137--R153, March 2024{\natexlab{a}}.
\newblock ISSN 0016-8033.
\newblock \doi{10.1190/geo2023-0035.1}.
\newblock URL \url{https://library.seg.org/doi/abs/10.1190/geo2023-0035.1}.

\bibitem[Corrales et~al.(2022)Corrales, Izzatullah, Ravasi, and Hoteit]{corrales_bayesian_2022}
Miguel Corrales, Muhammad Izzatullah, Matteo Ravasi, and Hussein Hoteit.
\newblock Bayesian {RockAVO}: {Direct} petrophysical inversion with hierarchical conditional {GANs}.
\newblock 2022.
\newblock URL \url{https://dx.doi.org/10.1190/image2022-3745255.1}.

\bibitem[Izzatullah et~al.(2024{\natexlab{b}})Izzatullah, Alali, Ravasi, and Alkhalifah]{izzatullah_physics-reliable_2024}
Muhammad Izzatullah, Abdullah Alali, Matteo Ravasi, and Tariq Alkhalifah.
\newblock Physics-reliable frugal local uncertainty analysis for full waveform inversion.
\newblock \emph{Geophysical Prospecting}, n/a\penalty0 (n/a), June 2024{\natexlab{b}}.
\newblock ISSN 1365-2478.
\newblock \doi{10.1111/1365-2478.13528}.
\newblock URL \url{https://onlinelibrary.wiley.com/doi/abs/10.1111/1365-2478.13528}.

\bibitem[Zhuo et~al.(2018)Zhuo, Liu, Shi, Zhu, Chen, and Zhang]{zhuo_message_2018}
Jingwei Zhuo, Chang Liu, Jiaxin Shi, Jun Zhu, Ning Chen, and Bo~Zhang.
\newblock Message {Passing} {Stein} {Variational} {Gradient} {Descent}.
\newblock In \emph{Proceedings of the 35th {International} {Conference} on {Machine} {Learning}}, pages 6018--6027. PMLR, July 2018.
\newblock URL \url{https://proceedings.mlr.press/v80/zhuo18a.html}.
\newblock ISSN: 2640-3498.

\bibitem[Kullback and Leibler(1951)]{kullback_information_1951}
S.~Kullback and R.~A. Leibler.
\newblock On {Information} and {Sufficiency}.
\newblock \emph{The Annals of Mathematical Statistics}, 22\penalty0 (1):\penalty0 79--86, 1951.
\newblock ISSN 0003-4851.
\newblock URL \url{https://www.jstor.org/stable/2236703}.
\newblock Publisher: Institute of Mathematical Statistics.

\bibitem[Gorham and Mackey(2017)]{gorham_measuring_2017}
Jackson Gorham and Lester Mackey.
\newblock Measuring {Sample} {Quality} with {Kernels}.
\newblock In \emph{Proceedings of the 34th {International} {Conference} on {Machine} {Learning}}, pages 1292--1301. PMLR, July 2017.
\newblock URL \url{https://proceedings.mlr.press/v70/gorham17a.html}.
\newblock ISSN: 2640-3498.

\bibitem[Ba et~al.(2021)Ba, Erdogdu, Ghassemi, Sun, Suzuki, Wu, and Zhang]{ba_understanding_2021}
Jimmy Ba, Murat~A. Erdogdu, Marzyeh Ghassemi, Shengyang Sun, Taiji Suzuki, Denny Wu, and Tianzong Zhang.
\newblock Understanding the {Variance} {Collapse} of {SVGD} in {High} {Dimensions}.
\newblock October 2021.
\newblock URL \url{https://openreview.net/forum?id=Qycd9j5Qp9J}.

\bibitem[D'Angelo and Fortuin(2021)]{dangelo_annealed_2021}
Francesco D'Angelo and Vincent Fortuin.
\newblock Annealed {Stein} {Variational} {Gradient} {Descent}, March 2021.
\newblock URL \url{http://arxiv.org/abs/2101.09815}.

\bibitem[Hotelling(1933)]{hotelling_analysis_1933}
H.~Hotelling.
\newblock Analysis of a complex of statistical variables into principal components.
\newblock \emph{Journal of Educational Psychology}, 24\penalty0 (6):\penalty0 417--441, 1933.
\newblock ISSN 1939-2176.
\newblock \doi{10.1037/h0071325}.
\newblock Place: US Publisher: Warwick \& York.

\bibitem[Campello et~al.(2013)Campello, Moulavi, and Sander]{campello_density-based_2013}
Ricardo J. G.~B. Campello, Davoud Moulavi, and Joerg Sander.
\newblock Density-{Based} {Clustering} {Based} on {Hierarchical} {Density} {Estimates}.
\newblock In Jian Pei, Vincent~S. Tseng, Longbing Cao, Hiroshi Motoda, and Guandong Xu, editors, \emph{Advances in {Knowledge} {Discovery} and {Data} {Mining}}, pages 160--172, Berlin, Heidelberg, 2013. Springer.
\newblock ISBN 978-3-642-37456-2.
\newblock \doi{10.1007/978-3-642-37456-2_14}.

\bibitem[Ester et~al.(1996)Ester, Kriegel, Sander, Xu, et~al.]{ester1996density}
Martin Ester, Hans-Peter Kriegel, J{\"o}rg Sander, Xiaowei Xu, et~al.
\newblock A density-based algorithm for discovering clusters in large spatial databases with noise.
\newblock In \emph{kdd}, volume~96, pages 226--231, 1996.

\bibitem[Brougois et~al.(1990)Brougois, Bourget, Lailly, Poulet, Ricarte, and Versteeg]{brougois_marmousi_1990}
Aline Brougois, Marielle Bourget, Patrick Lailly, Michel Poulet, Patrice Ricarte, and Roelof Versteeg.
\newblock Marmousi, model and data.
\newblock page~cp. European Association of Geoscientists \& Engineers, May 1990.
\newblock ISBN 978-90-73781-01-6.
\newblock \doi{10.3997/2214-4609.201411190}.
\newblock URL \url{https://www.earthdoc.org/content/papers/10.3997/2214-4609.201411190}.
\newblock ISSN: 2214-4609.

\bibitem[Richardson(2023)]{richardson_alan_2023}
Alan Richardson.
\newblock Deepwave, September 2023.
\newblock URL \url{https://doi.org/10.5281/zenodo.8381177}.

\bibitem[Bunks et~al.(1995)Bunks, Saleck, Zaleski, and Chavent]{bunks_multiscale_1995}
C.~Bunks, F.~Saleck, S.~Zaleski, and G.~Chavent.
\newblock Multiscale seismic waveform inversion.
\newblock \emph{Geophysics}, 60\penalty0 (5):\penalty0 1457--1473, 1995.

\bibitem[Chen and Ghattas(2020)]{chen_projected_2020}
Peng Chen and Omar Ghattas.
\newblock Projected {Stein} {Variational} {Gradient} {Descent}.
\newblock In \emph{Advances in {Neural} {Information} {Processing} {Systems}}, volume~33, pages 1947--1958. Curran Associates, Inc., 2020.
\newblock URL \url{https://proceedings.neurips.cc/paper_files/paper/2020/hash/14faf969228fc18fcd4fcf59437b0c97-Abstract.html}.

\bibitem[Liu et~al.(2022)Liu, Zhu, Ton, Wynne, and Duncan]{liu_grassmann_2022}
Xing Liu, Harrison Zhu, Jean-François Ton, George Wynne, and Andrew Duncan.
\newblock Grassmann {Stein} {Variational} {Gradient} {Descent}, March 2022.
\newblock URL \url{http://arxiv.org/abs/2202.03297}.

\bibitem[Berti et~al.(2024{\natexlab{c}})Berti, Aleardi, and Stucchi]{berti_probabilistic_2024}
Sean Berti, Mattia Aleardi, and Eusebio Stucchi.
\newblock A probabilistic full waveform inversion of surface waves.
\newblock \emph{Geophysical Prospecting}, 72\penalty0 (9):\penalty0 3448--3473, 2024{\natexlab{c}}.
\newblock ISSN 1365-2478.
\newblock \doi{10.1111/1365-2478.13595}.

\bibitem[Sun and Williamson(2024)]{sun2024invertible}
Yen Sun and Paul Williamson.
\newblock Invertible neural networks for uncertainty quantification in refraction tomography.
\newblock \emph{The Leading Edge}, 43\penalty0 (6):\penalty0 358--366, 2024.

\end{thebibliography}






\newpage
\clearpage
\appendix
\renewcommand{\thesection}{\Alph{section}} 
\renewcommand{\thefigure}{\thesection\arabic{figure}} 
\setcounter{figure}{0}

\section{\textsc{Supplementary Results for Single-scale Experiments}}\label{appendix_A}

\begin{figure}[!h]
    \centering
    \includegraphics[width=\textwidth]{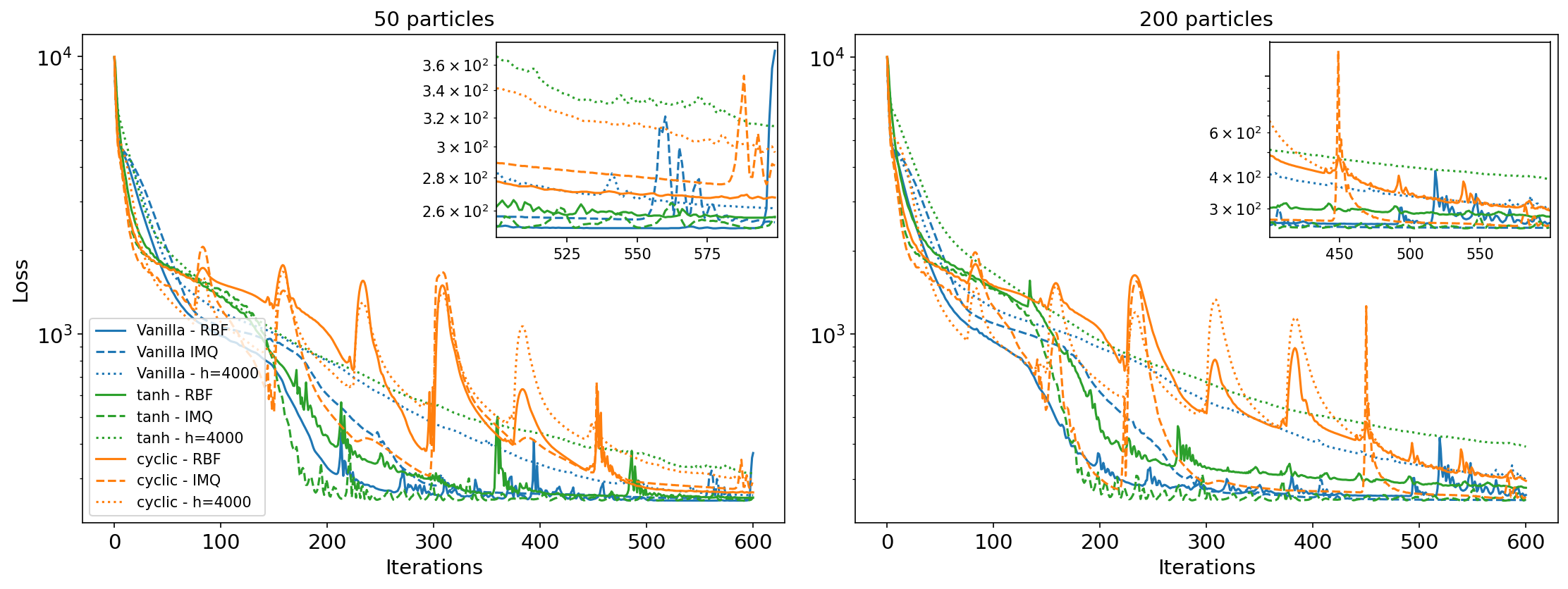}
    \vspace{-0.5cm}
    \caption{Data misfit for 50 particles (left) and 200 particles (right) across different experiments in the single-frequency scenario. The zoom window provides a clearer misfit comparison of the final 150 iterations.}
    \label{fig:loss_comparison}
\end{figure}

\begin{figure}
    \centering
    \includegraphics[width=\textwidth]{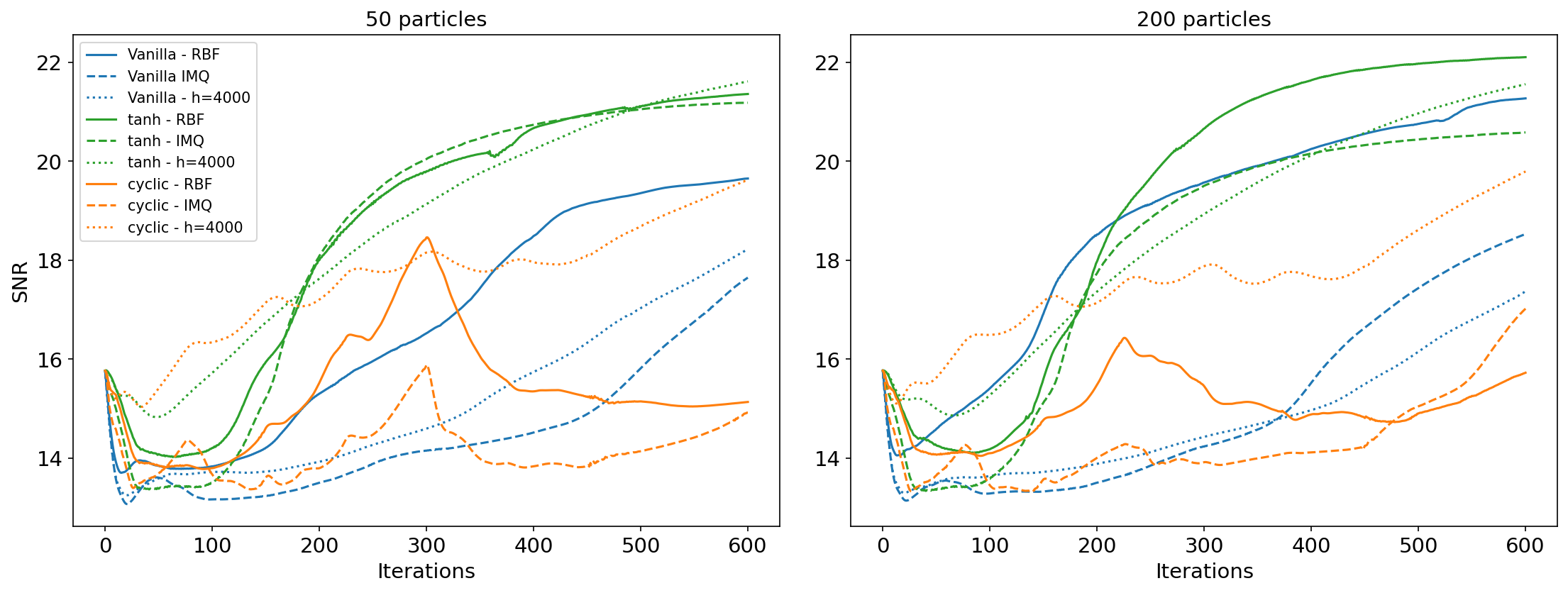}
    \vspace{-0.5cm}
    \caption{SNR for 50 particles (left) and 200 particles (right) across different experiments in the single-frequency scenario. }
    \label{fig:snr_comparison}
\end{figure}

\begin{figure}
    \centering
    \includegraphics[width=\textwidth]{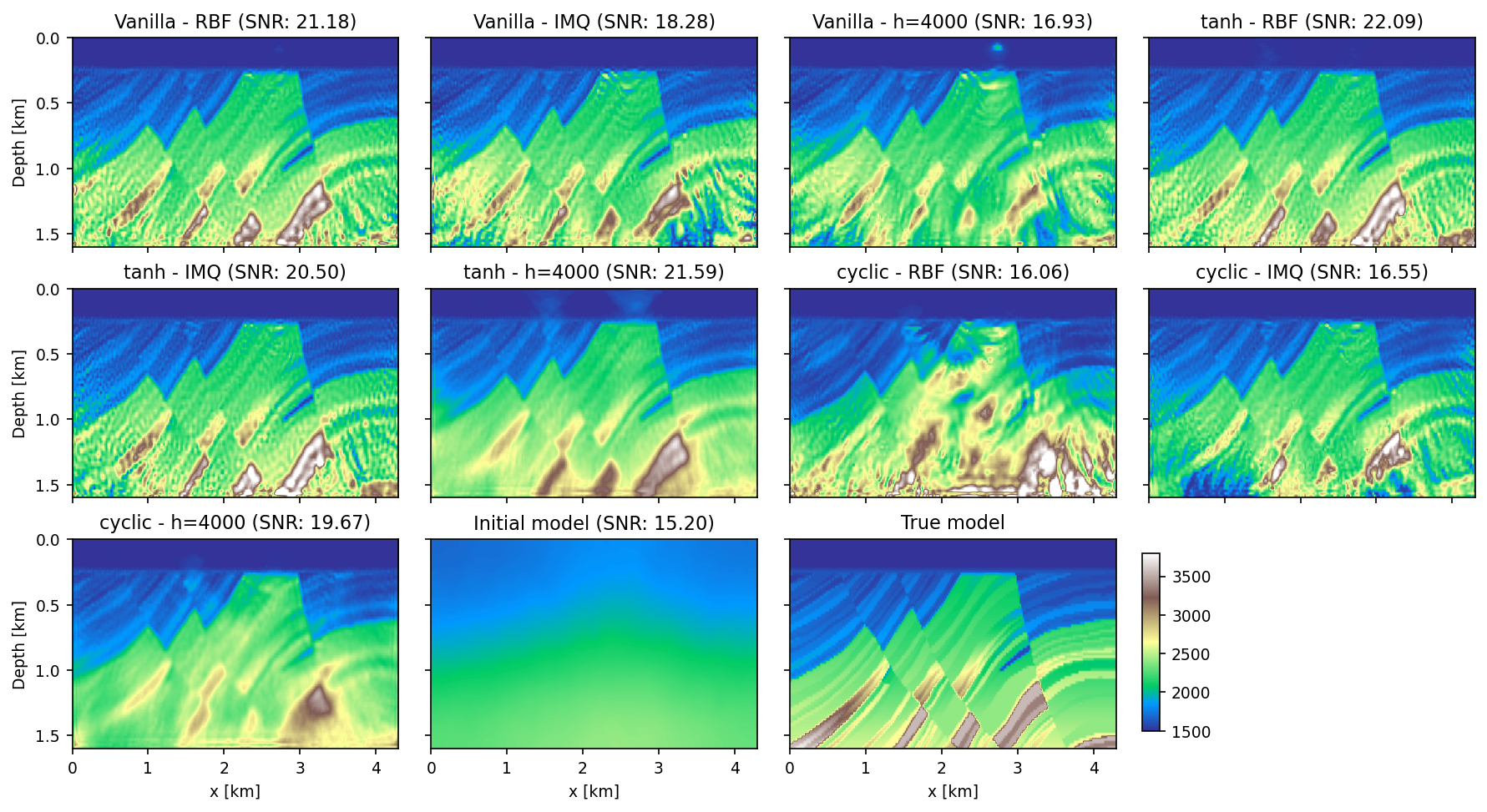}
    \vspace{-0.5cm}
    \caption{Comparison of the mean models from different experiments using 200 particles, highlighting various SVGD variants and hyperparameters after 600 iterations in the single-frequency scenario. The velocity values are expressed in m/s}
    \label{fig:mean_comparison_200p}
\end{figure}

\begin{figure}
    \centering
    \includegraphics[width=\textwidth]{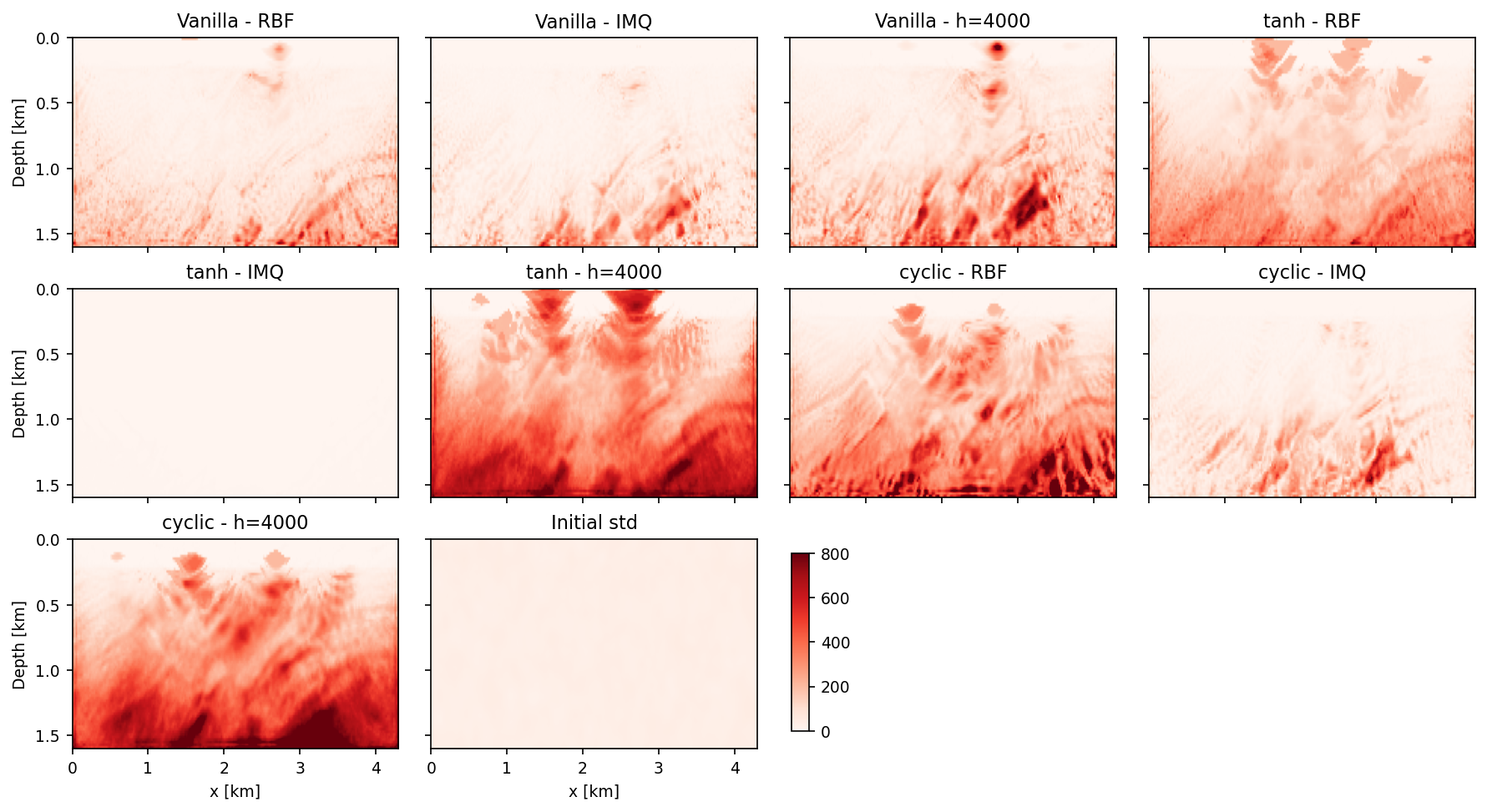}
    \vspace{-0.5cm}
    \caption{Comparison of standard deviation from different experiments using 200 particles, highlighting various SVGD variants and hyperparameters after 600 iterations in the single-frequency scenario. The velocity values are expressed in m/s}
    \label{fig:std_comparison_200p}
\end{figure}

\begin{figure}
    \centering
    \includegraphics[width=\textwidth]{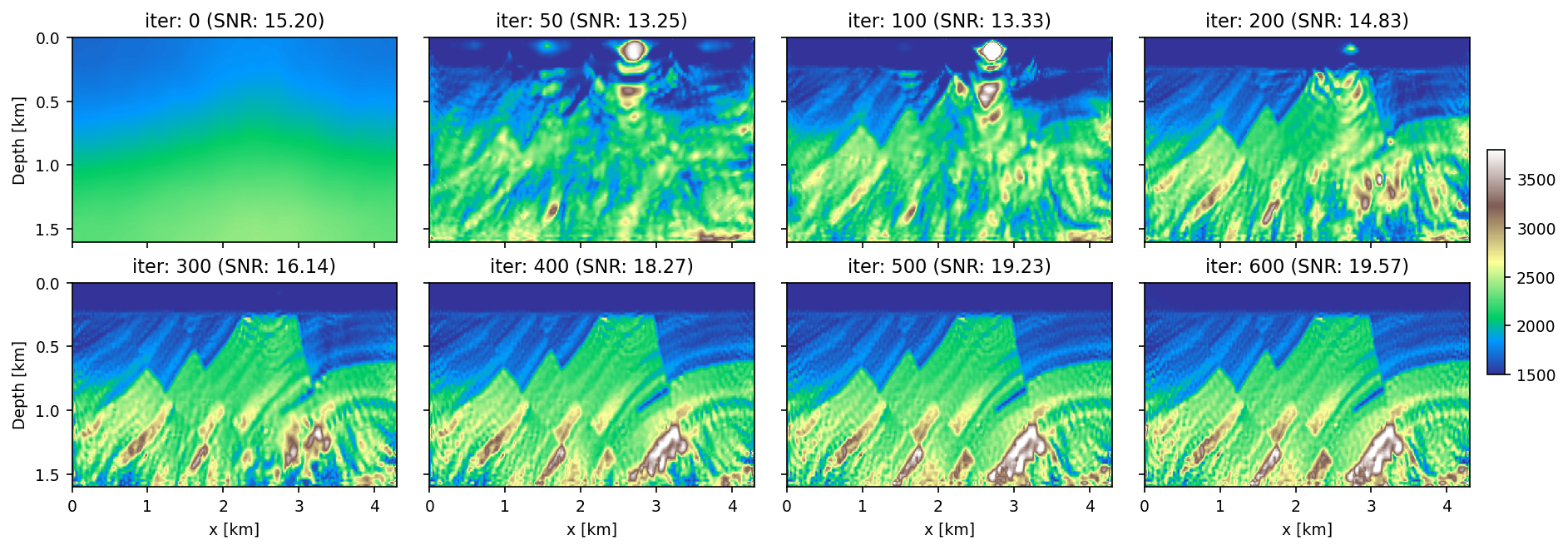}
    \vspace{-0.5cm}
    \caption{Evolution of the mean for the 200-particle experiment using Vanilla SVGD with RBF kernel.}
    \label{fig:mean_evol_200p_VSVGD_RBF}
\end{figure}

\begin{figure}
    \centering
    \includegraphics[width=\textwidth]{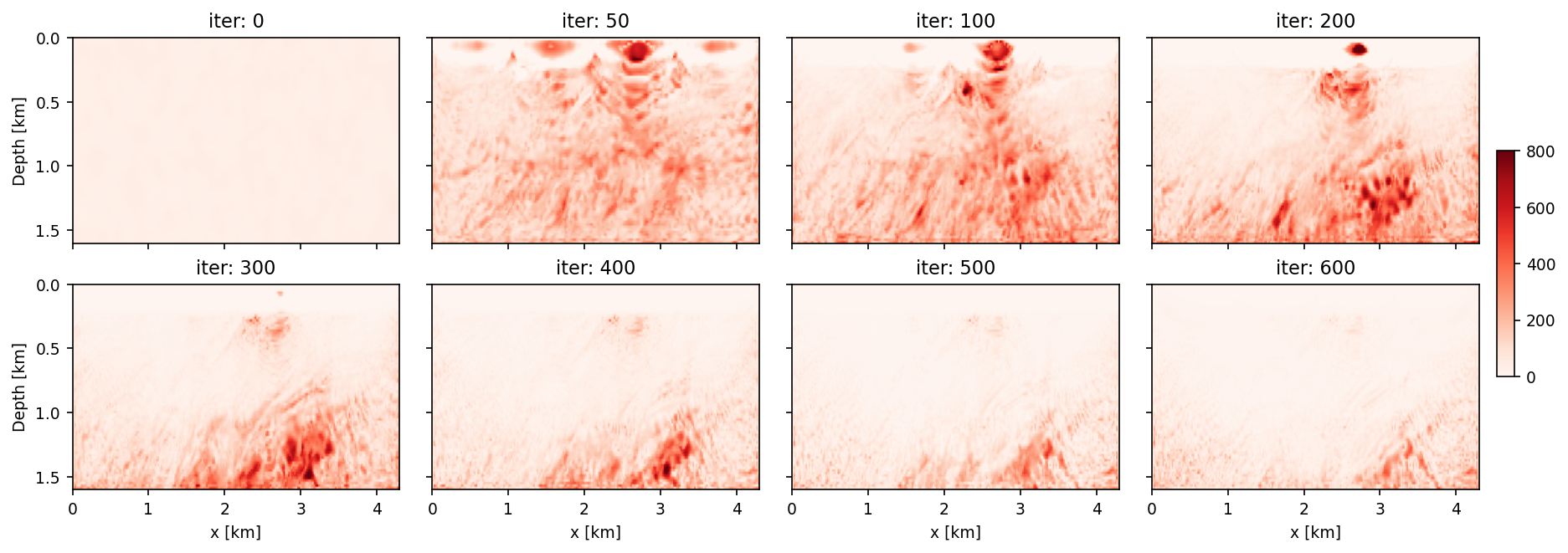}
    \vspace{-0.5cm}
    \caption{Evolution of the standard deviation for the 200-particle experiment using Vanilla SVGD with RBF kernel.}
    \label{fig:std_evol_200p_VSVGD_RBF}
\end{figure}

\begin{figure}
    \centering
    \includegraphics[width=\textwidth]{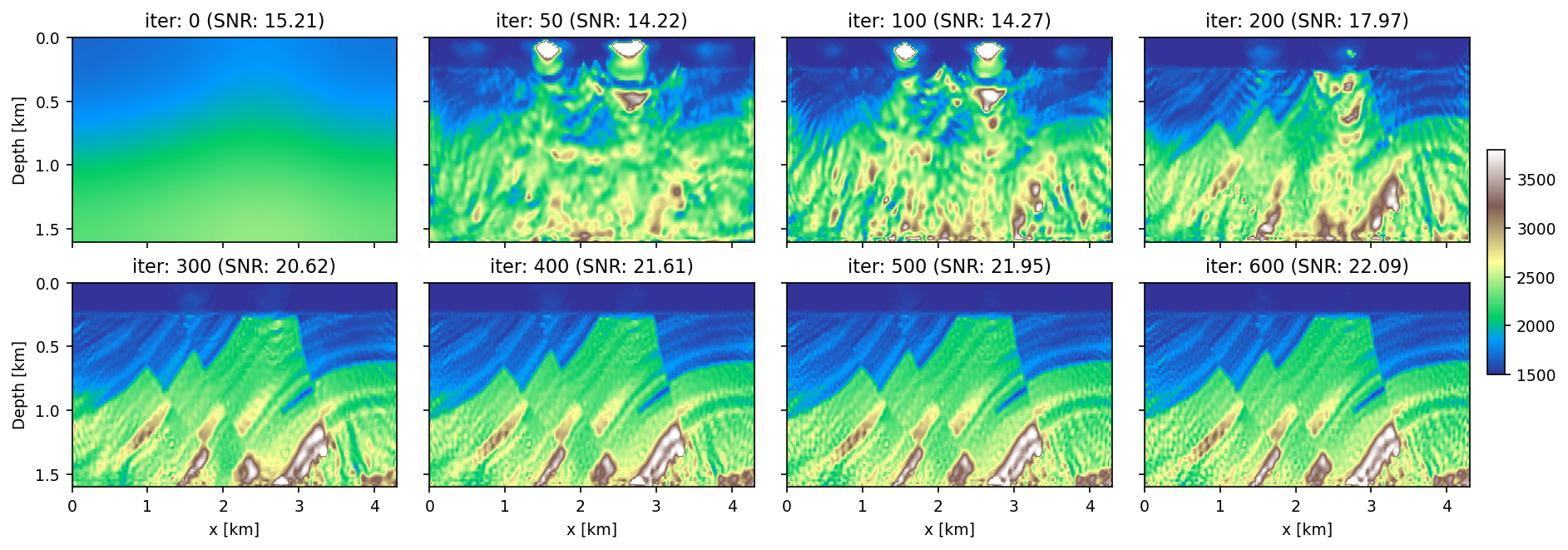}
    \vspace{-0.5cm}
    \caption{Evolution of the mean for the 200-particle experiment using Annealed SVGD (tanh) with RBF kernel.}
    \label{fig:mean_evol_200p_tanh_RBF}
\end{figure}

\begin{figure}
    \centering
    \includegraphics[width=\textwidth]{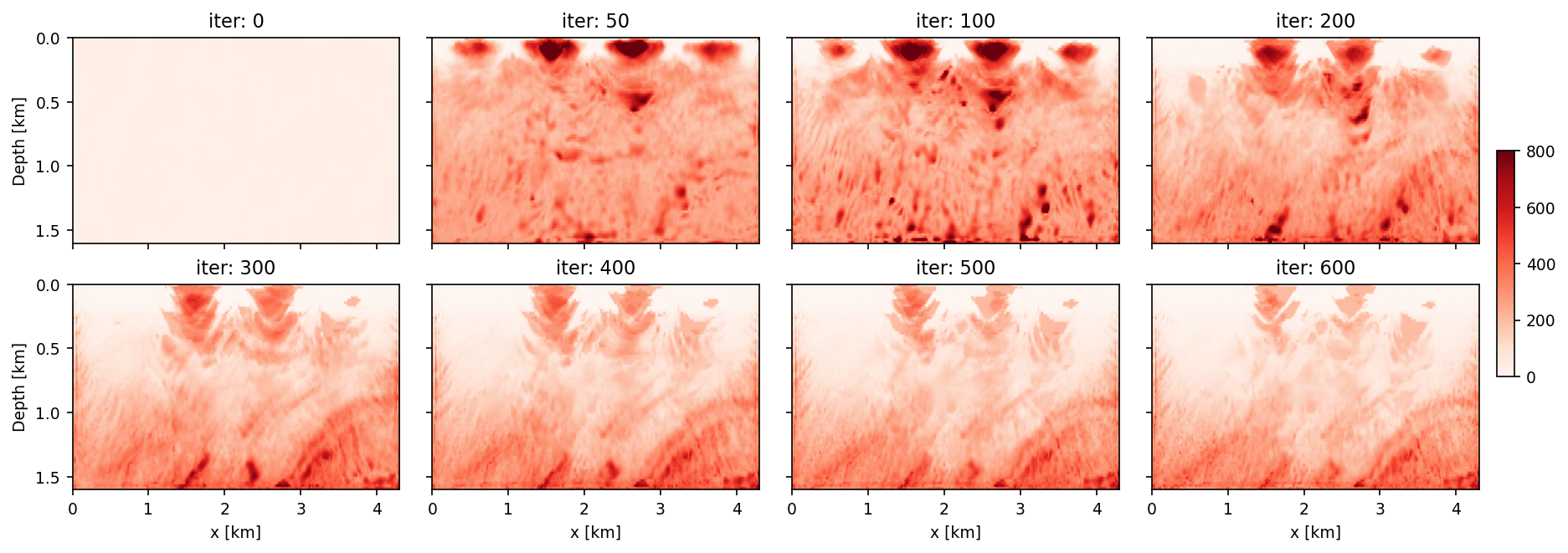}
    \vspace{-0.5cm}
    \caption{Evolution of the standard deviation for the 200-particle experiment using Annealed SVGD (tanh) with RBF kernel.}
    \label{fig:std_evol_200p_tanh_RBF}
\end{figure}



\begin{figure}
    \centering
    \includegraphics[width=\textwidth]{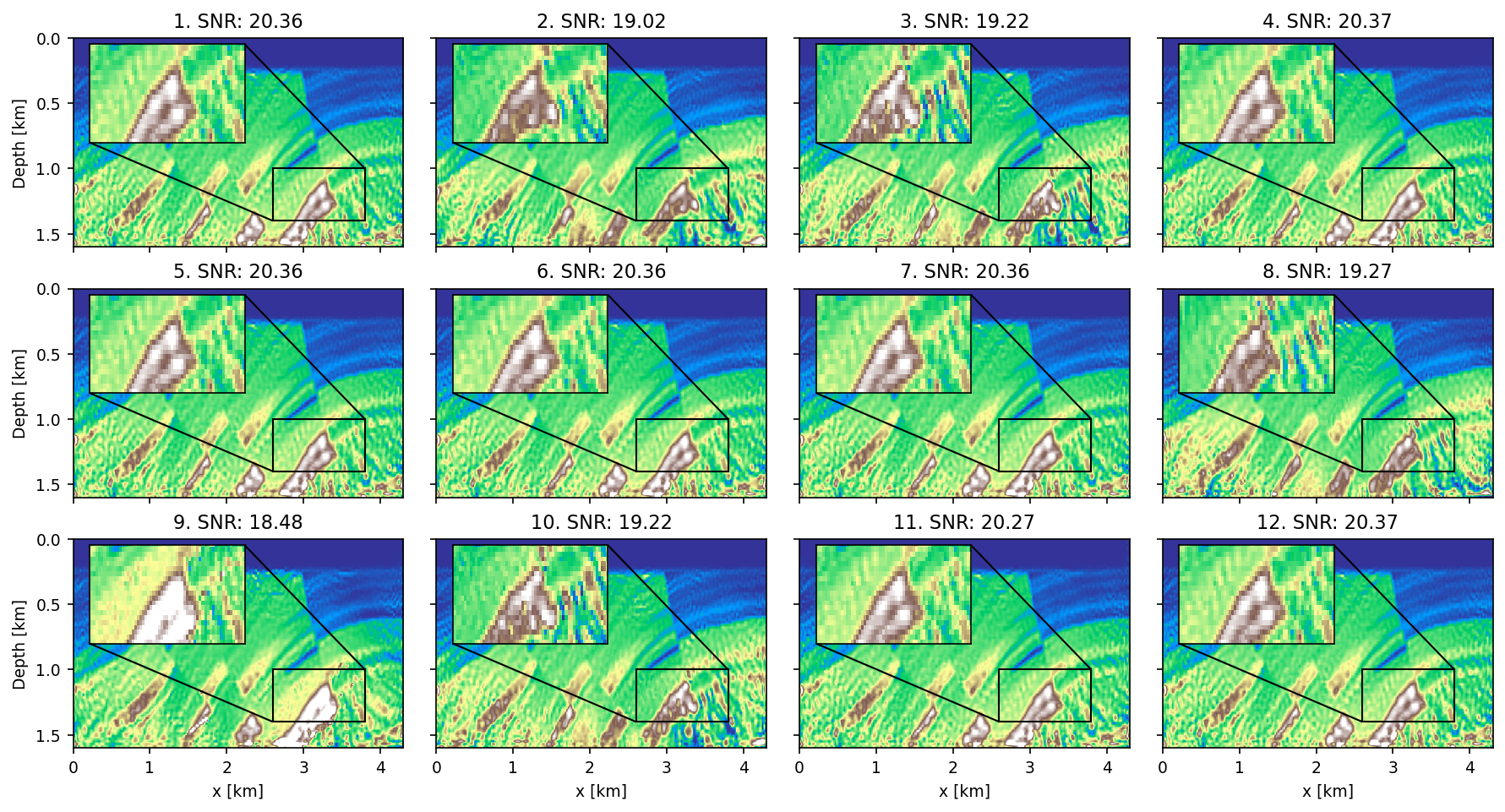}
    \vspace{-0.5cm}
    \caption{Single-frequency scenario: visualization of 12 particles from a 200-particle experiment using vanilla SVGD with RBF Kernel after 600 iterations.}
    \label{fig:appx_particle_VSVGD_200p}
\end{figure}

\begin{figure}
    \centering
    \includegraphics[width=\textwidth]{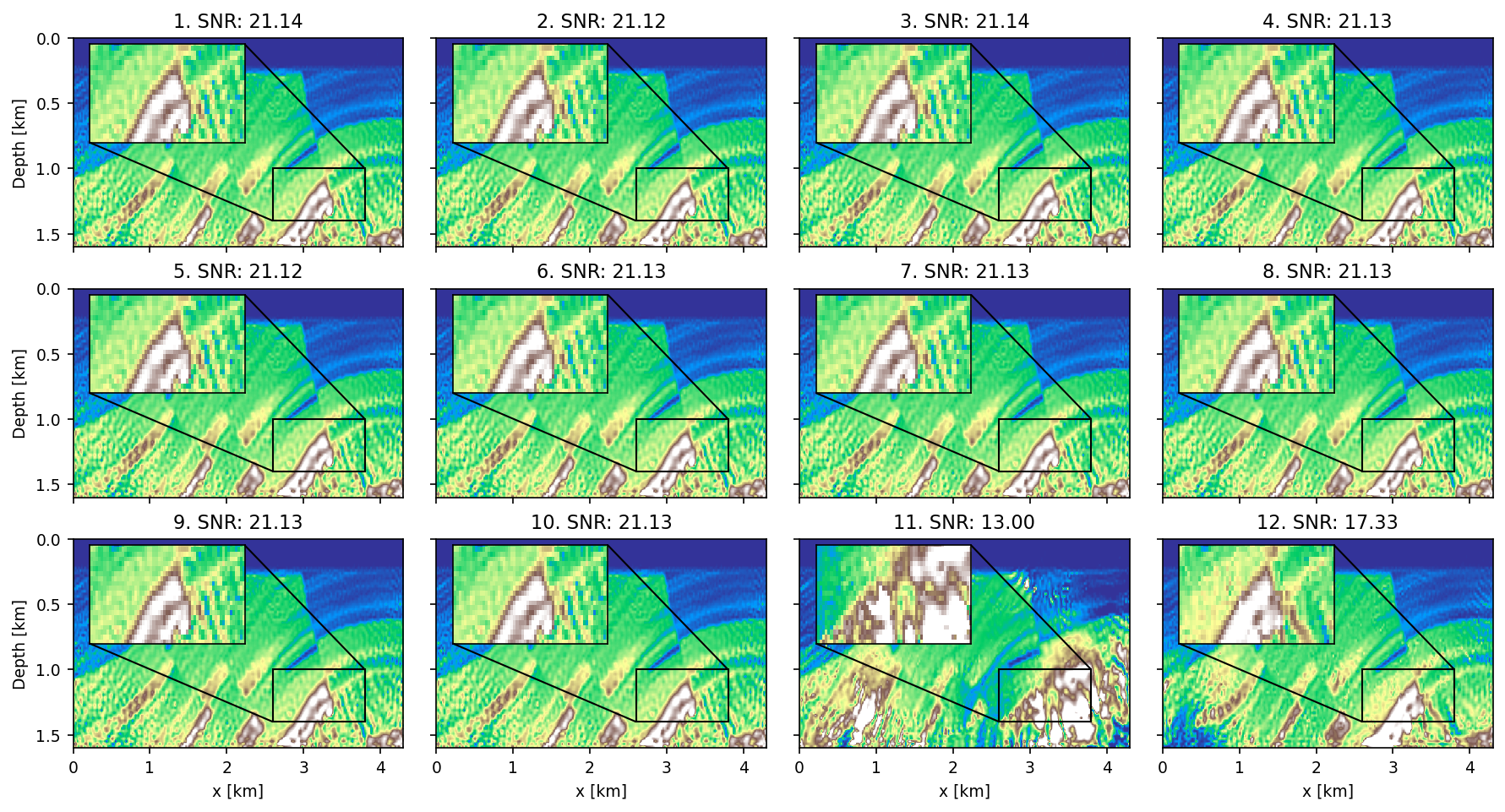}
    \vspace{-0.5cm}
    \caption{Single-frequency scenario: visualization of 12 particles from a 200-particle experiment using annealed SVGD (tanh) with RBF Kernel after 600 iterations.}
    \label{fig:appx_particle_ASVGD_tanh_200p}
\end{figure}

\begin{figure}
    \centering
    \includegraphics[width=\textwidth]{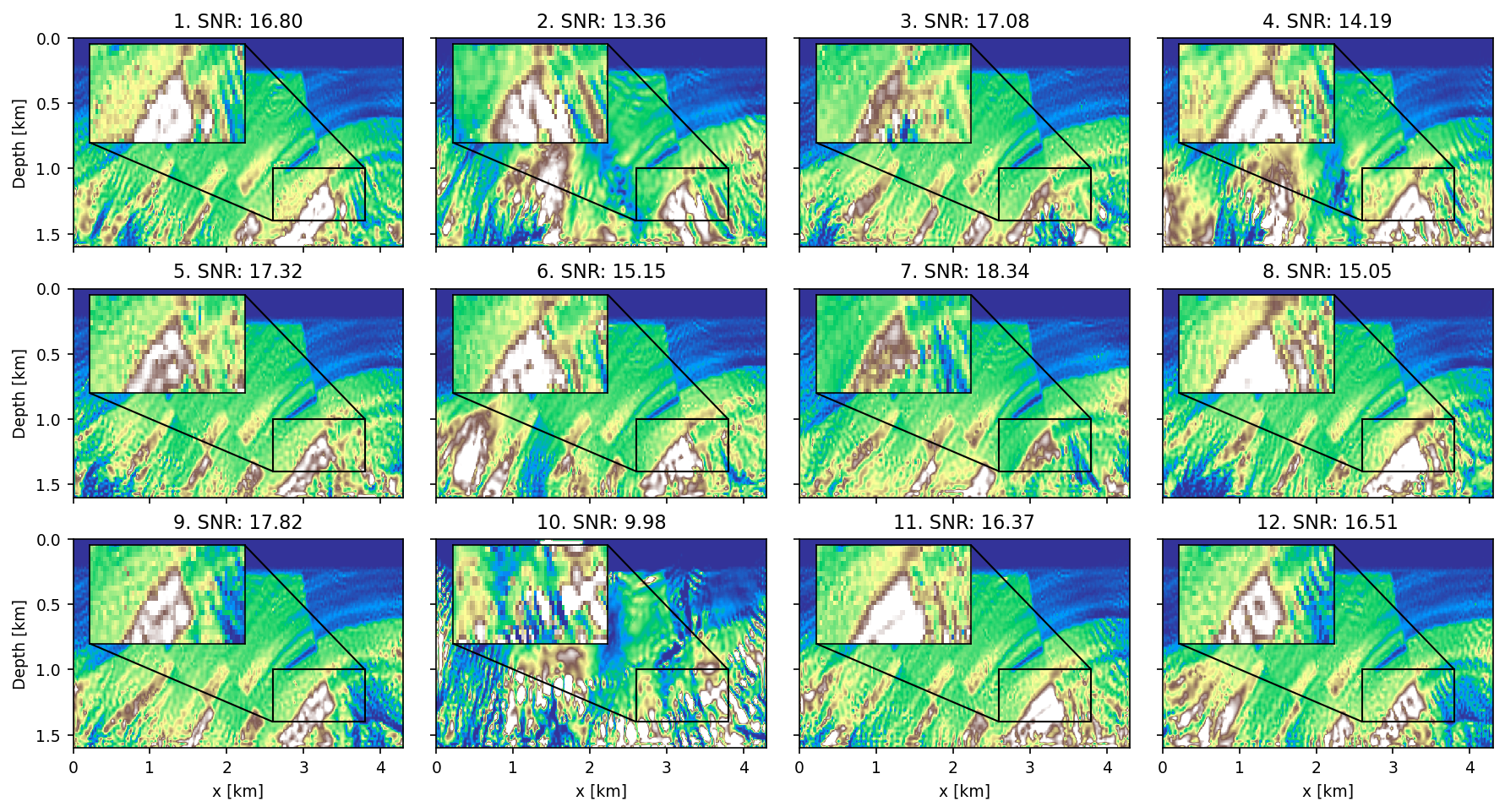}
    \vspace{-0.5cm}
    \caption{Single-frequency scenario: Visualization of 12 particles from a 200-particle experiment using Annealed SVGD, (tanh) with RBF kernel and constant bandwidth ($h=4000$) after 600 iterations}
    \label{fig:appx_particle_ASVGD_tanh_h4000_200p}
\end{figure}

\begin{figure}
    \centering
    \includegraphics[width=0.5\textwidth]{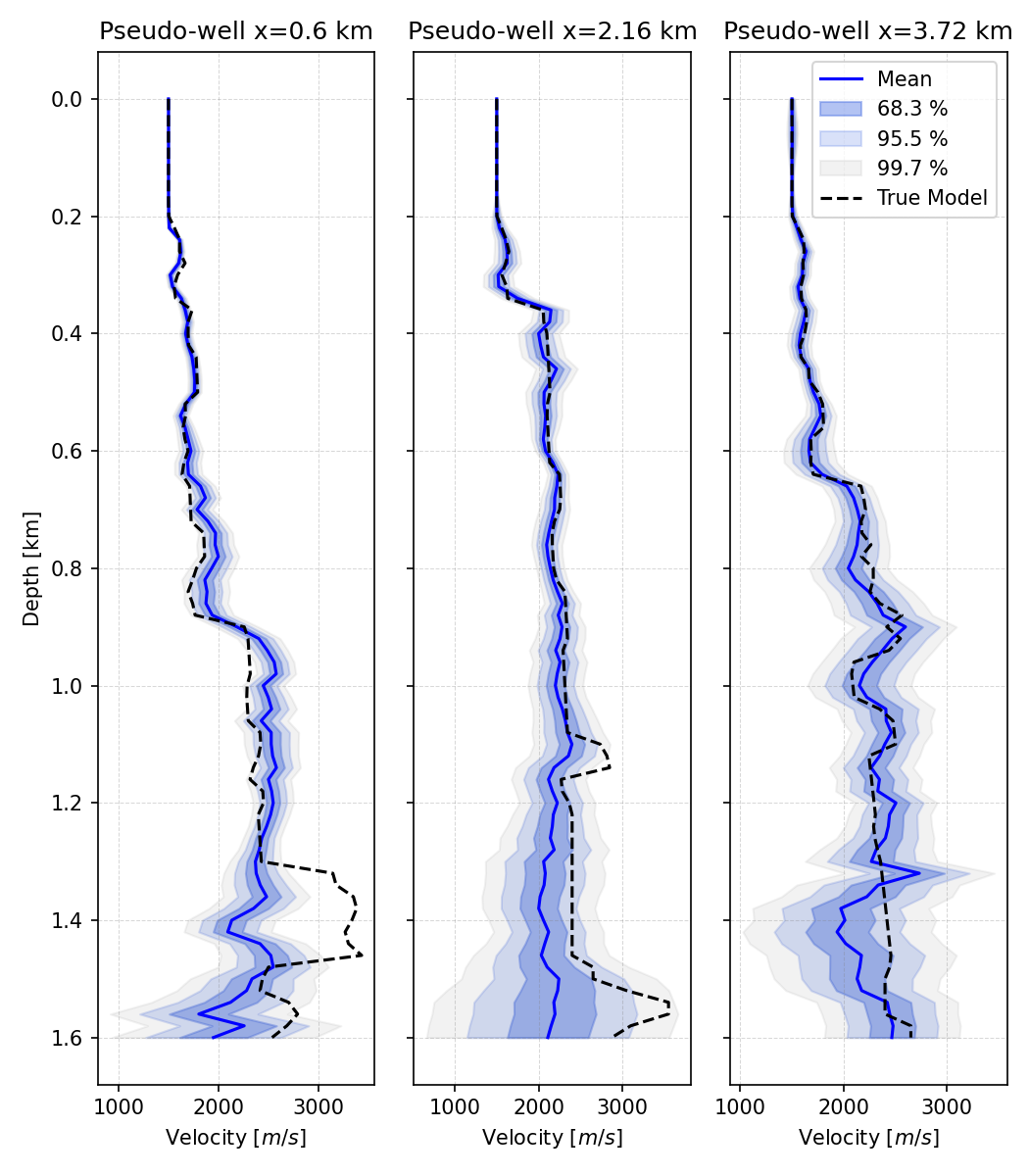}
    \vspace{-0.5cm}
    \caption{Single frequency scenario: pseudo-wells marginals from a 200-particle experiment using vanilla SVGD with RBF kernel and constant bandwidth ($h=4000$) 200 particles. }
    \label{fig:appx_marginal_wells_vanilla_h_4000}
\end{figure}

\begin{figure}
    \centering
    \includegraphics[width=0.5\textwidth]{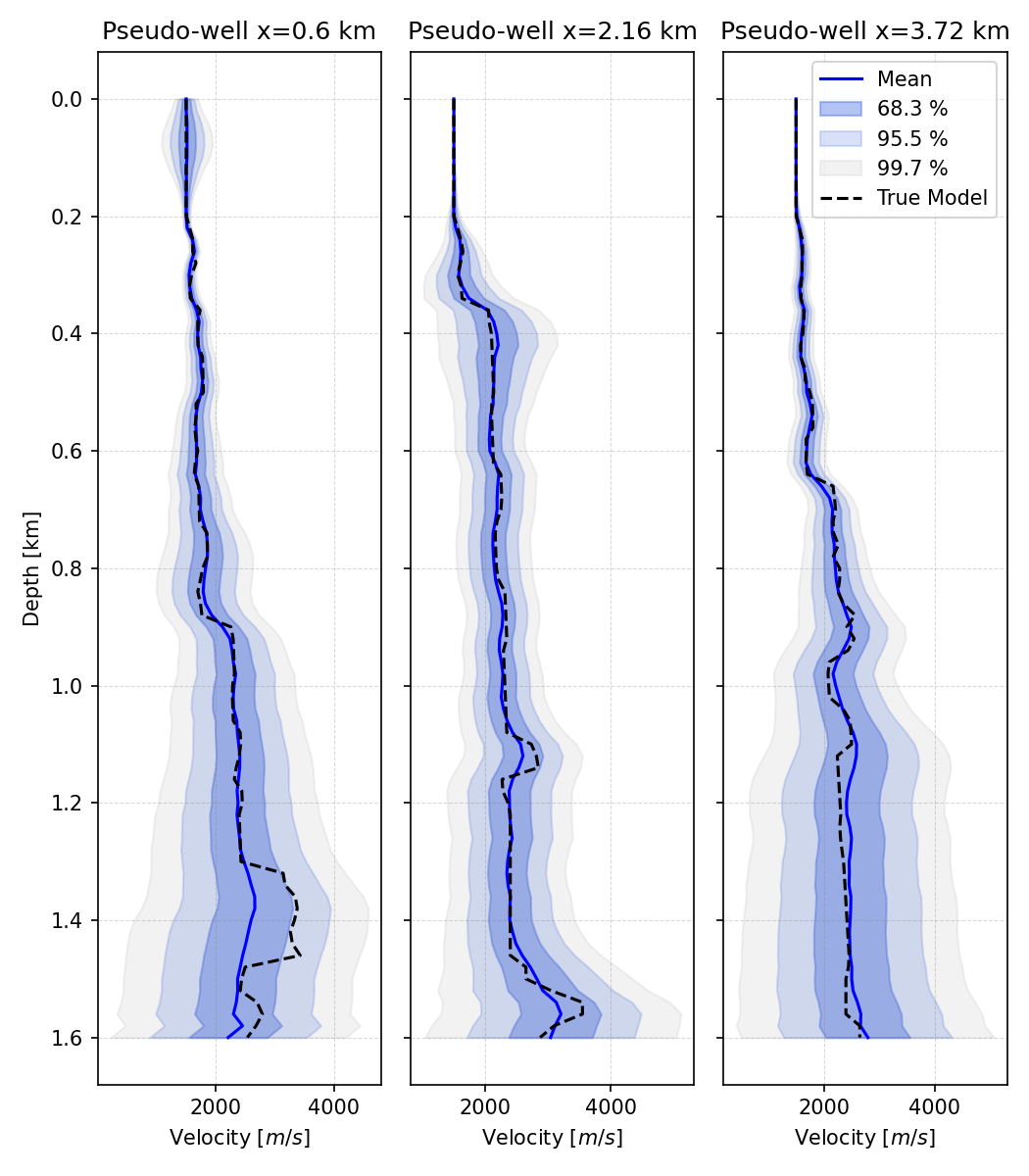}
    \vspace{-0.5cm}
    \caption{Single frequency scenario: pseudo-wells marginals from a 200-particle experiment using annealed SVGD (tanh) with RBF kernel and constant bandwidth ($h=4000$) 200 particles. }
    \label{fig:appx_marginal_wells_tanh_h_4000}
\end{figure}


\begin{figure}
    \centering
    \includegraphics[width=0.9\textwidth]{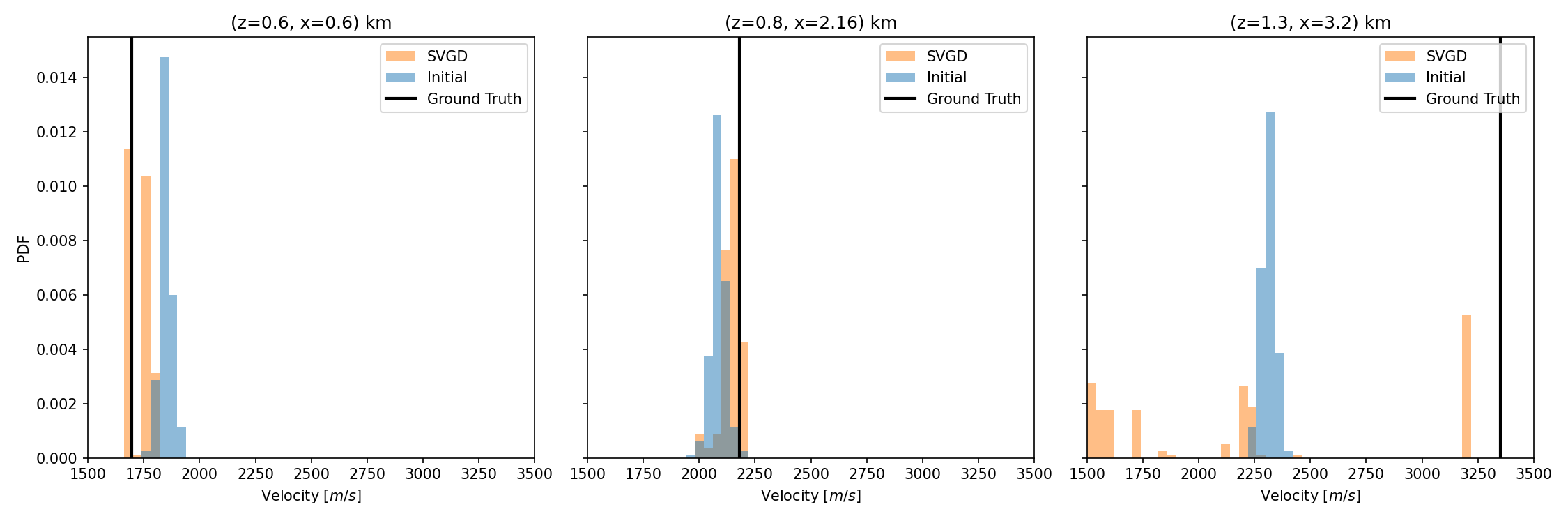}
    \vspace{-0.5cm}
    \caption{Single frequency scenario: pixels marginals for vanilla SVGD with RBF kernel and constant bandwidth ($h=4000$), and 200 particles. }
    \label{fig:appx_marginal_pixels_vanilla_h_4000}
\end{figure}


\begin{figure}
    \centering
    \includegraphics[width=0.9\textwidth]{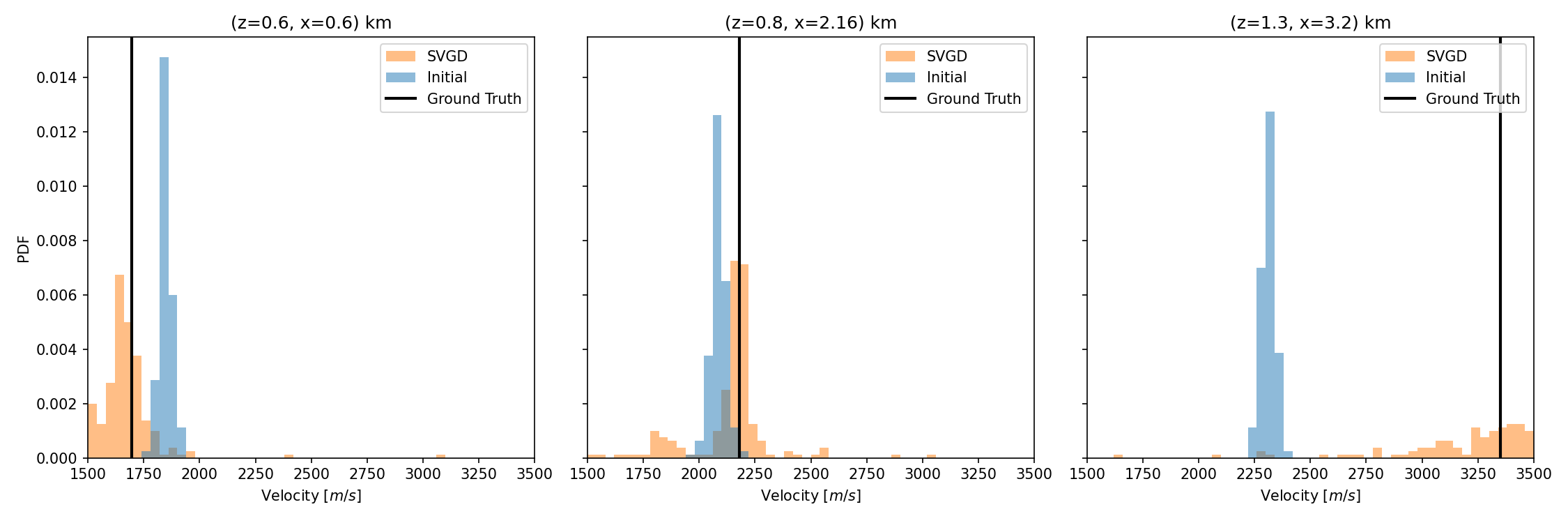}
    \vspace{-0.5cm}
    \caption{Single frequency scenario: pixels marginals for annealed SVGD (tanh) with RBF kernel and constant bandwidth ($h=4000$), and 200 particles.  }
    \label{fig:appx_marginal_pixels_tanh_h_4000}
\end{figure}

\pagebreak
\clearpage
\setcounter{figure}{0}
\section{\textsc{Supplementary Results for Multiscale Experiments}}\label{appendix_B}

\begin{figure}[!h]
    \centering
    \includegraphics[width=\textwidth]{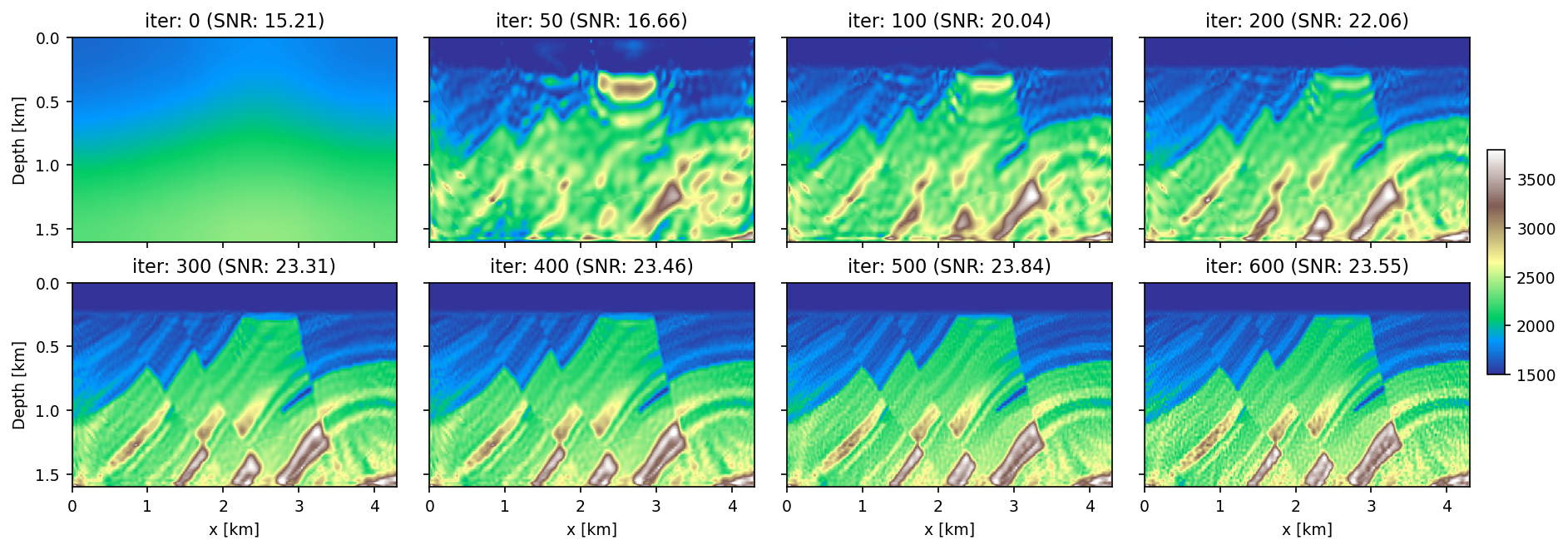}
    \vspace{-0.5cm}
    \caption{Multiscale scenario: Mean evolution for the 200-particle experiment using vanilla SVGD with RBF kernel.}
    \label{fig:Multi_scale_mean_evolution_200p_Vanilla}
\end{figure}

\begin{figure}
    \centering
    \includegraphics[width=\textwidth]{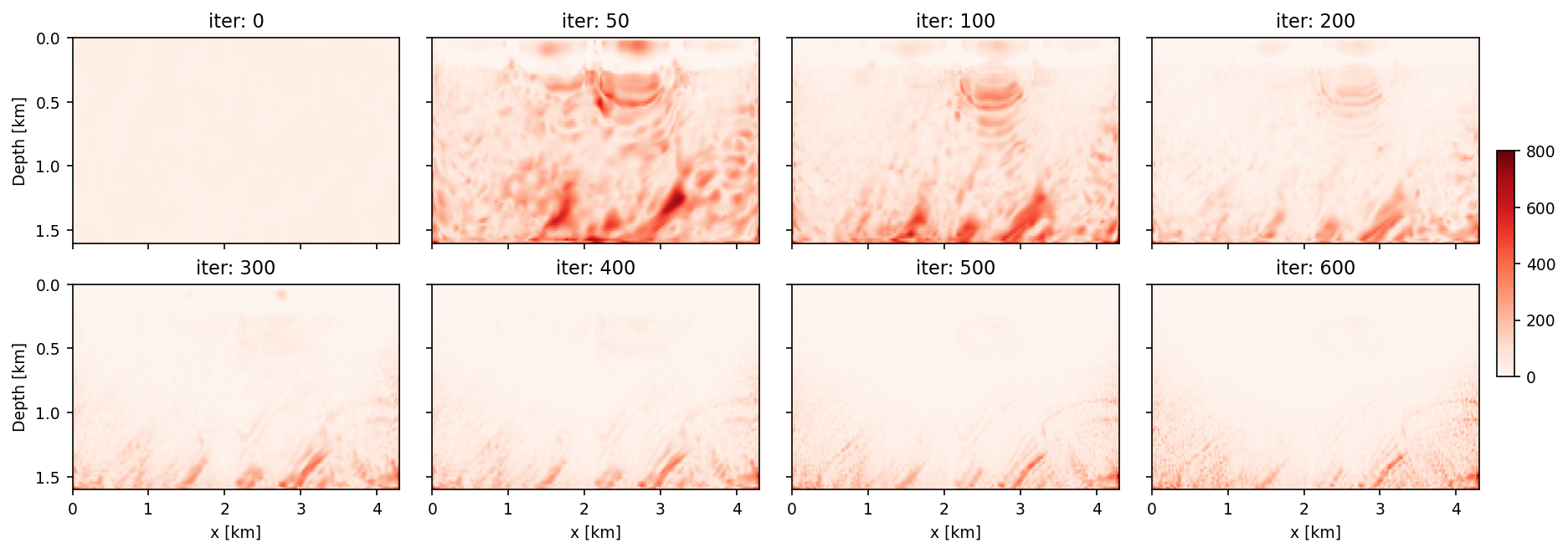}
    \vspace{-0.5cm}
    \caption{Multiscale scenario: Standard deviation evolution for the 200-particle experiment using vanilla SVGD with RBF kernel.}
    \label{fig:Multi_scale_std_evolution_200p_Vanilla_RBF}
\end{figure}

\begin{figure}
    \centering
    \includegraphics[width=\textwidth]{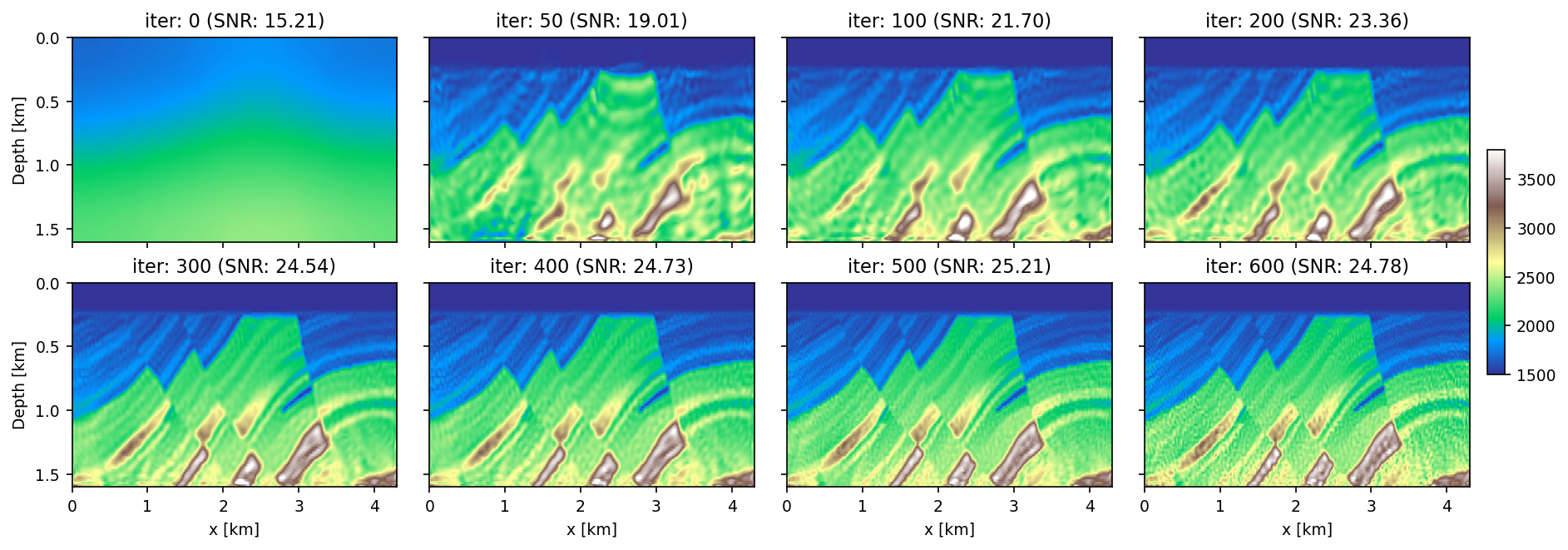}
    \vspace{-0.5cm}
    \caption{Multiscale scenario: Mean evolution for the 200-particle experiment using annealed SVGD (tanh) and RBF kernel.}
    \label{fig:Multi_scale_mean_evolution_200p_tanh_RBF}
\end{figure}

\begin{figure}
    \centering
    \includegraphics[width=\textwidth]{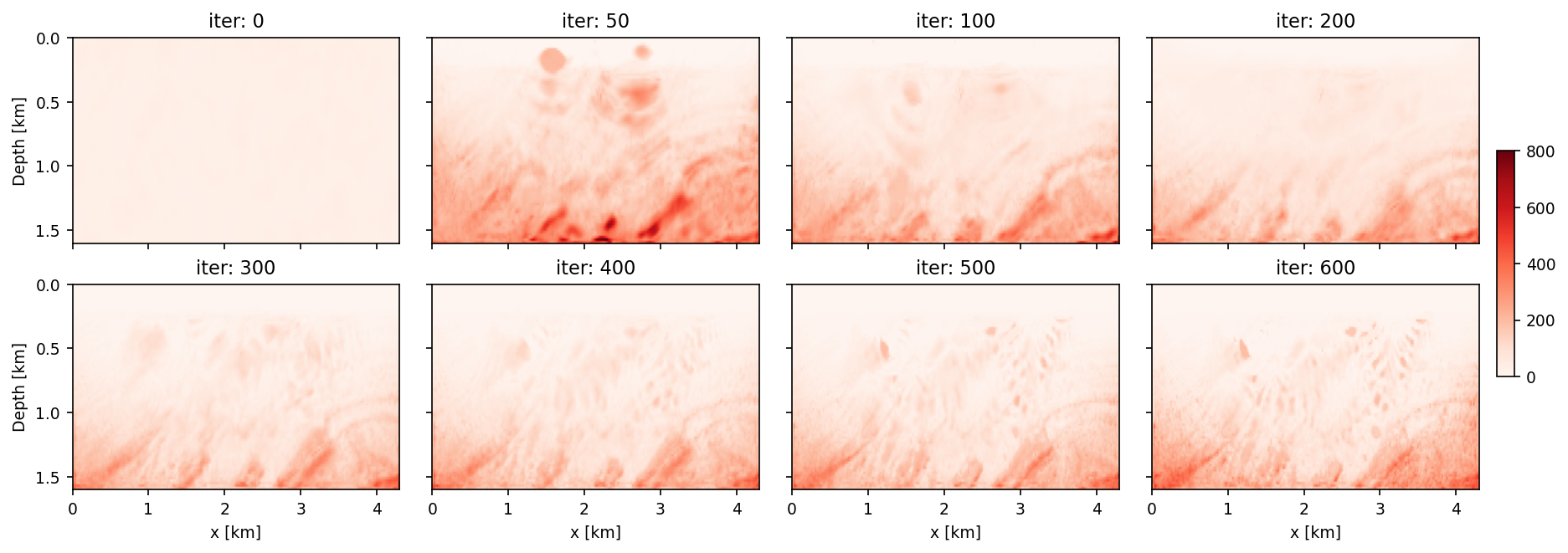}
    \vspace{-0.5cm}
    \caption{Multiscale scenario: Standard deviation evolution for the 200-particle experiment using annealed SVGD (tanh) and RBF kernel.}
    \label{fig:Multi-scale_std_evol_200p_tanh_RBF}
\end{figure}

\begin{figure}
    \centering
    \includegraphics[width=\textwidth]{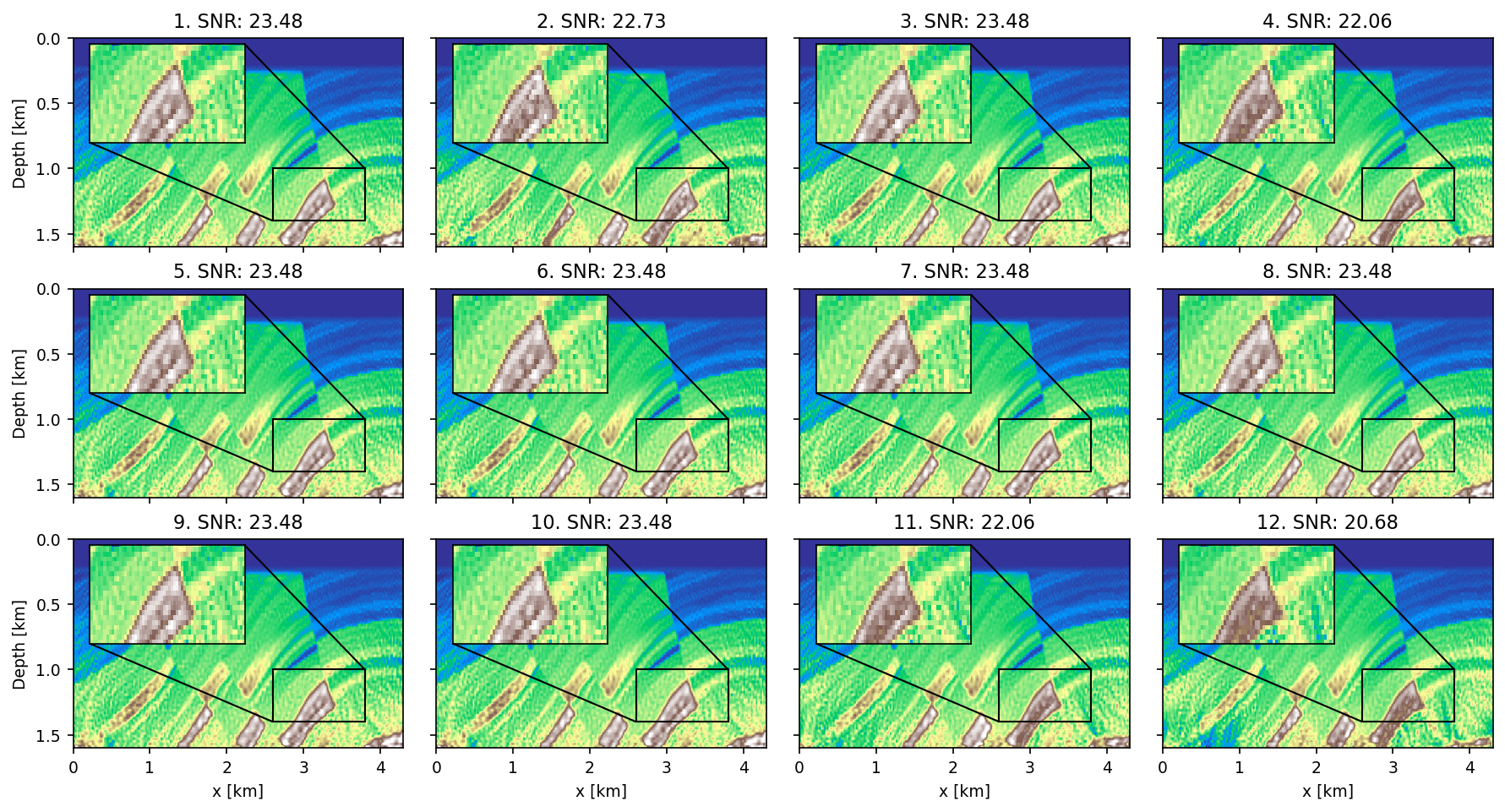}
    \vspace{-0.5cm}
    \caption{Multiscale scenario: visualization of 12 particles from a 200-particle experiment using vanilla SVGD with RBF Kernel after 600 iterations.}
    \label{fig:appx_particle_VSVGD_200p_multiscale}
\end{figure}

\begin{figure}
    \centering
    \includegraphics[width=\textwidth]{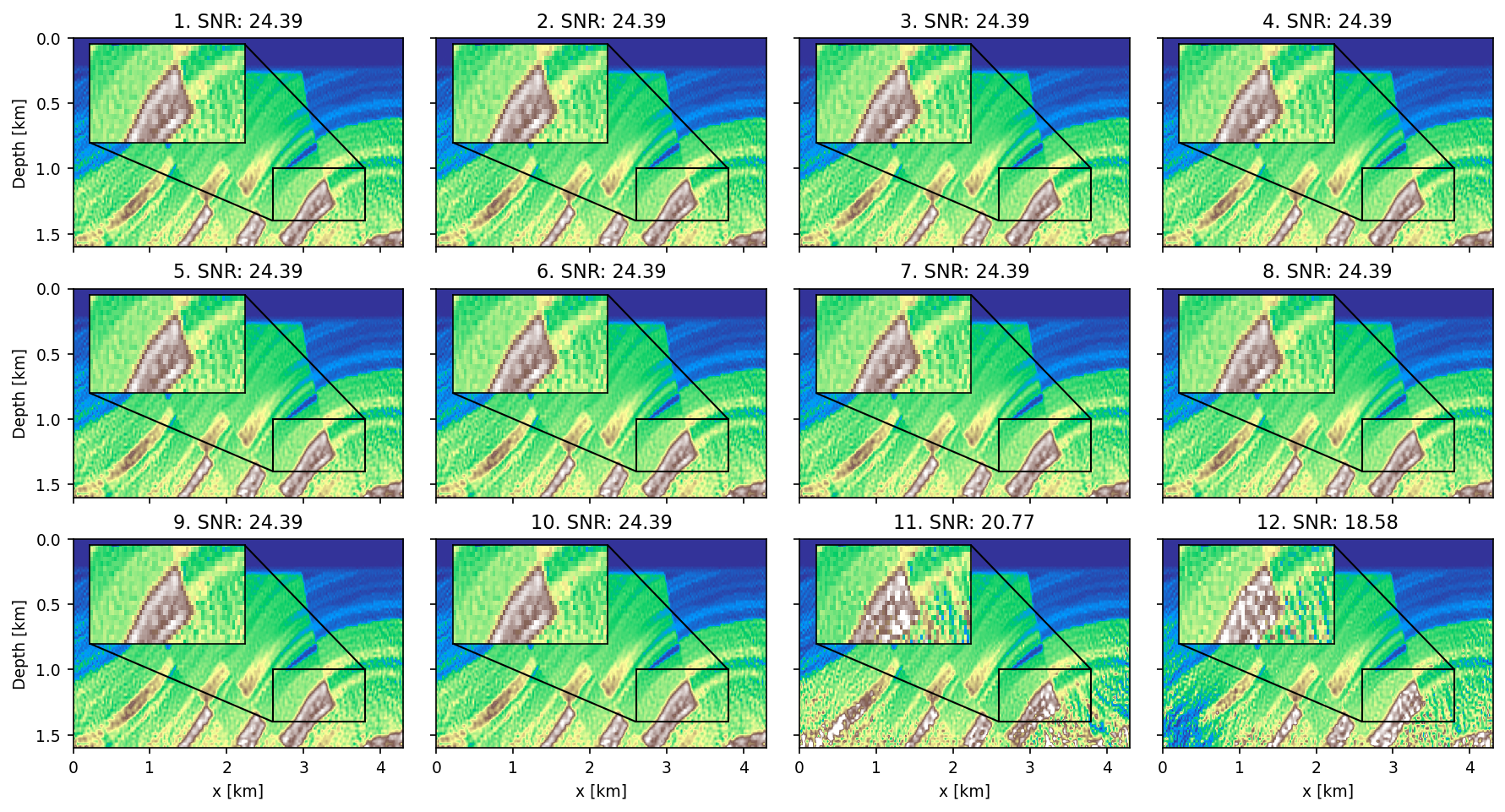}
    \vspace{-0.5cm}
    \caption{Multiscale scenario: visualization of 12 particles from a 200-particle experiment using annealed SVGD (tanh) with RBF Kernel after 600 iterations.}
    \label{fig:appx_particle_ASVGD_tanh_200p_multiscale}
\end{figure}

\begin{figure}
    \centering
    \includegraphics[width=\textwidth]{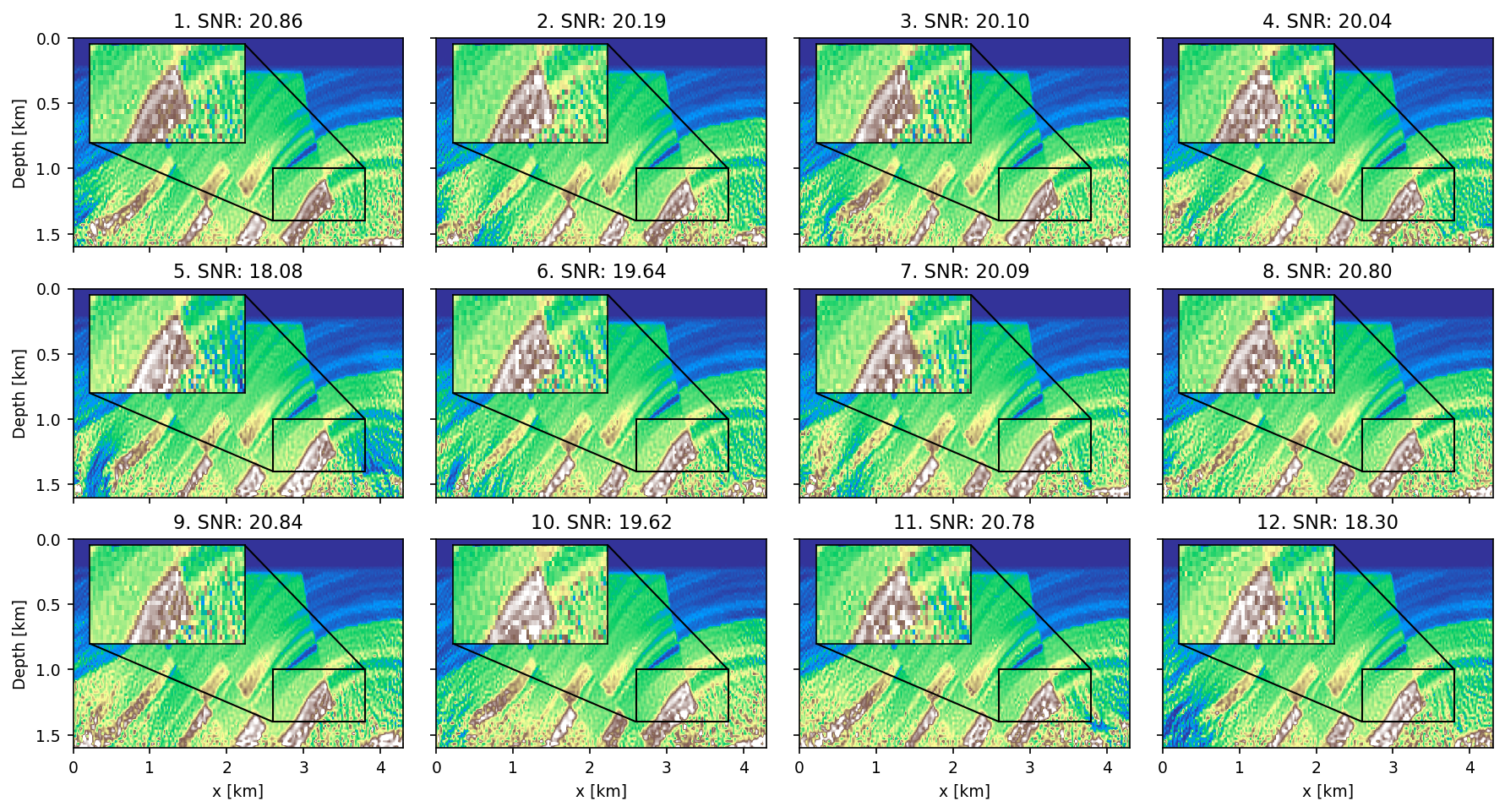}
    \vspace{-0.5cm}
    \caption{Multiscale scenario: visualization of 12 particles from a 200-particle experiment using annealed SVGD (tanh) with RBF kernel and constant bandwidth ($h=2500$) after 600 iterations.}
    \label{fig:appx_particle_ASVGD_tanh_h_2500_200p_multiscale}
\end{figure}

\begin{figure}
    \centering
    \includegraphics[width=0.5\textwidth]{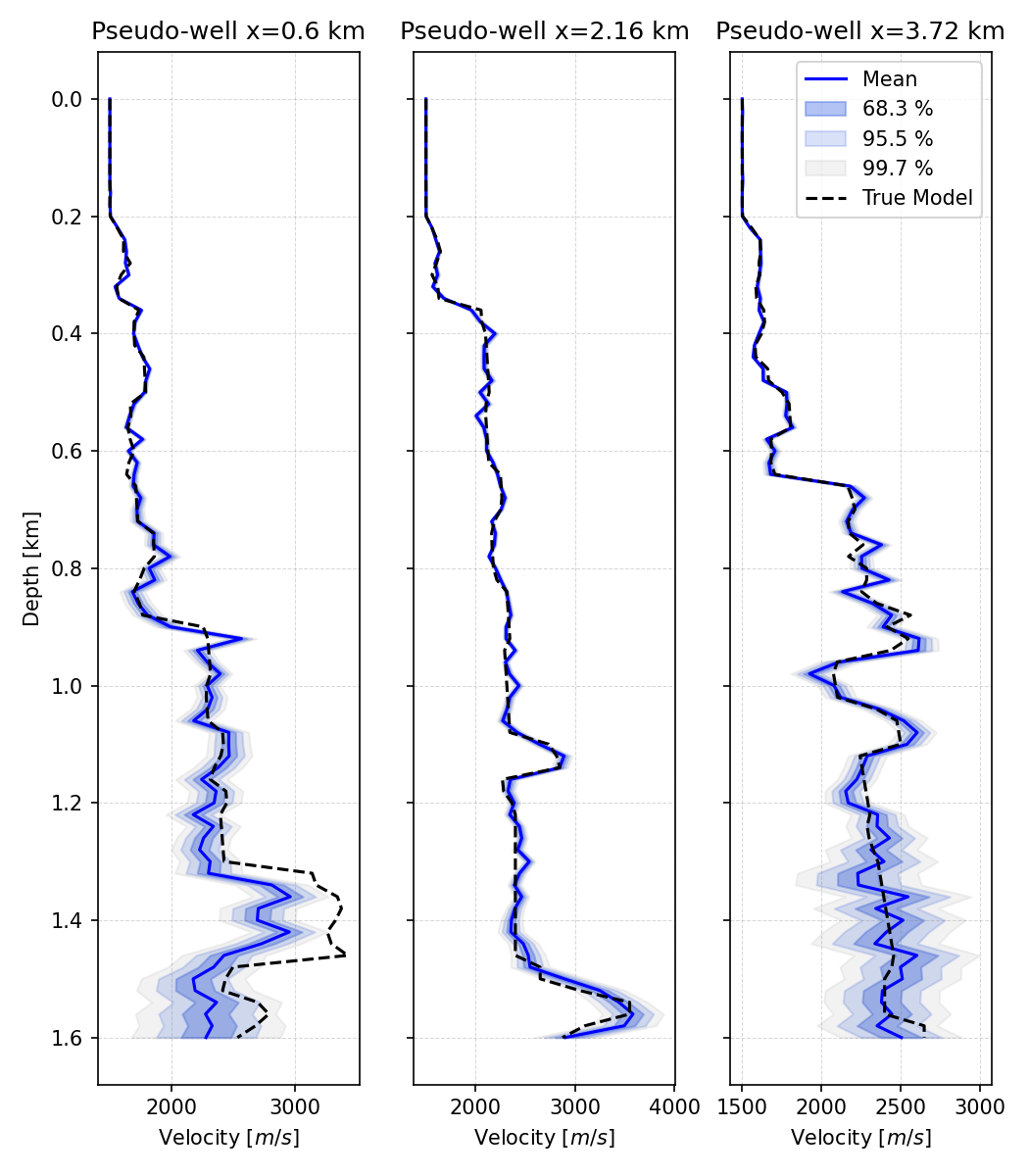}
    \vspace{-0.5cm}
    \caption{Multiscale scenario: pseudo-wells marginals from a 200-particle experiment using vanilla SVGD with RBF kernel.}
    \label{fig:multiscale_marginal_wells_vanilla_rbf}
\end{figure}

\begin{figure}
    \centering
    \includegraphics[width=0.5\textwidth]{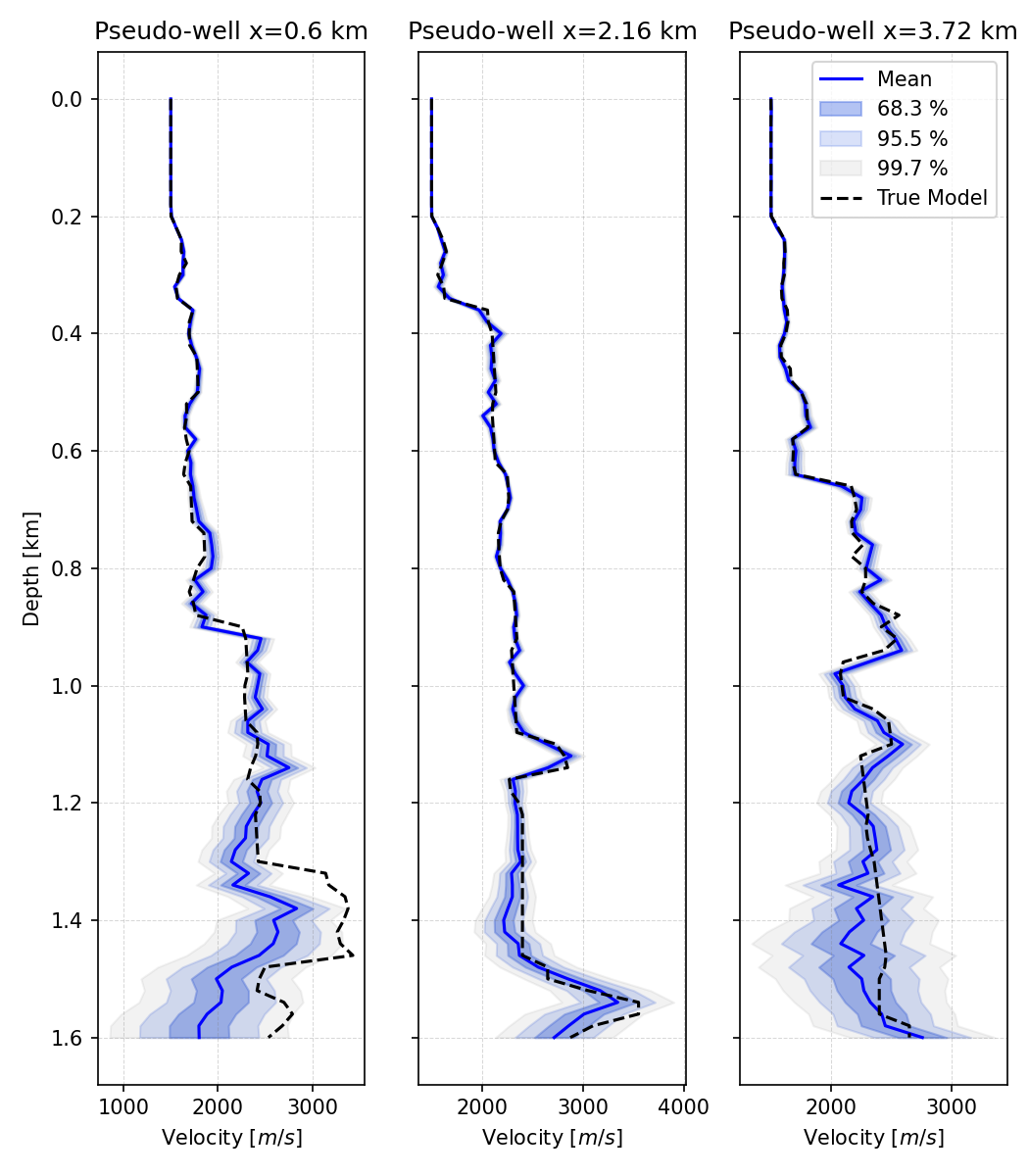}
    \vspace{-0.5cm}
    \caption{Multiscale scenario: pseudo-wells marginals from a 200-particle experiment using vanilla SVGD with RBF kernel and constant bandwidth ($h=2500$).}
    \label{fig:multiscale_marginal_wells_vanilla_h_2500}
\end{figure}

\begin{figure}
    \centering
    \includegraphics[width=0.5\textwidth]{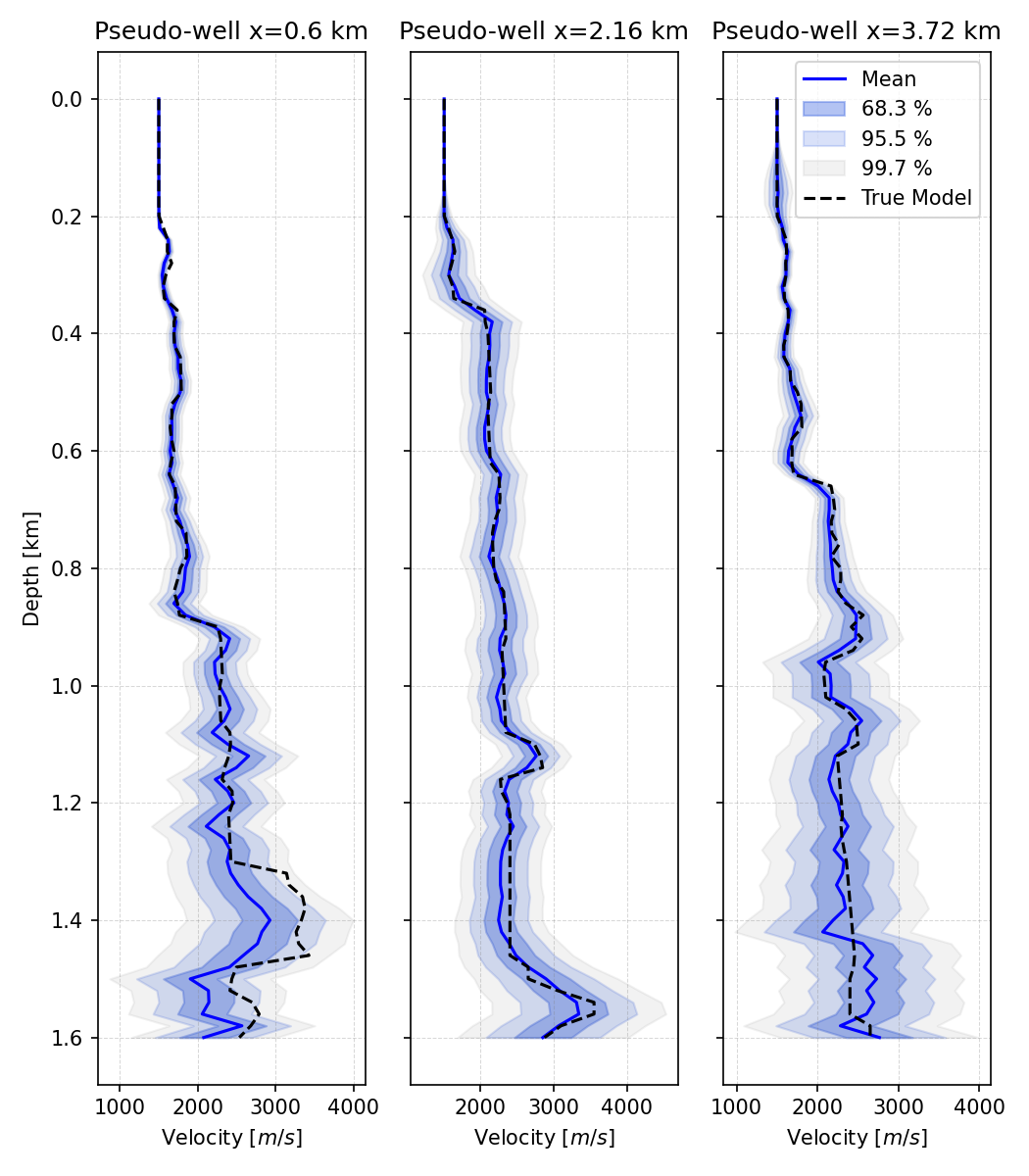}
    \vspace{-0.5cm}
    \caption{Multiscale scenario: pseudo-wells marginals from a 200-particle experiment using annealed SVGD (tanh) with RBF kernel. }
    \label{fig:multiscale_marginal_wells_tanh_rbf}
\end{figure}

\begin{figure}
    \centering
    \includegraphics[width=0.5\textwidth]{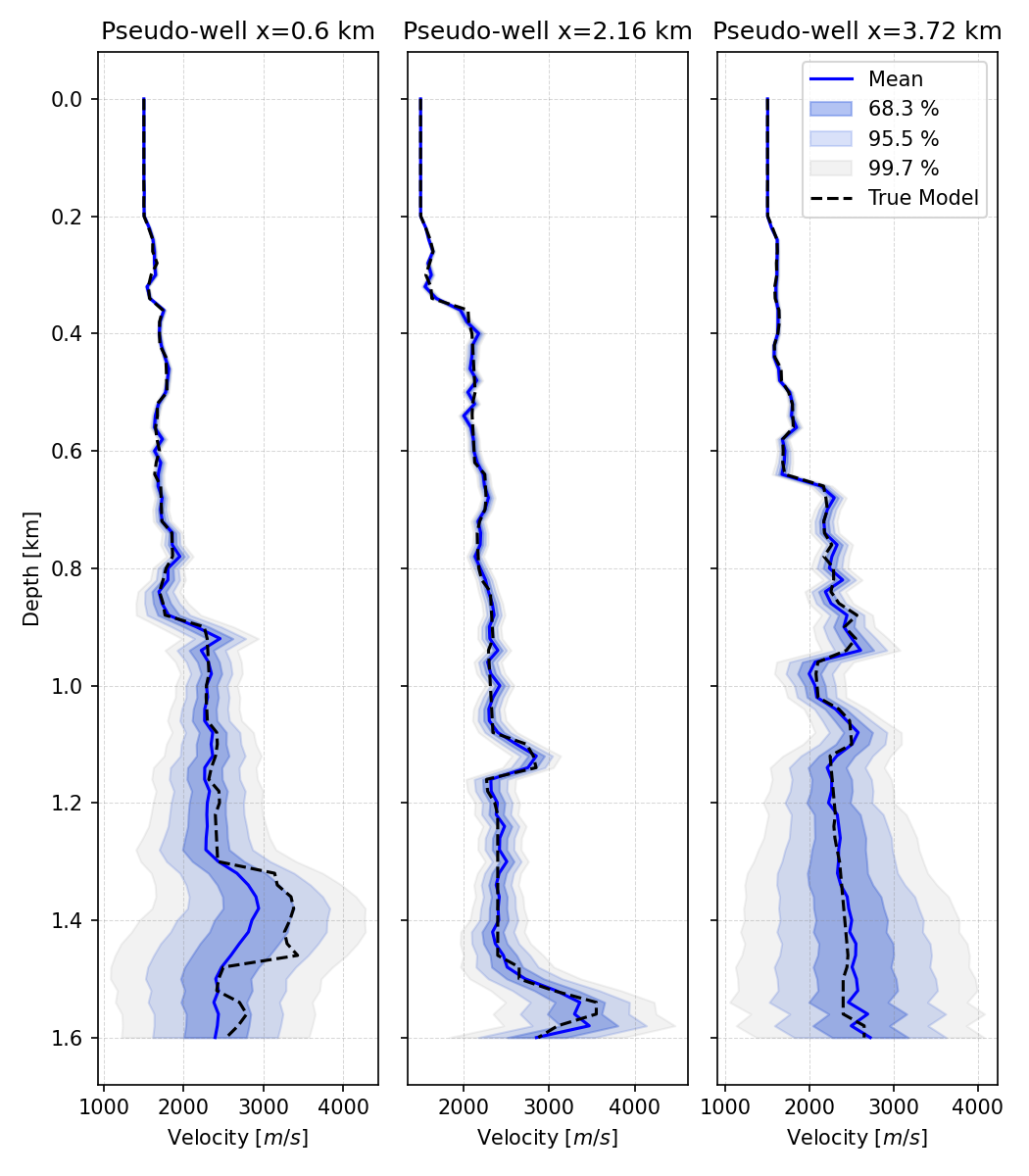}
    \vspace{-0.5cm}
    \caption{Multiscale scenario: pseudo-wells marginals from a 200-particle experiment using annealed SVGD (tanh) with RBF kernel and constant bandwidth ($h=2500$).}
    \label{fig:multiscale_marginal_wells_tanh_rbf_h_2500}
\end{figure}

\begin{figure}
    \centering
    \includegraphics[width=0.9\textwidth]{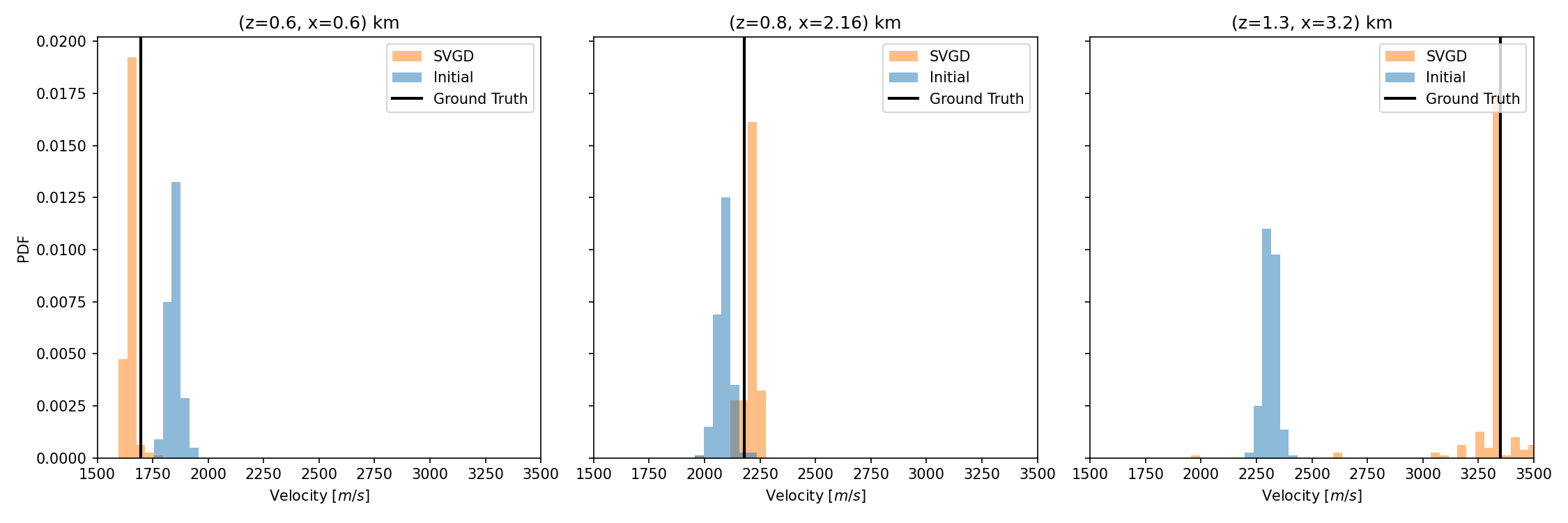}
    \vspace{-0.5cm}
    \caption{Multiscale scenario:, pixels marginals for vanilla SVGD with RBF kernel, 200 particles.}
    \label{fig:multiscale_appx_marginal_pixels_vanilla_rbf}
\end{figure}

\begin{figure}
    \centering
    \includegraphics[width=0.9\textwidth]{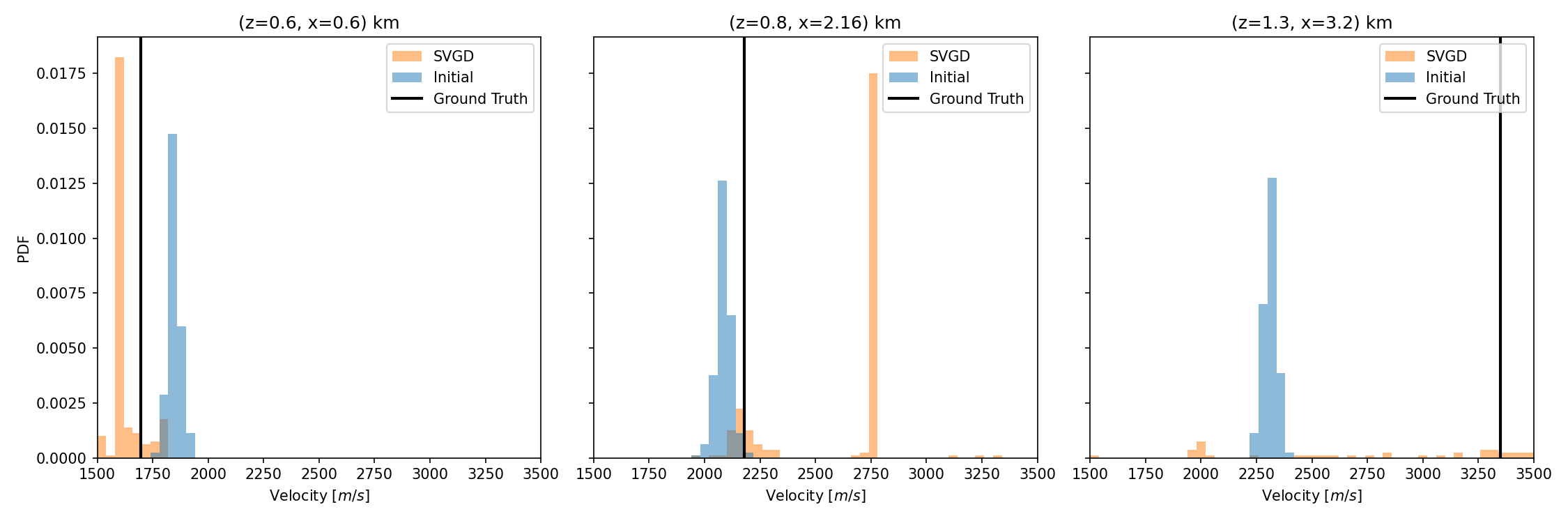}
    \vspace{-0.5cm}
    \caption{Multiscale scenario: pixels marginals for vanilla SVGD with RBF kernel and constant bandwidth ($h=2500$), 200 particles.}
    \label{fig:multiscale_appx_marginal_pixels_vanilla_h_2500}
\end{figure}

\begin{figure}
    \centering
    \includegraphics[width=0.9\textwidth]{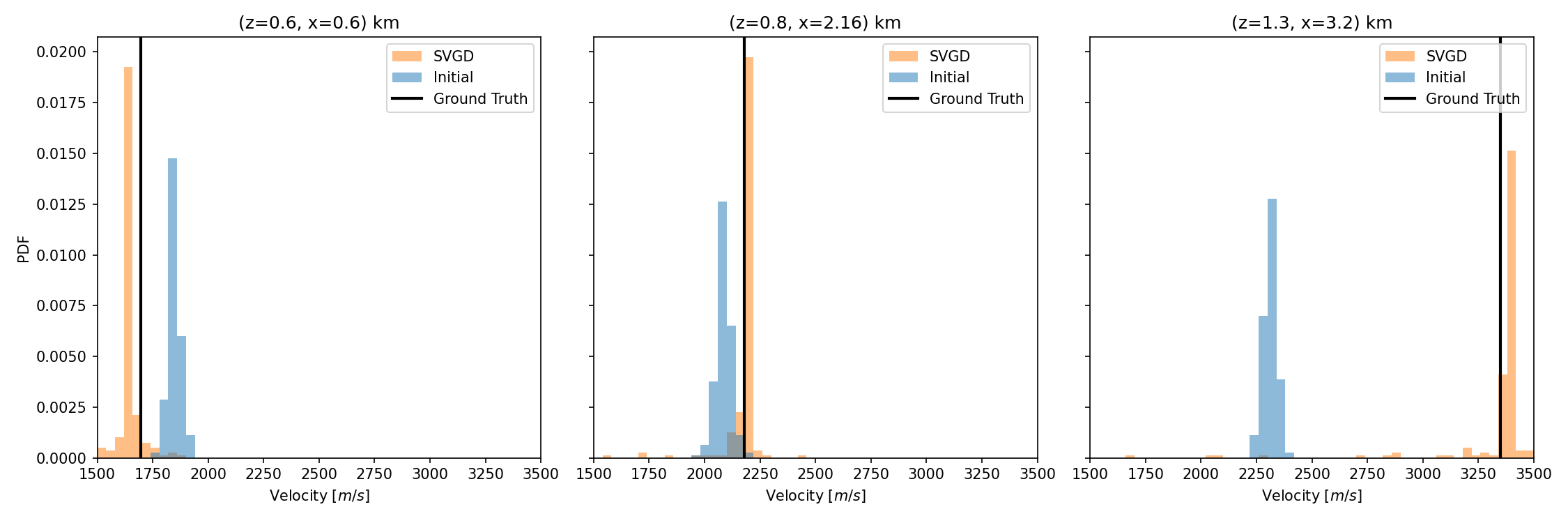}
    \vspace{-0.5cm}
    \caption{Multiscale scenario: pixels marginals for annealed SVGD (tanh) with RBF kernel, 200 particles.}
    \label{fig:multiscale_appx_marginal_pixels_tanh_rbf}
\end{figure}

\begin{figure}
    \centering
    \includegraphics[width=0.9\textwidth]{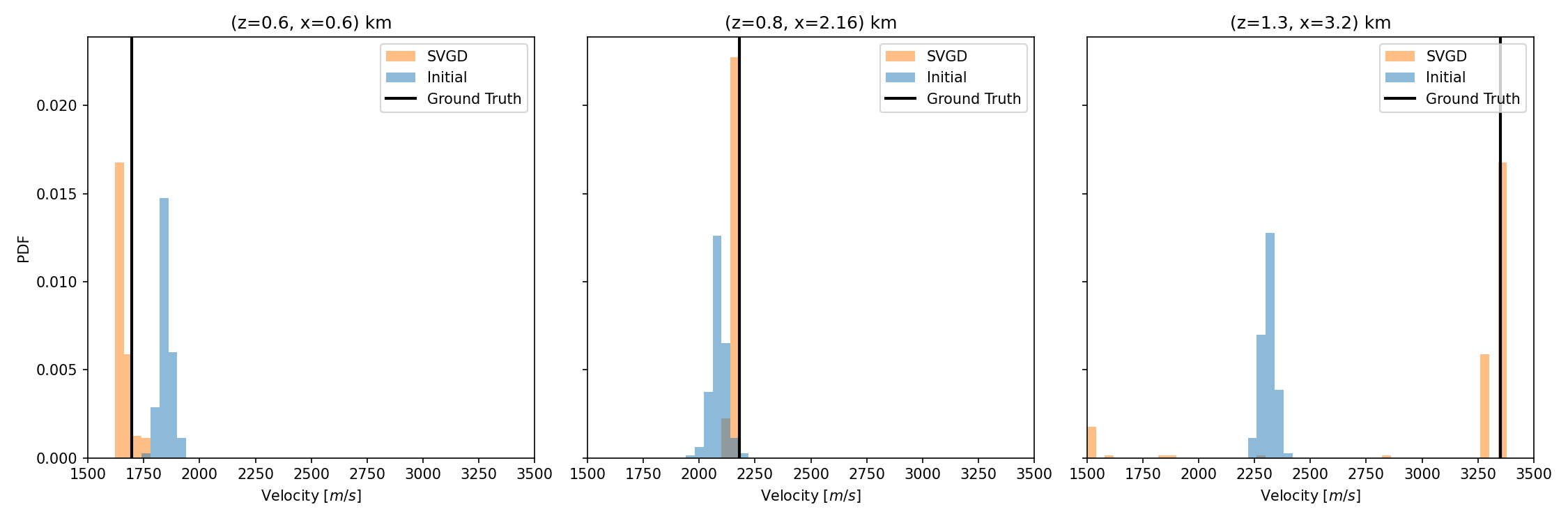}
    \vspace{-0.5cm}
    \caption{Multiscale scenario: pixels marginals for annealed SVGD (tanh) with RBF kernel and constant bandwidth ($h=2500$), 200 particles.}
    \label{fig:multiscale_appx_marginal_pixels_tanh_rbf_h_2500}
\end{figure}

\end{document}